\documentclass[fleqn,usenatbib]{mnras}

\usepackage{newtxtext,newtxmath}

\usepackage[T1]{fontenc}

\DeclareRobustCommand{\VAN}[3]{#2}
\let\VANthebibliography\thebibliography
\def\thebibliography{\DeclareRobustCommand{\VAN}[3]{##3}\VANthebibliography}

\usepackage{graphicx}	% Including figure files
\usepackage{amsmath}	% Advanced maths commands
\usepackage{amssymb}	% Extra maths symbols

\usepackage{xcolor}
\usepackage{hyperref}
\usepackage{siunitx}
\usepackage{xspace}
\usepackage{bm}

\usepackage{booktabs}

\usepackage{enumitem}
\usepackage{soul}
\usepackage{comment}

\newcommand{\tC}{\mathrm{C}}
\newcommand{\tA}{\mathrm{A}}
\newcommand{\bdw}{\boldsymbol{w }}
\newcommand{\bda}{\boldsymbol{a }}
\newcommand{\bdb}{\boldsymbol{b }}
\newcommand{\bde}{\boldsymbol{e }}
\newcommand{\bdT}{\boldsymbol{T }}
\newcommand{\bdE}{\boldsymbol{E }}

\newcommand{\diff}{\mathop{}\!\mathrm{d}}
\newcommand\Diff[1]{\mathop{}\!\mathrm{d^#1}}

\renewcommand{\vec}[1]{{\bm#1}}
\newcommand{\vers}[1]{\hat{\vec{#1}}}

\newcommand{\Planck}{\textit{Planck}\xspace}

\newcommand{\SW}{Sachs-Wolfe\xspace}

\newcommand{\fnl}{f_\text{NL}}
\newcommand{\fNL}{\fnl}

\title[$\mu E$ cross-correlations and non-Gaussianity]{Leverage on small-scale primordial non-Gaussianity through cross-correlations between CMB $E$-mode and $\mu$-distortion anisotropies}

\author[Remazeilles, Ravenni, Chluba]{
Mathieu Remazeilles$^{1,2}$\thanks{E-mail: \href{mailto:remazeilles@ifca.unican.es}{remazeilles@ifca.unican.es}},
Andrea Ravenni$^{2,3,4}$\thanks{E-mail: \href{mailto:andrea.ravenni@unipd.it}{andrea.ravenni@unipd.it}}
and Jens Chluba$^{2}$\thanks{E-mail: \href{mailto:jens.chluba@manchester.ac.uk}{jens.chluba@manchester.ac.uk}}
\\
$^1$Instituto de Física de Cantabria (CSIC-Universidad de Cantabria), Avda. de los Castros s/n, 39005 Santander, Spain.\\
$^2$Jodrell Bank Centre for Astrophysics, Department of Physics and Astronomy,
The University of Manchester, Manchester, M13 9PL, U.K.\\
$^3$Dipartimento di Fisica e Astronomia “G. Galilei”, Universit{\`a} degli Studi di Padova, via~Marzolo~8, I-35131, Padova, Italy.\\
$^4$INFN, Sezione di Padova, via~Marzolo~8, I-35131, Padova, Italy.
}

\date{\vspace{-5mm} Accepted XXX. Received YYY; in original form ZZZ}

\pubyear{2021}

\begin{document}
\label{firstpage}
\pagerange{\pageref{firstpage}--\pageref{lastpage}}
\maketitle

\begin{abstract}
Multi-field inflation models and non-Bunch-Davies vacuum initial conditions both predict sizeable \textit{non-Gaussian} primordial perturbations and \textit{anisotropic} $\mu$-type spectral distortions of the cosmic microwave background (CMB) blackbody. While CMB anisotropies allow us to probe non-Gaussianity at wavenumbers $k\simeq 0.05\,{\rm Mpc^{-1}}$, $\mu$-distortion anisotropies are related to non-Gaussianity of primordial perturbation modes with much larger wavenumbers, $k\simeq 740\,{\rm Mpc^{-1}}$. Through cross-correlations between CMB and $\mu$-distortion anisotropies, one can therefore shed light on the aforementioned inflation models. We investigate the ability of a future CMB satellite imager like \textit{LiteBIRD} to measure $\mu T$ and $\mu E$ cross-power spectra between anisotropic $\mu$-distortions and CMB temperature and $E$-mode polarization anisotropies in the presence of foregrounds, and derive \textit{LiteBIRD} forecasts on ${f_{\rm NL}^\mu(k\simeq 740\,{\rm Mpc^{-1}})}$. We show that $\mu E$ cross-correlations with CMB polarization provide more constraining power on $f_{\rm NL}^\mu$ than $\mu T$ cross-correlations in the presence of foregrounds, and the joint combination of $\mu T$ and $\mu E$ observables adds further leverage to the detection of small-scale primordial non-Gaussianity. For multi-field inflation, we find that \textit{LiteBIRD} would detect ${f_{\rm NL}^\mu}=4500$ at $5\sigma$ significance after foreground removal, and achieve a minimum error of ${\sigma(f_{\rm NL}^\mu=0) \simeq 800}$ at 68\% CL by combining CMB temperature and polarization. Due to the huge dynamic range of wavenumbers between CMB and $\mu$-distortion anisotropies, such large $f^\mu_{\rm NL}$ values would still be consistent with current CMB constraints in the case of very mild scale-dependence of primordial non-Gaussianity. Anisotropic spectral distortions thus provide a new path, complementary to CMB $B$-modes, to probe inflation with \textit{LiteBIRD}.
\end{abstract}

\begin{keywords}
cosmic background radiation -- polarization -- inflation -- early Universe -- methods: analytical
\end{keywords}

%%%%%%%%%%%%%%%%%%%%%%%%%%%%%%%%%%%%%%%%%%%%%%%%%%

%%%%%%%%%%%%%%%%% BODY OF PAPER %%%%%%%%%%%%%%%%%%

\section{Introduction}
Observing the Cosmic Microwave Background (CMB) has provided us precise information about the primordial perturbation field.
Its non-Gaussianity is tightly constrained by measurements of the 3- and 4-point correlation functions of the CMB Primary Anisotropies (PA), i.e. temperature and $E$-mode polarization fluctuations in the sky generated at the time of recombination \citep{Planck_2018_results_IX}.
However, another property of the CMB also tracks primordial perturbations: its energy spectrum.
Damping of acoustic modes in the pre-recombination era introduces heat in the photon-baryon plasma \citep{Sunyaev1970, Hu:1994bz, Daly1991, Chluba20122x2}.
After double Compton scattering and Bremsstrahlung become ineffective at $z\simeq\SI{2e6}{}$ the plasma lacks the ability to equilibrate the number density of photons.
This leads to a distortion of the CMB spectrum that can be, in principle, observed today \citep{Chluba:2019nxa}.

While other physical processes such as recombination \citep{Hart:2020voa} and photon injection \citep{Bolliet:2020ofj} imprint the CMB spectrum with peculiar spectral signatures, the effect of heat dissipation (after $z\simeq\SI{2e6}{}$) can be successfully described by the amplitude of two specific spectral shapes: $\mu$- and $y$-distortions.
The first is generated solely in the primordial universe, before $z \simeq \SI{5e4}{}$, when the photons, having acquired an \emph{effective} chemical potential $\mu$, distribute like a Bose-Einstein rather than a Planck distribution \citep[e.g.,][]{Sunyaev1970, Burigana1991, Hu1993, Chluba:2011hw, Sunyaev2013, Lucca2020}.
Compton-$y$ distortions, achieved when energy is not effectively redistributed, are generated throughout the rest of the cosmic history, both pre- and post-recombination.
While in the pre-recombination era $y$-distortions can be produced by the same mechanism generating $\mu$-distortions, they also receive important contributions in the late universe from the Sunyaev-Zeldovich effect \citep{Mroczkowski:2018nrv} and from reionization \citep{Hu:1993tc, Pitrou:2009bc}.
Here we only consider $\mu$-distortions because they are both a more powerful probe of primordial non-Gaussianity and because one does not need to account for biases due to late-time secondary sources.

If the primordial perturbation field is non-Gaussian, spectral distortions can be spatially modulated, which leads, after projection on the sphere, to Spectral Distortion Anisotropies (SDA).
If we consider the local model of non-Gaussianity, which peaks in the squeezed limit, the (small-scale) modulation in power is imprinted by a long-wavelength mode.
CMB PA are a high signal-to-noise tracer of said long-wavelength mode, thus cross-correlating PA and SDA one can hope to achieve strong constraints on all models of primordial non-Gaussianity with a large squeezed contribution.
This avenue was opened by \cite{Pajer:2012vz, Ganc:2012ae} and further extended and quantitatively refined in \cite{Emami:2015xqa, Ota:2016mqd, Chluba:2016aln, Ravenni:2017lgw, Cabass:2018jgj}. Recently it has been shown how the same idea can be applied to models involving primordial black holes \citep{Ozsoy:2021qrg, Zegeye2021} and to models in which scalar and tensor perturbations are correlated \citep{Orlando:2021nkv}.
While fainter than the monopole signal, spectral distortions anisotropies do not require \citep{Ganc:2012ae} the use of a purposefully-built absolutely-calibrated spectrometer \citep{Chluba:2019nxa}, but can be extracted from differential measurement carried out by an imager \citep[e.g.][]{Litebird2019}.
Moreover, cross-correlating them with much more intense signals such as the CMB temperature and polarization, drastically increases the signal-to-noise ratio, making a detection possible even with the next generation of satellites \citep{Remazeilles2018}.
As is well known \citep{Pajer:2012vz, Ganc:2012ae}, bulk of the information is contained in the lowest multipoles, with the signal-to-noise ratio decaying quickly towards smaller angular scales.
Thus, for a fixed sensitivity, ample sky coverage is more important than high angular resolution, motivating our focus on satellite rather than ground-based telescope observations.

In \cite{Remazeilles2018} for the first time we performed realistic forecasts by analysing synthetic datasets including foregrounds and the effect of detectors, focusing the analysis on CMB temperature and $\mu$ maps.
As shown in \cite{Ravenni:2017lgw}, $E$-mode polarization adds valuable information, and in fact its use provides tighter constraints than the temperature.
The reasons for this are twofold. 
$E$-modes do not receive sizeable contributions in the late universe beside the low-$\ell$ bump due to reionization, whereas the integrated Sachs-Wolfe effect does not correlate with primordial signals but still adds to the temperature anisotropy.
Moreover, temperature foregrounds are more complex than polarized ones.
While this is not a challenge if we wish to recover just temperature or polarization anisotropies, it makes a difference when targeting a non-trivial spectral component (i.e., $\mu$ distortions).
In this paper we investigate these claims, showing how $\mu E$ cross-correlations offer tighter and more reliable estimates of non-Gaussianity compared to $\mu T$, and forecast the constraint given by the combination of both probes. 

This paper is organised as follows. In Sect.~\ref{sec:theory}, we introduce the theoretical model that we wish to test and we model the observables. In Sec.~\ref{sec:sky_sim} we describe the sky simulations that we analyse in this work. In Sec.~\ref{sec:clarification} we dedicate a detailed discussion on the expected noise curves for $\mu$-distortion anisotropies. In Sec.~\ref{sec:forecast} we describe our pipeline and component separation methodology, and show our results, that we further discuss in Sec.~\ref{sec:conclusions}, where we conclude.

\vspace{-4mm}
\section{Theoretical modelling}
\label{sec:theory}
The primordial perturbation field can be described in terms of its $n$-point correlation functions in Fourier space.
The two- and three-point functions, the power spectrum $P(k)$ and bispectrum $B(k_1, k_2, k_3)$, are defined as 
\begin{align}
    \langle
        \zeta_\vec{k_1} \zeta_\vec{k_2}
    \rangle
    = \ &
    (2\pi)^3
    \delta^{(3)}(\vec{k_1} + \vec{k_2}) 
    P_\zeta(k)\, ,
\\
    \langle
        \zeta_\vec{k_1} \zeta_\vec{k_2}  \zeta_\vec{k_3}
    \rangle
    = \ &
    (2\pi)^3
    \delta^{(3)}(\vec{k_1} + \vec{k_2} + \vec{k_3}) 
    B_\zeta(k_1, k_2, k_3) \, .
\end{align}
The bispectrum shape varies depending on the class of models one considers.
Here we are interested in shapes that peak on squeezed configurations ($k_1 \ll k_2 \approx k_3$), as those are the ones PA-SDA correlations are most sensitive to.
The most studied model peaking in this limit is the local bispectrum \citep[e.g.,][]{Gangui:1993tt, Verde:1999ij, Komatsu:2001rj}, 
\begin{equation}
    B_\zeta^\text{loc}
    (k_1, k_2, k_3)
    = 
    \frac{6}{5} \fnl
    \left[
        P_\zeta(k_1) \, P_\zeta(k_2)
        + 2 \text{ perms.}
    \right] ,
\end{equation}
which naturally arises from non linearity of the perturbations in real space.
It is especially important because it could allow us to discern multi-field (which predict $\fnl \gtrsim 1$) from single-field inflation \citep[see][for an extended list of references]{Planck_2018_results_IX}.

The local non-Gaussianity parameter constraint set by \cite{Planck_2018_results_IX} --- $\fnl=-0.9 \pm 5.1$ --- has been achieved measuring the bispectrum of CMB PA, which probe primordial perturbation modes with typical wavenumbers $k_0\simeq 0.05\,{\rm Mpc}^{-1}$.
Therefore, the \Planck constraint can be thought as being valid on those scales.
In contrast, $\mu$-distortion anisotropies are generated by non-Gaussian perturbation modes with much larger wavenumbers $k\simeq 740\,{\rm Mpc}^{-1}$, therefore they would lead to $\fnl$ constraints in a vastly different regime \citep{Emami:2015xqa}.
In fact, $\fNL$ does not need to be constant in general; various models predict some scale dependence \citep{Dimastrogiovanni:2016aul, Byrnes:2010ft, 2011JCAP...03..017S, Chen:2005fe}.
While this is well established, explicit expressions are generally provided only through expansions in powers of $\ln(k/k_{\rm p})$ about some pivot scale $k_{\rm p}$.
The coefficient of this expansion are related to the hierarchy of slow roll parameters, and thus are small.
However, when we consider cross-correlations of PA and SDA, we are effectively integrating over an extremely large range of scales ($k \in  [\simeq \SI{d-4}{Mpc}^{-1}, \simeq \SI{d4}{Mpc}^{-1}]$) so that expansions in $\ln(k/k_{\rm p}) = \mathcal{O}(10)$ might not converge in general.
Bearing in mind this caveat \citep[see also the discussion in][]{Planck_2018_results_IX}, it is still instructive to notice that assuming even a mild scale-dependence of primordial non-Gaussianity, e.g. $f_{\rm NL}(k)= f_{\rm NL}(k_0)\left(k/k_0\right)^{n_{\rm NL}}$ with $n_{\rm NL}\lesssim 0.7$, leads to $f_{\rm NL}(k\simeq 740\,{\rm Mpc}^{-1}) > 4000$ at $\mu$-distortion scales, while still being consistent with $f_{\rm NL}(k_0\simeq 0.05\,{\rm Mpc}^{-1})\simeq 5$ at CMB scales.
As customary \citep{Pajer:2012vz, Ganc:2012ae, Emami:2015xqa}, here we use a phenomenological approach and assume $\fNL(k_1, k_2, k_3) = \fnl^\mu = \text{constant}$ when $k_1$ spans the range of PA scales, and $k_2, k_3$ span the SD scales.

The recipe to calculate power spectra and cross-correlations taking properly into account transfer effects of both PA and SDA is well established \citep{Pajer:2012vz, Ganc:2012ae, Chluba:2016aln, Shiraishi:2015lma, Ravenni:2017lgw}.
To fix the notation, let us define the harmonic coefficients of the PA field $X = T, E$ as
\begin{equation}
    a_{\ell m}^{X}
    =
    (-i)^\ell 4\pi \!
    \int \frac{\Diff{3} \vec{k}}{(2\pi)^3}
    \mathcal{T}_\ell^{X/\zeta}(k)
    \zeta_\vec{k}
    Y_{\ell m}^*(\vers{k}) \, ,
\end{equation}
and similarly the harmonic coefficients of the $\mu$ field as
\begin{equation}
\begin{split}
    a_{\ell m}^{\mu}
    =
    (-i)^\ell 4\pi \!
    \int 
    &
    \frac{\Diff{3} \vec{k_1}}{(2\pi)^3}
    \frac{\Diff{3} \vec{k_2}}{(2\pi)^3}
    \Diff{3} \vec{k_3} \,
    \delta^{(3)}(\vec{k_1}+\vec{k_2}+\vec{k_3})
\\
    &
    \zeta_\vec{k_1} \zeta_\vec{k_2}
    f^\mu(k_1, k_2, k_3)
    j_\ell(k \, r_\text{ls})
    Y_{\ell m}^*(-\vers{k_3}) \, .
\end{split}
\end{equation}
In the last two equations $\mathcal{T}_\ell^{X/\zeta}(k)$ and $f^\mu(k_1, k_2, k_3)$ are respectively the PA transfer function and the SDA window function that relate observables to primordial perturbations.
We calculated the PA transfer functions using CLASS \citep{Blas:2011JCAP...07..034B}.
The spherical Bessel functions $j_\ell(k \, r_\text{ls})$ account for the angular projection of SDA's on the last scattering surface, at a comoving distance $r_\text{LS}$.
Close to exact expressions of the $\mu$ window function were provided in \cite{Chluba:2016aln}, and a reasonable approximation is \citep{Pajer:2012vz, Ganc:2012ae, Chluba:2016aln}
\begin{gather}
    f^\mu(k_1, k_2, k_3)
    =
    2.27 
    \left[
        e^{(k_1^2 + k_2^2)/k_{\rm D}(z)}
    \right]_{z_{\mu y}}^{z_\mu}
    \Pi(k_3 / k_{\rm D}(z_{\mu y}))
\\
\nonumber
    \Pi(x)
    =
    \frac{3 j_1 (x)}
    {x},
\end{gather}
where $k_{\rm D}(z_{\mu})=\SI{12000}{Mpc^{-1}}$ and $k_{\rm D}(z_{\mu y})=\SI{46}{Mpc^{-1}}$ are the diffusion damping scales at the beginning and end of the $\mu$-distortion era.
It is then straightforward to calculate the $\mu X$ cross-correlation
\begin{align}\label{eq:xmu}
    C_\ell^{\,\mu X}
    \approx
    &
    12 \fNL^{\mu}
    \int \diff k \,
    \frac{2}{\pi} \,
    \frac{1}{5} \,
     k^2 \, 
    \mathcal{T}_\ell^{X/\zeta}(k) \,
    j_\ell(k \, r_\text{ls})\,
    P_\zeta(k)
\nonumber\\
    &
    \times 2.27 \, 
    \Pi \left(\frac{k}{k_{\rm D}(z_{\mu y})}\right)
    \int
    \frac{\Diff{3} \vec{k'}}{(2\pi)^{3}}
    \left[
        e^{(2k^2/k_{\rm D}(z)}
    \right]_{z_{\mu y}}^{z_\mu}
    P_\zeta(k')
    \nonumber \\
     &\simeq 
     12 f_{\rm NL}^\mu\, \langle \mu \rangle\, \frac{2\pi}{25}\frac{A_\text{s}}{\ell(\ell+1)}\,.
\end{align}
Notice that the first integral would match the temperature power spectrum in the \SW limit if one were to take $X=T$ and approximate the transfer function $\mathcal{T}_\ell^{X/\zeta}(k) \approx j_\ell(k \, r_\text{ls})/5$.
The integral in the second line is approximately equivalent to the sky averaged $\mu$ distortion sourced by dissipation of acoustic modes.
It is then obvious that for fixed $\fnl^\mu$, comparatively higher values of the $\mu$ monopole strengthen the detection of this cross-correlation \citep{Chluba:2016aln}.

The Gaussian contribution to the auto-power spectrum of $\mu$-distortion anisotropies is extremely small \citep{Pajer:2012vz,Ganc:2012ae}, and thus negligible when compared to any realistic instrumental noise term.
As such, from the theoretical side, we only need to consider the non-Gaussian contribution to $C_\ell^{\mu \mu}$ \citep[see][]{Emami:2015xqa} which, being quadratic in $\fNL$, can be important for large values of $\fNL$:%
\footnote{We stress that very large values of this parameter might be breaking the perturbative expansions. Knowing this limitation, we still investigate these limits as a proof of concept: the tightest constraints we will show it is possible to set are anyway safe from this caveat.}
\begin{align}
C_\ell^{\mu \mu} & = 144\, \left(f_{\rm NL}^\mu\right)^2\, \langle \mu \rangle^2\, \frac{2\pi}{25}\frac{A_\text{s}}{\ell(\ell+1)}\,,
\end{align}
where ${A_\text{s}=2.4\times 10^{-9}}$ is the power spectrum amplitude of the primordial curvature perturbation, assuming scale-invariance (i.e. ${n_s\simeq 1}$), and $\langle \mu \rangle = 2.3\times 10^{-8}$ is the ${\Lambda \rm CDM}$ prediction for the average $\mu$-distortion \citep{Chluba:2016bvg}.

\section{Sky simulations}
\label{sec:sky_sim}

\subsection{Cosmological signals}

The covariance matrix of the correlated cosmological fields reads as
\begin{align}
\label{eq:covmat}
 \mathbb{C}_\ell = 
\begin{pmatrix}
C_\ell^{\,TT} & C_\ell^{\,TE} & C_\ell^{\,\mu T}   \\
C_\ell^{\,TE} & C_\ell^{\,EE} & C_\ell^{\,\mu E} \\
C_\ell^{\,\mu T} & C_\ell^{\,\mu E} & C_\ell^{\mu \mu}   \\
\end{pmatrix}\,,
\end{align}
where $C_\ell^{\,TT}, C_\ell^{\,TE}, C_\ell^{\,EE}$ are theoretical CMB power spectra calculated from the \textit{Planck} 2018 ${\Lambda \rm CDM}$ best-fit model \citep{Planck_2018_results_VI}, while the theoretical cross-power spectra $C_\ell^{\,\mu T}, C_\ell^{\,\mu E}$ between CMB and $\mu$-distortion anisotropies are given by Eq.~\eqref{eq:xmu} \citep{Ravenni:2017lgw}.

Following \cite{Remazeilles2018}, we perform a Cholesky decomposition of the covariance matrix Eq.~\eqref{eq:covmat} in order to simulate correlated maps of $\mu$-distortion anisotropies, CMB temperature anisotropies, and CMB $E$-mode anisotropies based on the theoretical auto- and cross-power spectra described above. Our simulated maps are in \texttt{HEALPix}\footnote{\url{https://healpix.jpl.nasa.gov}} format \citep{Gorski2005} using $N_{\rm side}=512$ pixelisation.  

Through non-zero $\mu T$ and $\mu E$ correlations, our analysis focuses on non-Gaussianity of the primordial field at very small scales with $f_{\rm NL}^\mu\equiv f_{\rm NL}(k\simeq 740\,{\rm Mpc}^{-1})\neq 0$, while we neglect non-Gaussian fluctuations in the maps at scales probed by CMB anisotropies, so $f_{\rm NL}$ is consistent with zero in lower $k$ ranges. Since we do not attempt at computing any bispectrum from the maps but only cross-power spectra $\mu T$ and $\mu E$ between maps, we do not inject $f_{\rm NL}$ in the primary anisotropy maps, but only use it to introduce $\mu T$ and $\mu E$ correlations.

%%%%%%%%%%%%%%%%%%%%%%%%%%%%%%%%%%%%%%%%%%%%%%%%%%%%%%
\begin{figure}
  \begin{center}
    \includegraphics[width=\columnwidth]{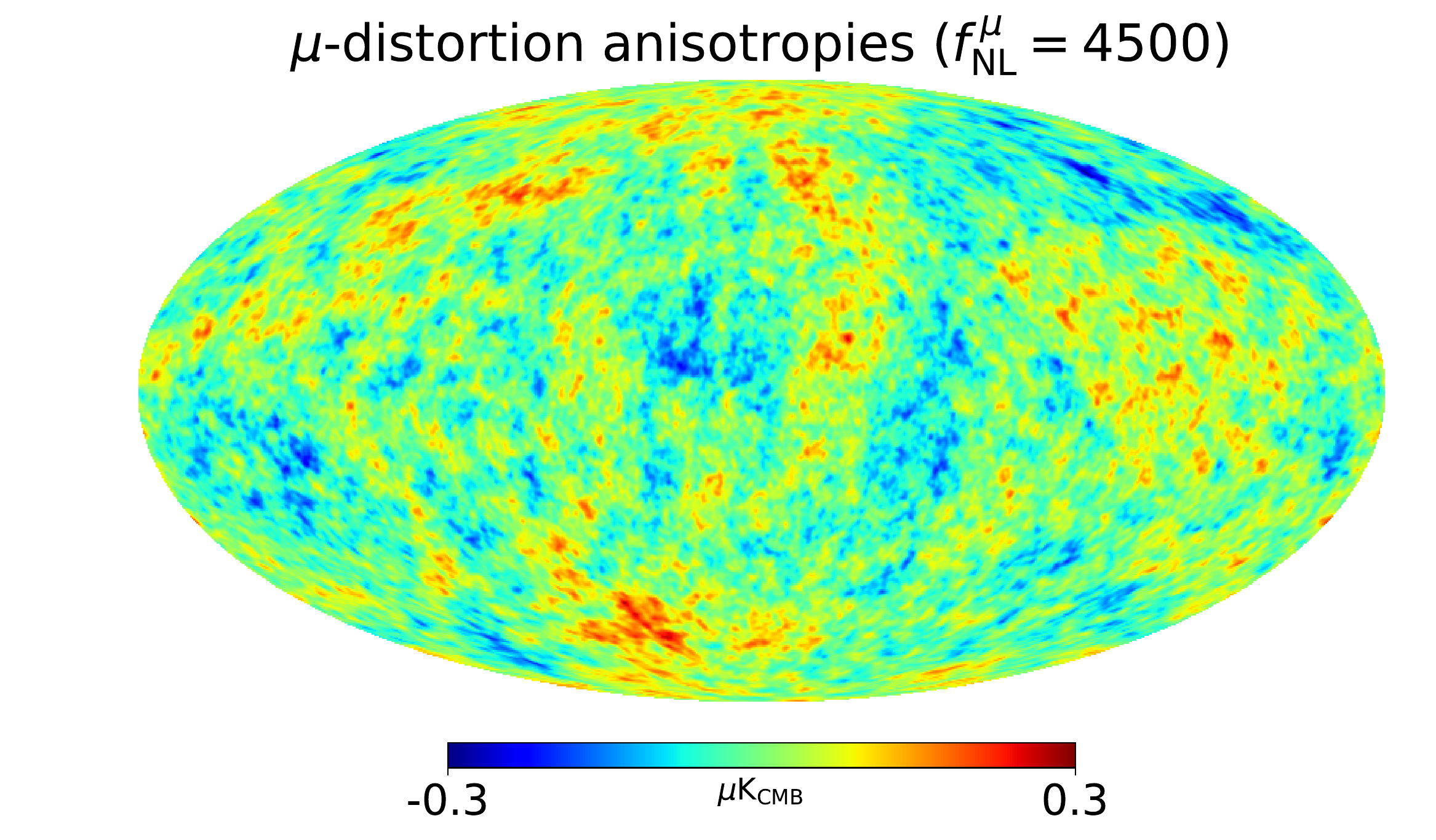}~
    \\[1.5mm]
     \includegraphics[width=\columnwidth]{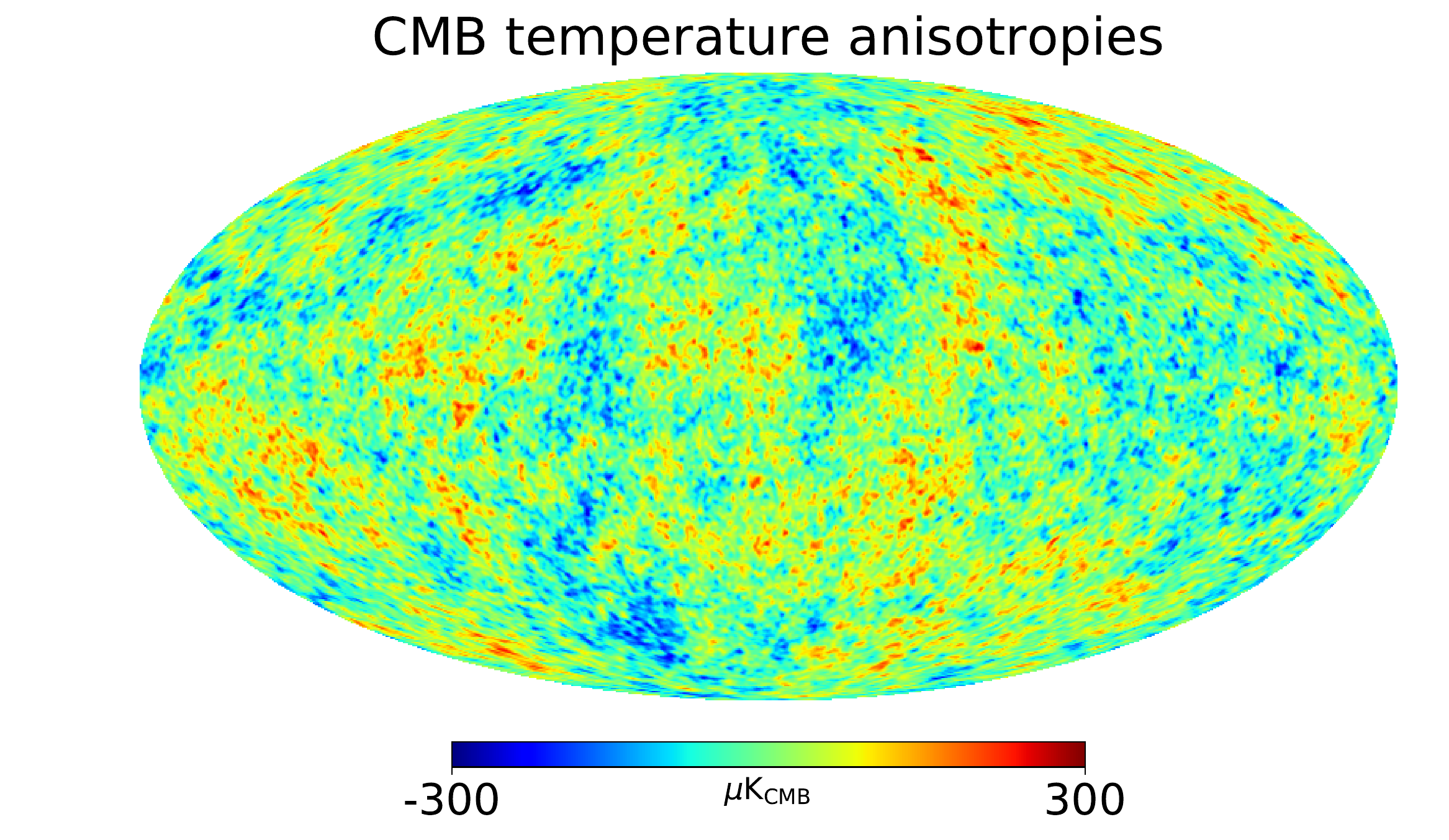}~
    \\[1.5mm]
        \includegraphics[width=\columnwidth]{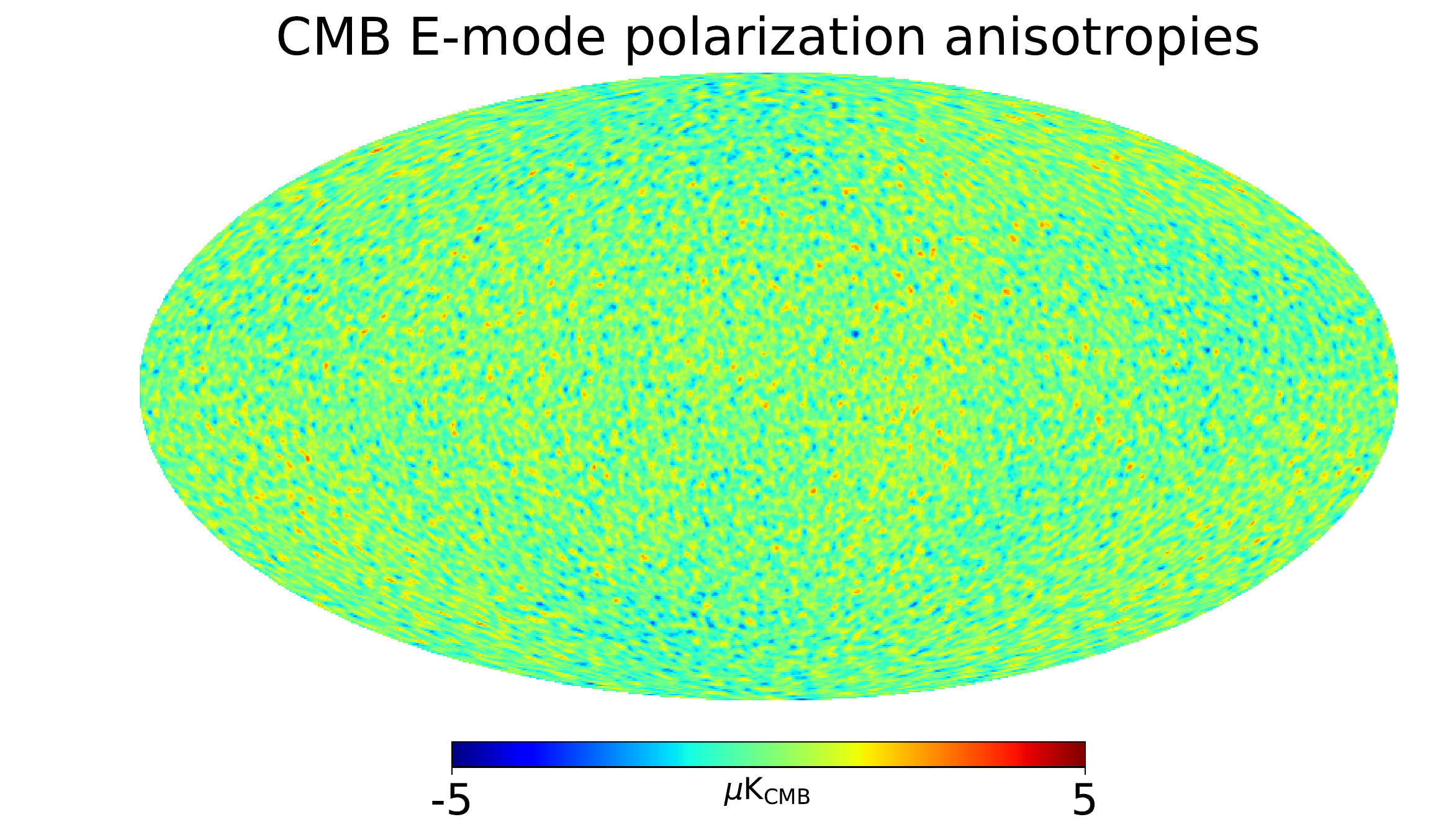}~
 \end{center}
\caption{Simulated maps at $60'$ resolution of correlated cosmological signals: $\mu$-distortion anisotropies for ${\langle \mu \rangle = 2.3\times 10^{-8}}$ and ${f_{\rm NL}^{\mu}(k\simeq 740\,{\rm Mpc}^{-1})=4500}$ (\textit{top}), CMB temperature anisotropies (\textit{middle}), CMB $E$-mode polarization anisotropies (\textit{bottom}). Large-scale anticorrelation between CMB and $\mu$-distortion anisotropies is visible on the maps.}
\label{Fig:cosmo_maps}
\end{figure}
%%%%%%%%%%%%%%%%%%%%%%%%%%%%%%%%%%%%%%%%%%%%%%%%%%%%%%

In the following, we will consider a set of three sky simulations, in which $f_{\rm NL}^\mu = 4500$, $f_{\rm NL}^\mu = 10^4$, and $f_{\rm NL}^\mu = 10^5$, respectively. In addition, we consider a sky simulation in which $f_{\rm NL}^\mu = 0$, i.e. without anisotropic $\mu$-distortion signal, for our null tests. The $\mu$-map realisation of each simulation is the same for all fiducial $f_{\rm NL}^\mu$ values considered.

\begin{figure}
  \begin{center}
    \includegraphics[width=\columnwidth]{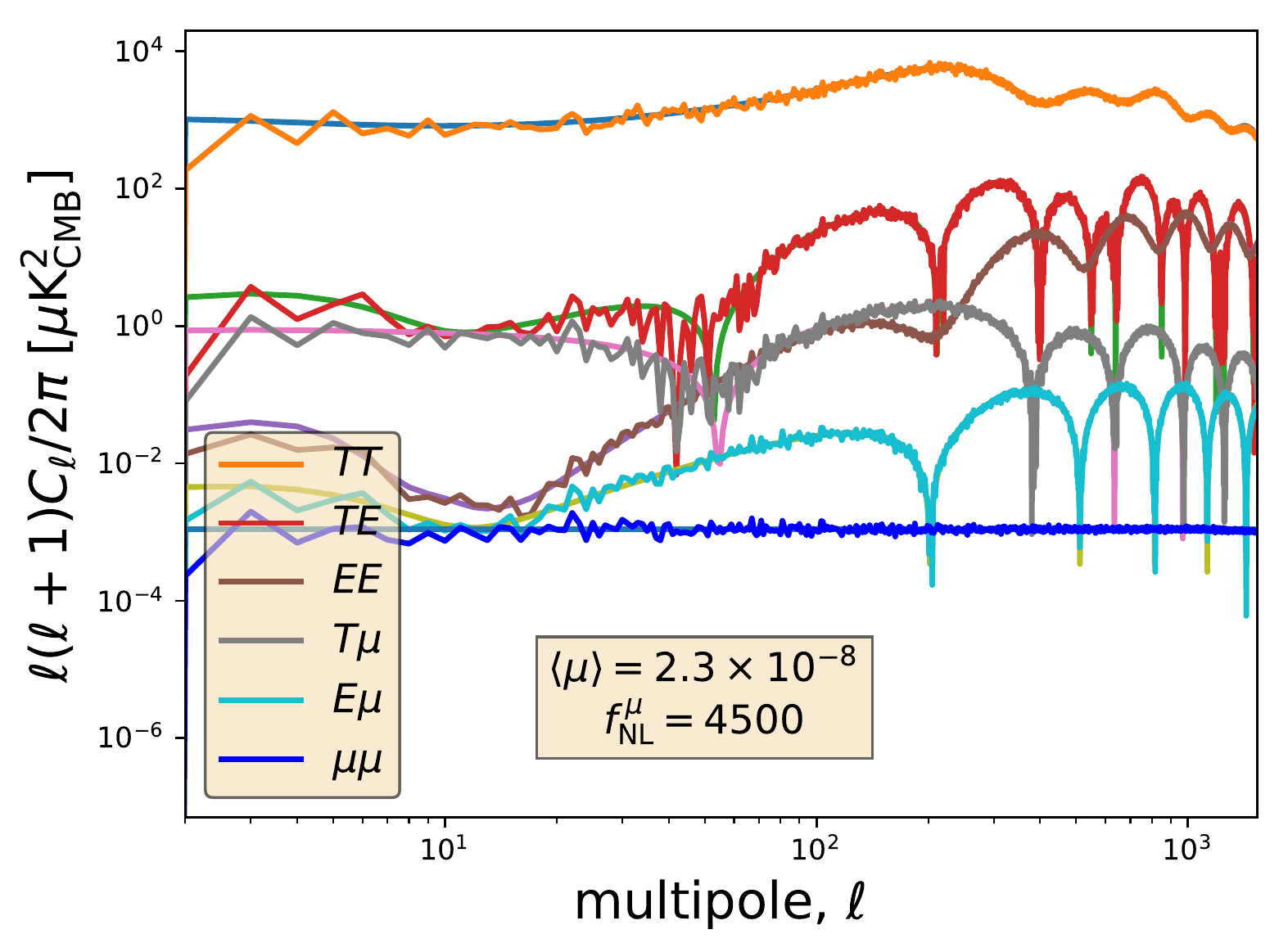}~
 \end{center}
\caption{Auto- and cross-power spectra of the simulated maps of Fig.~\ref{Fig:cosmo_maps}, as plotted on top of the theoretical power spectra from \citet{Ravenni:2017lgw} for $\langle \mu \rangle = 2.3\times 10^{-8}$ and $f_{\rm NL}^{\mu}(k\simeq 740\,{\rm Mpc}^{-1})=4500$.}
\label{Fig:cosmo_spectra}
\end{figure}

Our simulated maps of anisotropic $\mu$-type distortions, CMB temperature, and CMB $E$-mode polarization are shown in Fig.~\ref{Fig:cosmo_maps} for $\langle \mu \rangle = 2.3\times 10^{-8}$ and $f_{\rm NL}^{\mu}=4500$, at an angular resolution of $60'$.  The anticorrelation between CMB and $\mu$-distortion anisotropies at large angular scales, as expected from theory, is clearly visible on the simulated maps. The auto- and cross-power spectra of the simulated maps are plotted in Fig.~\ref{Fig:cosmo_spectra}, where they are shown to match the theoretical spectra for $f_{\rm NL}^{\mu}=4500$.

%%%%%%%%%%%%%%%%%%%%%%%%%%%%%%%%%%%%%%%%%%%%%%%%%%%%%%
%%%%%%%%%%%%%%%%%%%%%%%%%%%%%%%%%%%%%%%%%%%%%%%%%%%%%%

%%%%%%%%%%%%%%%%%%%%%%%%%%%%%%%%%%%%%%%%%%%%%%%%%%%%%%
\begin{figure}
  \begin{center}
    \includegraphics[width=\columnwidth]{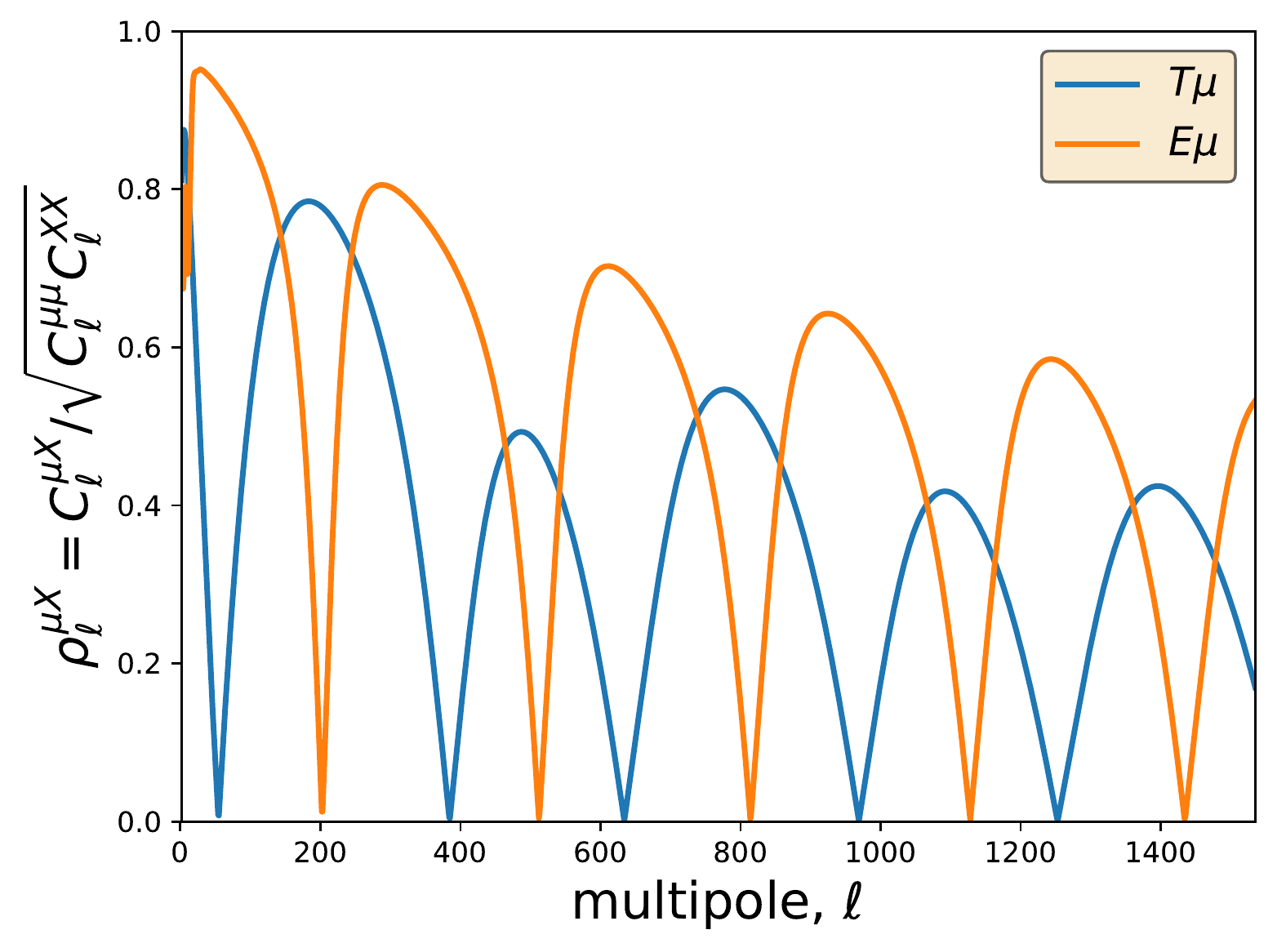}~
 \end{center}
\caption{Cross-correlation coefficient between the anisotropic $\mu$-distortion field and either CMB $T$ or $E$ fields. The intrinsic correlation between $\mu$-distortion and CMB $E$-modes is actually more significant than the intrinsic correlation between $\mu$-distortion and CMB temperature anisotropies.}
\label{Fig:pearson}
\end{figure}
%%%%%%%%%%%%%%%%%%%%%%%%%%%%%%%%%%%%%%%%%%%%%%%%%%%%%%

We can also see from Fig.~\ref{Fig:cosmo_spectra} that, for ${f_{\rm NL}^{\mu}(k\simeq 740\,{\rm Mpc}^{-1})\simeq 4500}$,  the $\mu T$ correlated signal (grey line) has a magnitude similar to that of the CMB $TE$ signal (red line) at low multipoles ${\ell < 60}$ and the CMB $EE$ signal (brown line) at higher multipoles ${60 < \ell < 300}$. Thus, if the amplitude of primordial non-Gaussianity is as large at high wavenumber ${k\simeq 740\,{\rm Mpc}^{-1}}$ then it makes $\mu$-distortion anisotropies an accessible signal to future CMB imagers like the \textit{LiteBIRD} satellite \citep{Litebird2019}, thanks to its amplification by cross-correlation with the more intense signal of CMB anisotropies. 

While the absolute amplitude of the $\mu E$ correlated signal is significantly lower than that of the $\mu T$ signal (Fig.~\ref{Fig:cosmo_spectra}), interestingly the degree of correlation between $\mu$-distortion and CMB $E$-mode anisotropies is significantly larger compared to the degree of correlation between $\mu$-distortion and CMB temperature anisotropies across the multipoles. This is evident from Fig.~\ref{Fig:pearson}, showing the Pearson correlation coefficients across multipoles for both $\mu T$ and $\mu E$. 

Because of better correlation with CMB polarization, the observable $C_\ell^{\,\mu E}$ should actually provide more constraining power than $C_\ell^{\,\mu T}$ on $f_{\rm NL}^{\mu}(k\simeq 740\,{\rm Mpc}^{-1})$  in the presence of foregrounds, as we are going to demonstrate in this work. In addition, the foreground signal is much weaker in polarization and somewhat less complex than in intensity since only few of the foregrounds are actually polarized. This should further facilitate the recovery of $C_\ell^{\,\mu E}$ as compared to $C_\ell^{\,\mu T}$ and further enhance the constraining power of $C_\ell^{\,\mu E}$.

CMB temperature and polarization anisotropies are achromatic in thermodynamic temperature (${\rm K}_{\rm CMB}$) units, while $\mu$-type spectral distortions have a peculiar spectral signature given by \citep[see][]{Sunyaev1970, Chluba2018, Lucca2020}: 
\begin{align}
\label{eq:mu_sed}
    a_\mu(\nu) 
    = 
    T_{\rm CMB}\left(\frac{\pi^2}{18\zeta(3)} - {\frac{1}{x}}\right)\,,
\end{align}
where $T_{\rm CMB}=2.7255$\,K  is the CMB blackbody temperature, ${x = h\nu /  k T_{\rm CMB}}$, and $\zeta(3)$ is the value of the Riemann zeta function at integer $n=3$.
We thus scale our simulated map of $\mu$-distortion anisotropies, $\mu(\hat{n})$ (top panel of Fig.~\ref{Fig:cosmo_maps}), across the frequency bands $\nu$ of our sky simulation by using the emission law Eq.~\eqref{eq:mu_sed} as
\begin{align}
I^{\mu}_\nu(\hat{n}) = a_\mu(\nu)\, \mu(\hat{n})\,.
\end{align}
Since the anisotropic $\mu$-distortion signal is unpolarised, it is absent from the $Q,U$ polarization channels of our sky simulations.

We ignore primordial $y$-distortion anisotropies in the current analysis as the amplitude of $y T$ and $y E$ correlations is anyway about an order of magnitude lower than the amplitude of $\mu T$ and $\mu E$ correlations \citep{Ravenni:2017lgw}. Moreover, the full spectral degeneracy between primordial $y$-distortions and thermal SZ effect from galaxy clusters in the late universe makes it challenging to disentangle the primordial signal from the SZ foreground in temperature channels, and thus this deserves further investigation in a future work.

\subsection{Galactic and extragalactic foregrounds}

We use the Planck Sky Model \citep[\textsc{PSM};][]{Delabrouille2013} to simulate maps of Galactic and extragalactic foreground emissions. For observations in temperature, the Galactic foreground components of our sky simulation include thermal dust, synchrotron, free-free, and anomalous microwave emission (AME), while the simulated extragalactic foregrounds include cosmic infrared background (CIB) anisotropies and thermal and kinetic Sunyaev-Zeldovich effects. For observations in polarization, our sky simulations include only thermal dust and synchrotron as main polarized Galactic foregrounds. 

Since we are interested in large angular scales, where the bulk of the $\mu T$ and $\mu E$ correlated signals lies, we do not include unresolved radio and infrared sources in our simulations.

\begin{figure*}
  \begin{center}
    \includegraphics[width=0.33\textwidth]{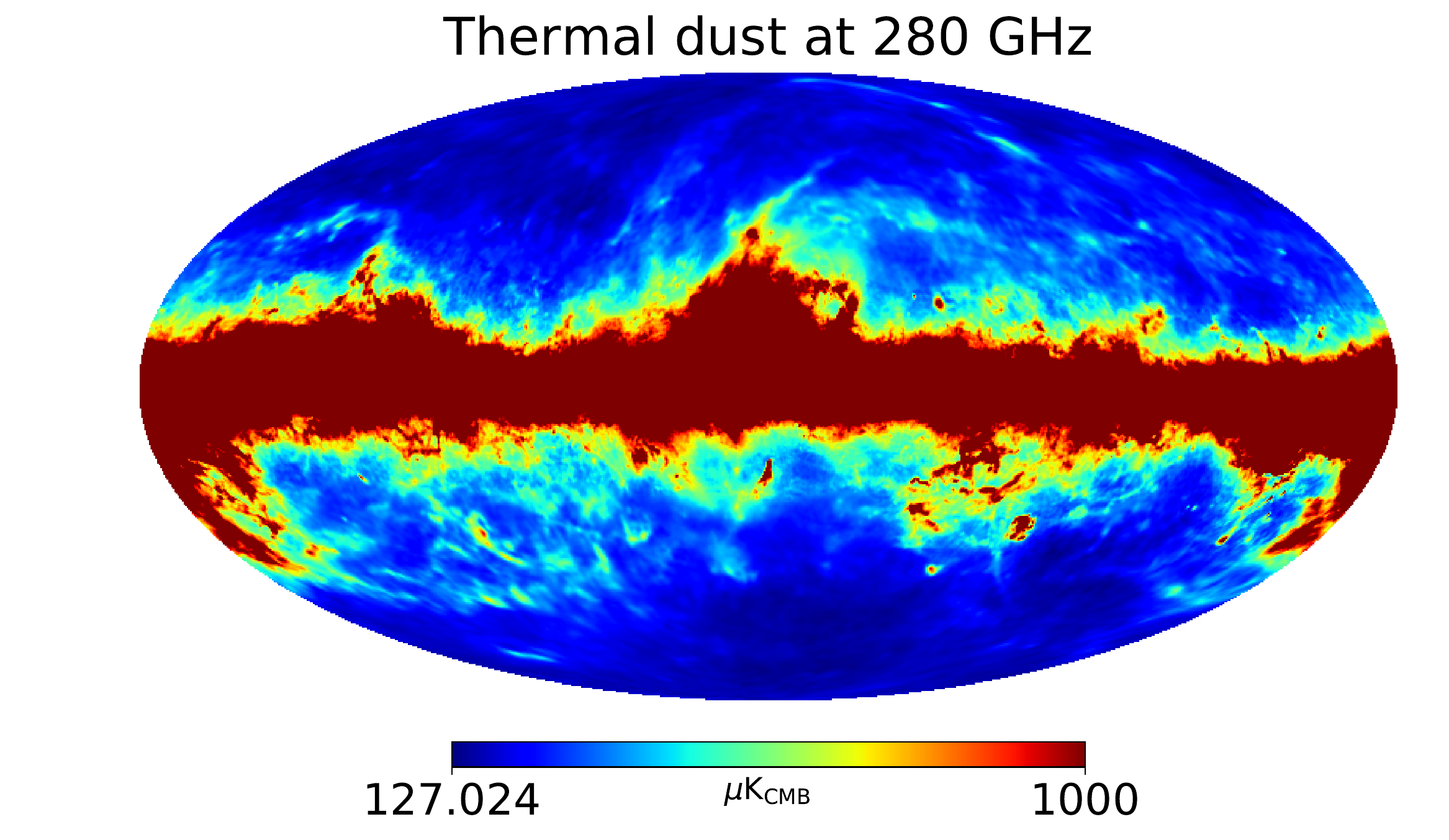}~
     \includegraphics[width=0.33\textwidth]{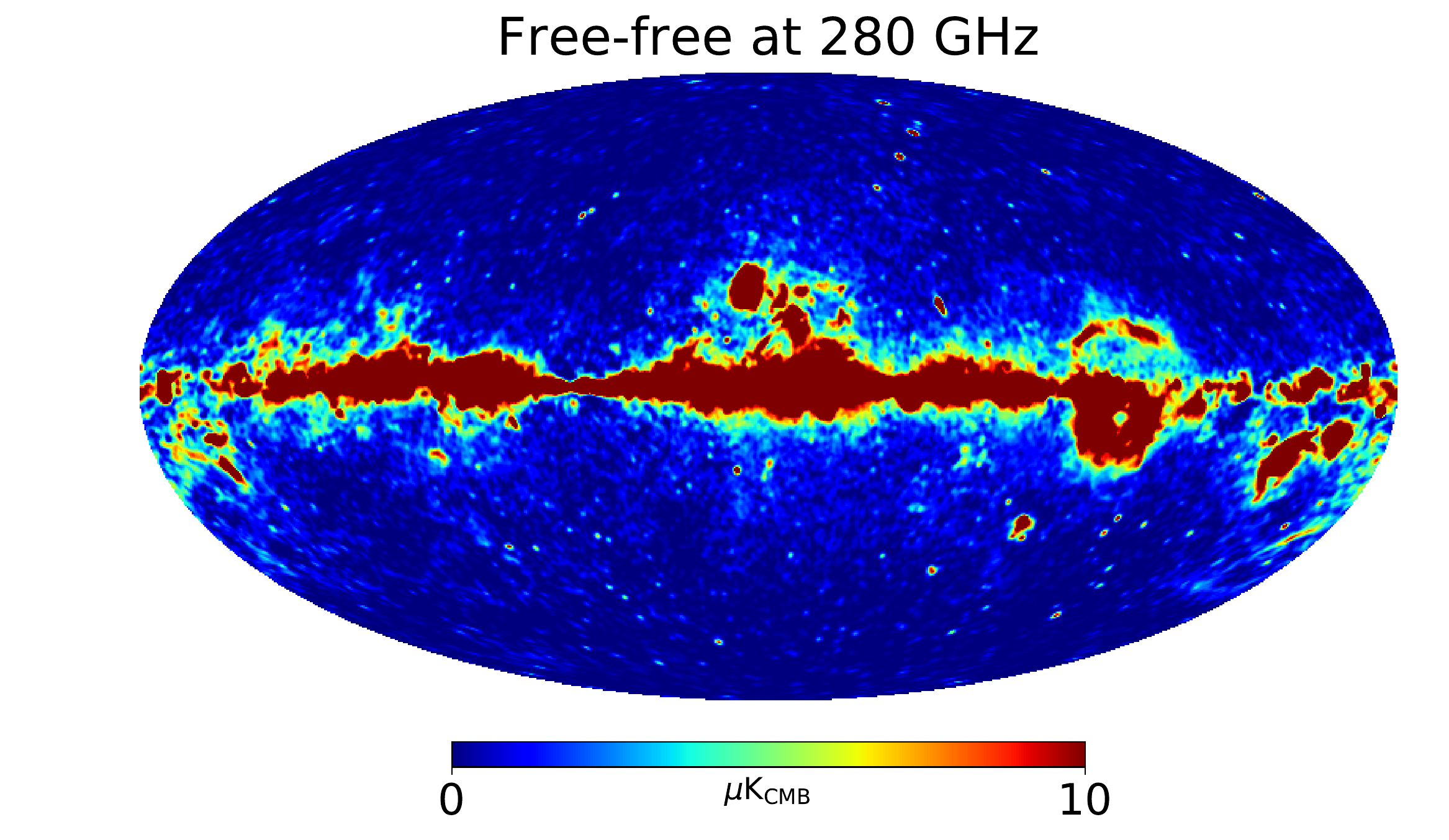}~
      \includegraphics[width=0.33\textwidth]{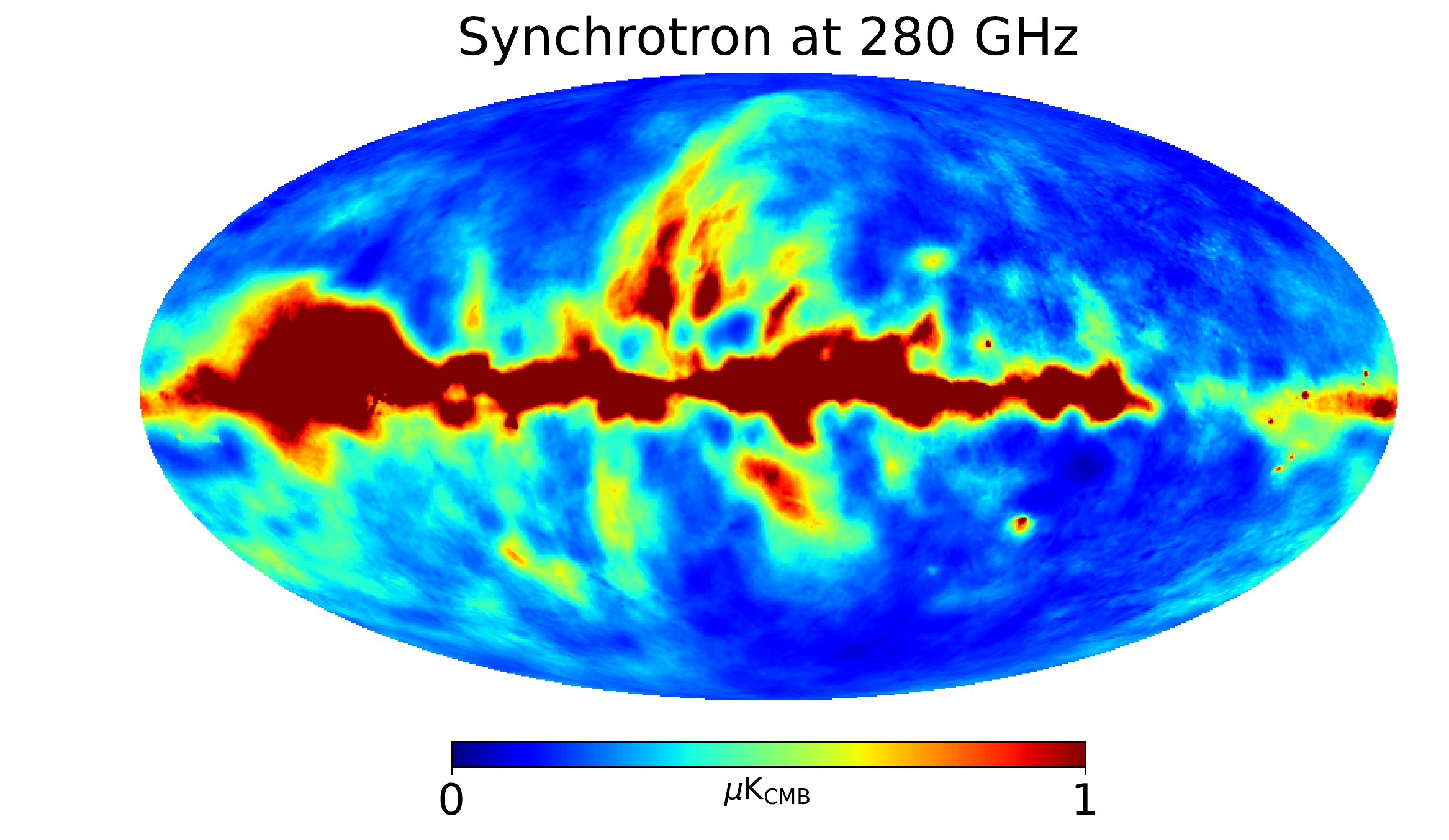}~
    \\[1.5mm]
      \includegraphics[width=0.33\textwidth]{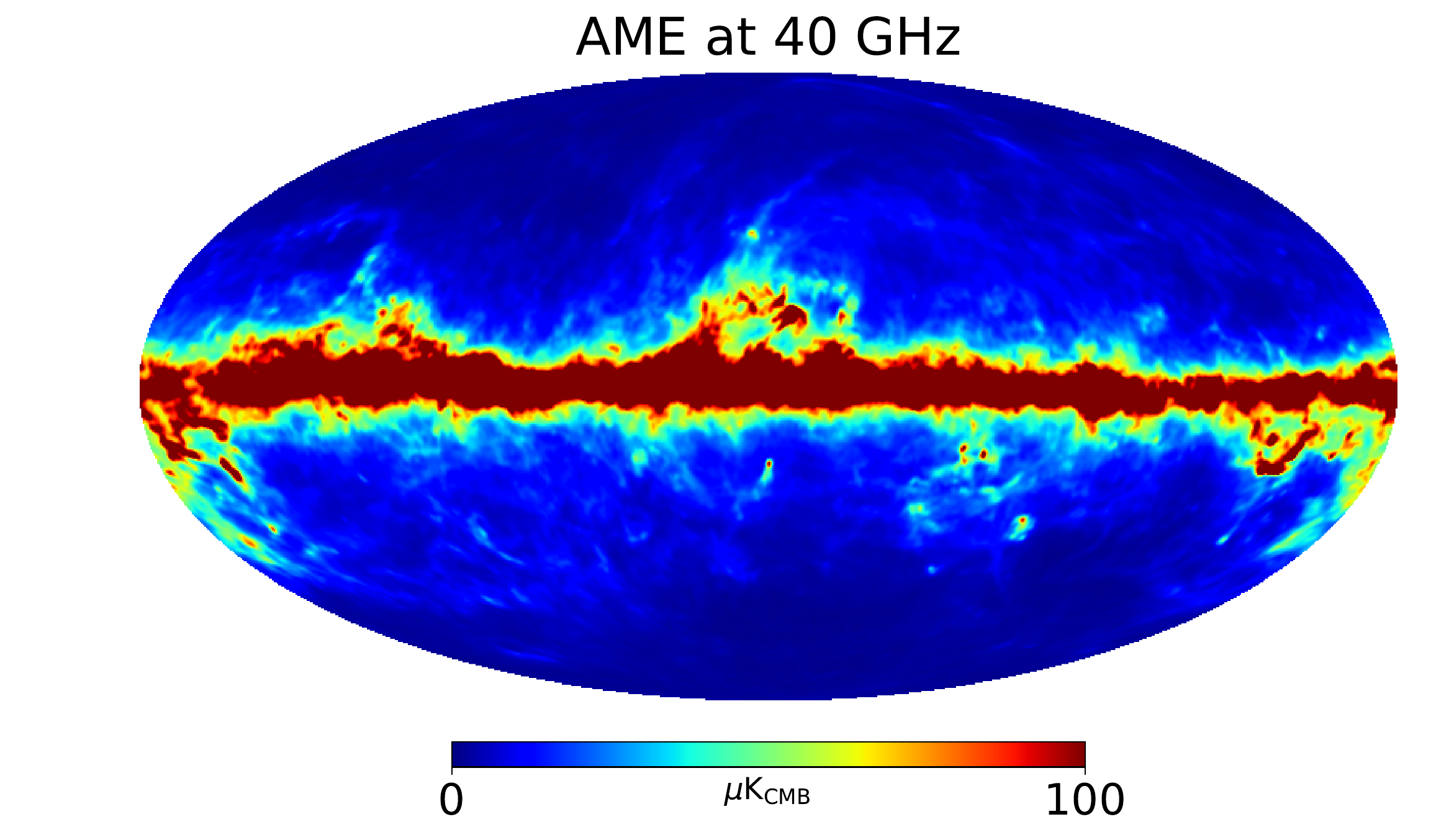}~
     \includegraphics[width=0.33\textwidth]{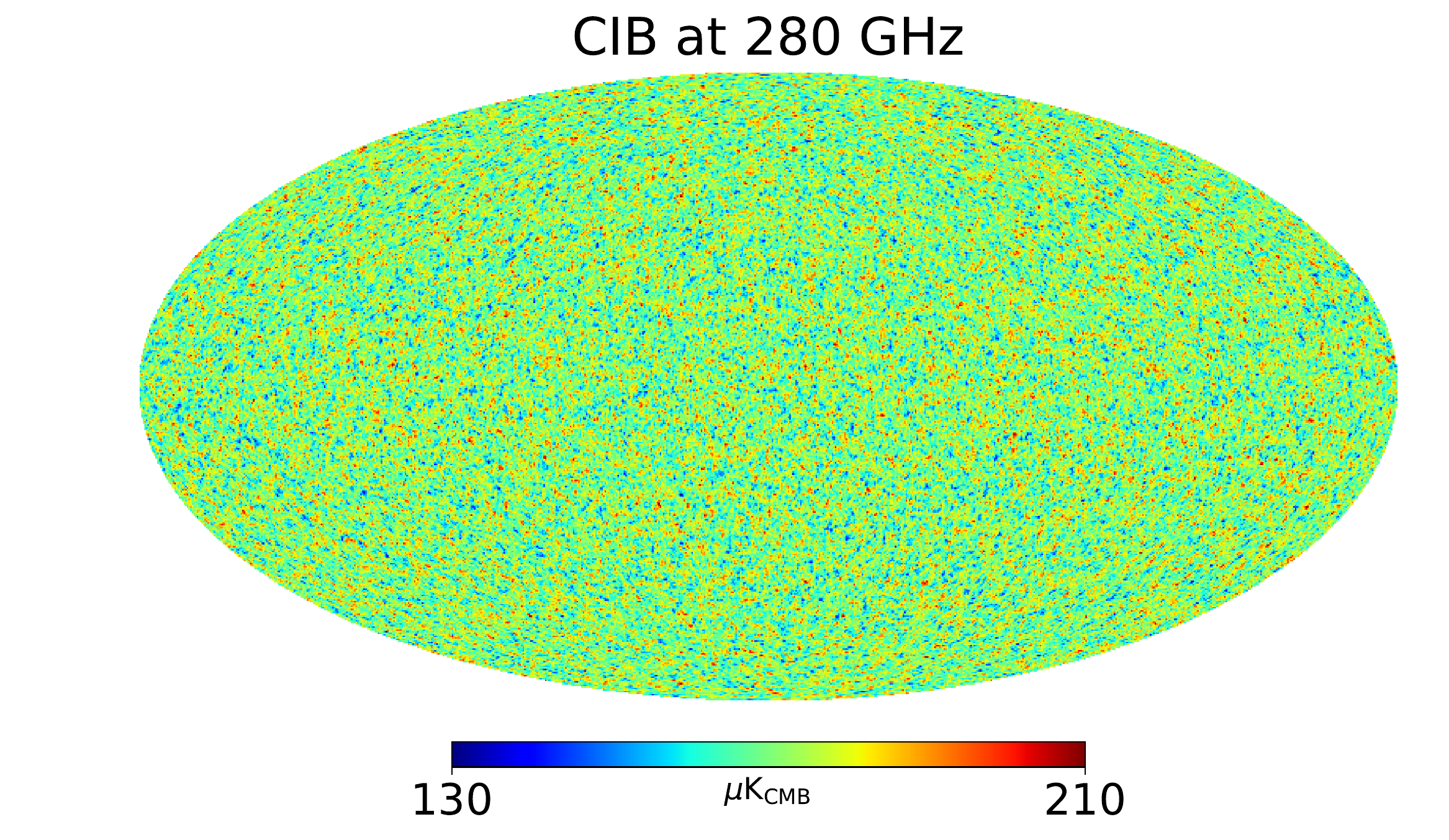}~
      \includegraphics[width=0.33\textwidth]{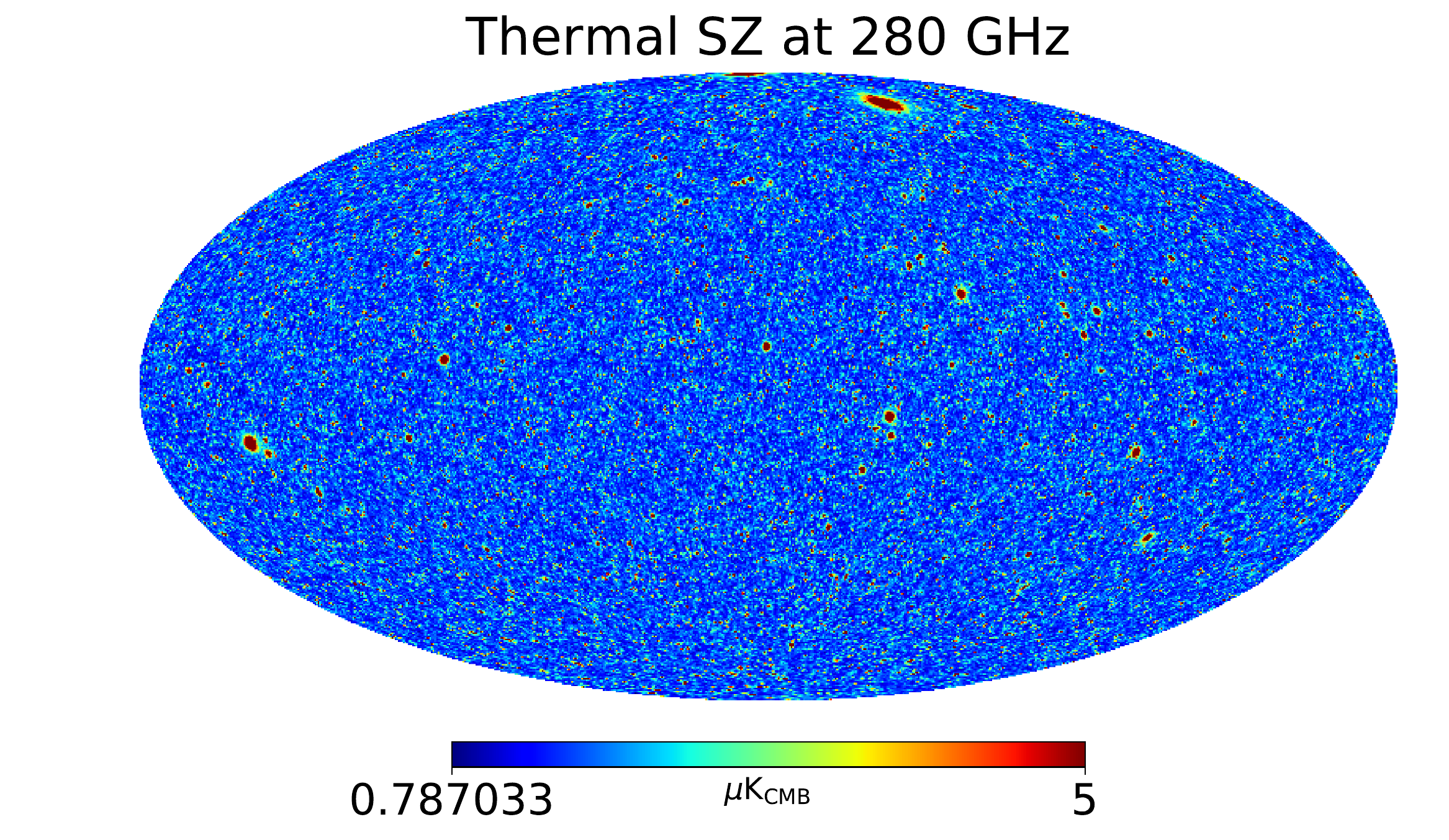}~
     \end{center}
\caption{Simulated maps of Galactic and extragalactic foreground emissions in temperature: thermal dust (\textit{upper left}),  free-free (\textit{upper middle}),  synchrotron (\textit{upper right}),  AME (\textit{lower left}),  CIB (\textit{lower middle}),  thermal SZ (\textit{lower right}). Emission in these maps is shown at $280$\, GHz, except for the AME which is shown at $40$\,GHz. 
}
\label{Fig:fg_maps}
\end{figure*}

\begin{figure}
  \begin{center}
    \includegraphics[width=0.4\textwidth]{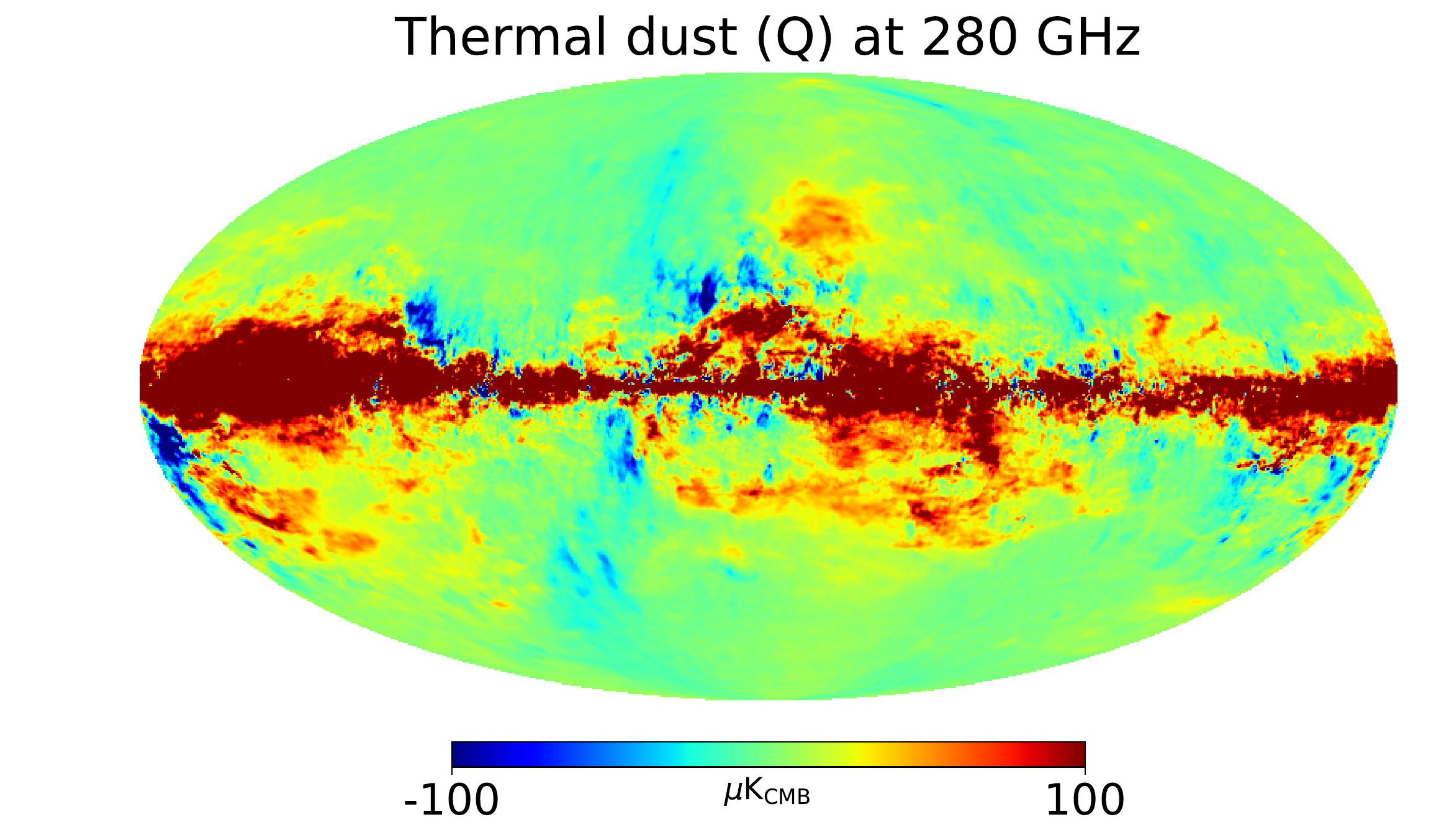}~
     \\[1.5mm]
         \includegraphics[width=0.4\textwidth]{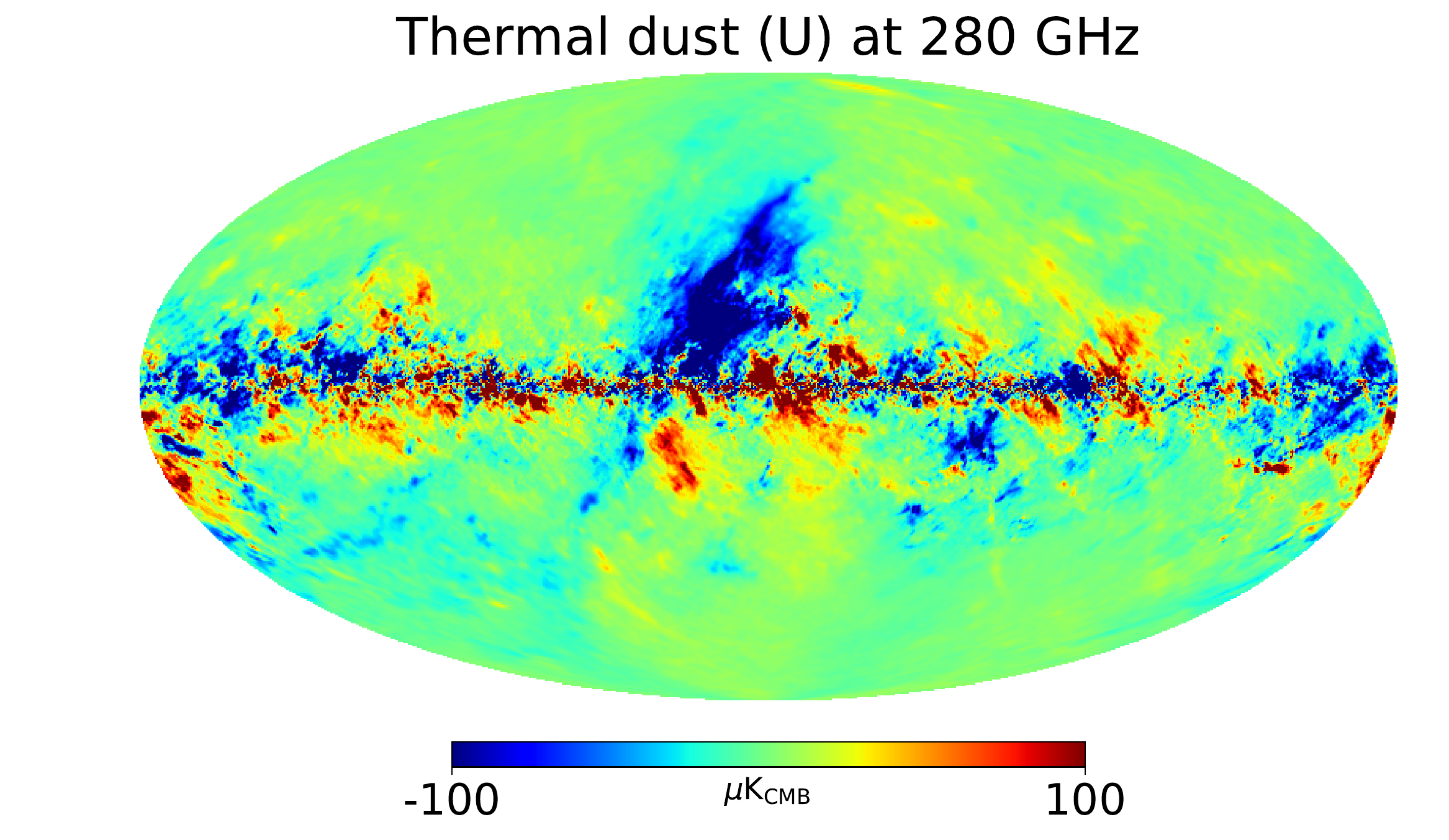}~
      \\[1.5mm]
      \includegraphics[width=0.4\textwidth]{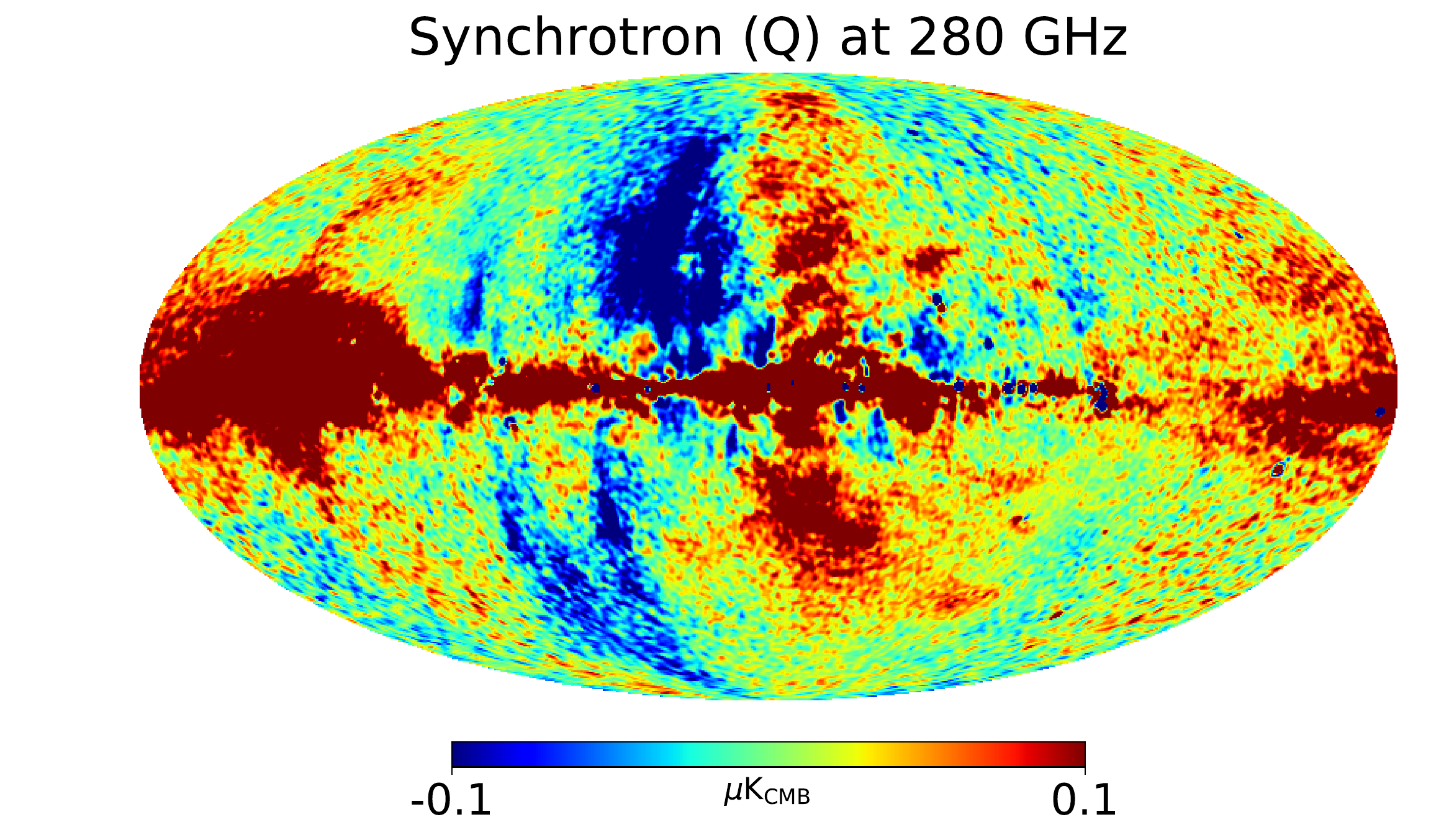}~
       \\[1.5mm]
            \includegraphics[width=0.4\textwidth]{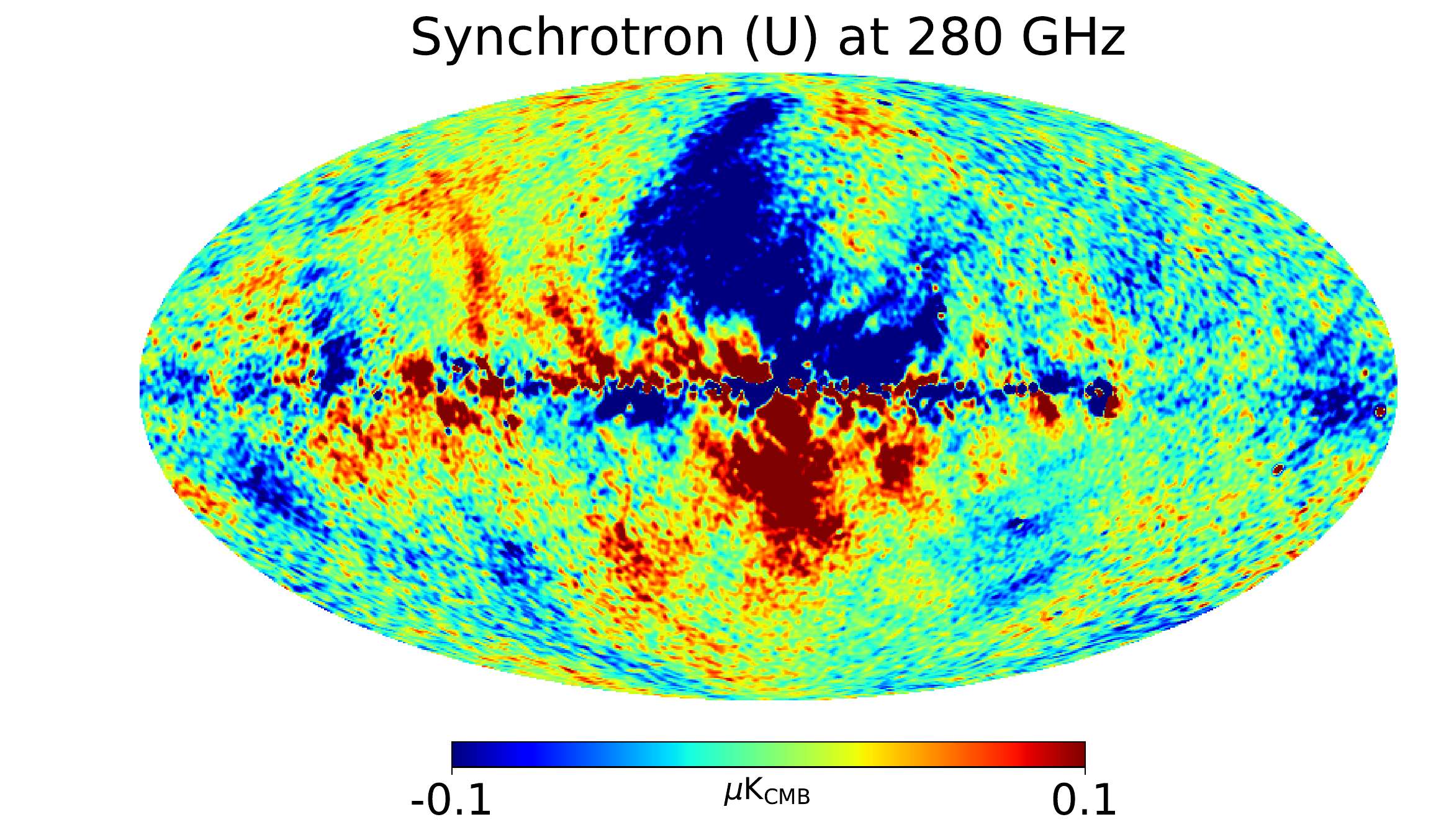}~
         \end{center}
\caption{Simulated maps of Galactic foreground emissions in polarization: thermal dust and synchrotron $Q$ and $U$ maps,  as observed at $280$\, GHz.}
\label{Fig:fg_maps_qu}
\end{figure}

Figure~\ref{Fig:fg_maps} displays the simulated foreground components as observed in temperature at $280$\,GHz, except for the AME component which is shown as observed at $40$\,GHz, while Fig.~\ref{Fig:fg_maps_qu} shows the polarized foreground components of the simulation as observed in Stokes $Q$, $U$ fields  at $280$\,GHz.

%%%%%%%%%%%%%%%%%%%%%%%%%%%%%%%%%%%%%%%%%%%%%%%%%%%%%%
%%%%%%%%%%%%%%%%%%%%%%%%%%%%%%%%%%%%%%%%%%%%%%%%%%%%%%

\subsubsection{Thermal dust} 

Thermal dust emission from our Galaxy originates from silicate and carbonaceous grains of nanometer size in the interstellar medium which, by absorbing the UV light from stars, heat themselves and thus re-emit light at submillimetre and infrared wavelengths.  This is the dominant foreground emission in the sky at frequencies $\gtrsim 100$\,GHz. 

Thermal dust emission is also polarised because dust grains are aspherical and spin around Galactic magnetic fields, while radiative torque by stellar radiation force the dust grains to align with their long axis perpendicular to the magnetic field lines. Since the cross-section is proportional to the size of an object, dust grains emit more radiation parallel to their long axis, which induces linear polarization of thermal dust emission that is orthogonal to the magnetic field.

We use the \textit{Planck} GNILC $I,Q,U$ maps at $\nu_0=353$\,GHz \citep{Planck_GNILC2016,Planck_GNILC2018} as the dust templates for our simulations. The dust template maps in MJy.${\rm sr^{-1}}$ units are scaled across frequencies through a modified blackbody function:
\begin{align}
\begin{pmatrix}
I^{\,\rm dust}_\nu(\hat{n})\\
Q^{\,\rm dust}_\nu(\hat{n})\\
U^{\,\rm dust}_\nu(\hat{n})\\ 
\end{pmatrix} 
= 
\begin{pmatrix}
I^{\,\rm GNILC}_{\nu_0}(\hat{n})\\
Q^{\,\rm GNILC}_{\nu_0}(\hat{n})\\
U^{\,\rm GNILC}_{\nu_0}(\hat{n})\\
\end{pmatrix} 
\left({\nu \over \nu_0}\right)^{\beta_d(\hat{n})}{B_\nu\left(T_d(\hat{n})\right) \over B_{\nu_0}\left(T_d(\hat{n})\right)}\,,
\end{align}
where the spectral index $\beta_d(\hat{n})$ and temperature $T_d(\hat{n})$ vary over the sky depending on the line-of-sight $\hat{n}$, and
\begin{align}
B_\nu\left(T\right) = {2h\nu^3\over c^2}{1\over {\rm e}^{{h\nu\over kT}}-1}
\end{align}
is the Planck's law for blackbody radiation.
The same \textit{Planck} GNILC templates of dust spectral index and dust temperature \citep{Planck_GNILC2016} are used for the $I$, $Q$, and $U$ fields.

\subsubsection{Synchrotron} 

Synchrotron emission is due to the relativistic cosmic-ray electrons from supernovae explosions which are accelerated by the magnetic field of our Galaxy. This is dominant foreground emission in the sky at the lowest frequencies.

Synchrotron emission is also polarised. When the electrons are spiralling around the magnetic fields, their orbit projected on the sky plane of observation is seen as an oscillation orthogonal to the magnetic fields, which induces linear polarization of the synchrotron emission in the orbit plane orthogonal to the magnetic field lines.

We use the reprocessed Haslam $408$\,MHz map \citep{Remazeilles2015} as the synchrotron template for intensity channels, while for polarization channels we use the \textit{Planck} Commander template of synchrotron polarization $Q$, $U$ maps at $30$\,GHz \citep{Planck_2018_results_IV}. Both templates in brightness Rayleight-Jeans temperature ($K_{\rm RJ}$) units  are scaled across the frequency channels with the same power law of spectral index varying over the sky and given by the synchrotron index template map $\beta_s(\hat{n})$ from \cite{Miville2008}:
\begin{align}
\begin{pmatrix}
I^{\,\rm sync}_\nu(\hat{n})\\
Q^{\,\rm sync}_\nu(\hat{n})\\
U^{\,\rm sync}_\nu(\hat{n})\\ 
\end{pmatrix} 
= 
\begin{pmatrix}
I^{\,\rm Haslam}_{\rm 408\,MHz}(\hat{n})\,\left({\nu \over 408\,{\rm MHz}}\right)^{\beta_s(\hat{n})}\\
Q^{\,Planck}_{\rm 30\,GHz}(\hat{n})\,\left({\nu \over 30\,{\rm GHz}}\right)^{\beta_s(\hat{n})}\\
U^{\,Planck}_{\rm 30\,GHz}(\hat{n})\,\left({\nu \over 30\,{\rm GHz}}\right)^{\beta_s(\hat{n})}\\
\end{pmatrix}\,.
\end{align}

\subsubsection{Free-free} 

Free electrons in ionised star-forming HII regions of our Galaxy are braked by Coulomb interactions with heavy ions, thus losing part of their kinetic energy which is converted in so-called free-free emission (or thermal bremsstrahlung).  Free-free emission is an important foreground to CMB temperature observations at low frequencies, although the bulk of free-free emission is mainly concentrated in the Galactic disk.

We use the \textit{Planck} Commander free-free template maps for the emission measure ${\rm EM}(\hat{n})$ and electronic temperature $T_{\rm e}(\hat{n})$ \citep{Planck_2015_X}, and we adopt the prescription in the aforementioned reference to scale free-free emission in brightness temperature units across the frequencies as follows:
\begin{align}
I^{\,\rm ff}_\nu(\hat{n}) = 10^6\, T_{\rm e}(\hat{n})\left(1-{\rm e}^{-\tau_{\rm ff}(\hat{n},\nu)}\right)\,,
\end{align}
where the free-free optical depth is given by
%-----------------------------
\begin{align}
\tau_{\rm ff}(\hat{n},\nu) = 0.05468\,T_{\rm e}(\hat{n})^{-3/2}\,\left({\nu\over 1\,{\rm GHz}}\right)^{-2}\,{\rm EM}(\hat{n})\,g_{\rm ff}(\hat{n},\nu)\,,
\end{align}
%-----------------------------
and the Gaunt correction factor is %-----------------------------
\citep{Draine2003}
\begin{align}
g_{\rm ff}(\hat{n},\nu) &=1+\log\left(1+ {\rm e}^{4.960+{\sqrt{3}\over \pi}\log\left[\left({\nu\over 1\,{\rm GHz}}\right)^{-1}\left({T_{\rm e}(\hat{n})\over 10^4\,{\rm K}}\right)^{3/2}\right]} \right)\,.
\end{align}
%-----------------------------
Galactic free-free emission is unpolarised because of the randomness of Coulomb interactions, and thus absent from the $Q,U$ polarization channels of our sky simulation.

\subsubsection{AME} 

Anomalous microwave emission (AME) from our Galaxy is now routinely detected at low frequencies $\sim 10$-$60$\,GHz \citep[e.g.][]{Planck_2015_X},  with a peculiar spectral signature that is inconsistent with that of synchrotron or free-free emissions. 

AME is strongly correlated with far-infrared thermal dust emission at $100$ microns \citep{Davies2006}, so the best explanation for AME so far is electric dipole radiation from spinning dust grains in our Galaxy \citep{Draine1998}. Since spinning dust polarization has been shown to be negligibly small \citep{Draine2016}, we consider unpolarized AME in our simulations.

We use the \textit{Planck} GNILC dust optical depth map at $353$\,GHz \citep{Planck_GNILC2016}, which we rescaled as a spinning dust map at $22.8$\,GHz using the thermal dust-AME correlation factor of $r=0.91$ measured by \cite{Planck_2015_XXV}. The spinning dust template is then scaled across frequencies using the model of \cite{Draine1998} for spinning dust emission law with 96.2\% warm neutral medium and $3.8$\% reflection nebulae.

\subsubsection{CIB} 

Cosmic infrared background (CIB) temperature anisotropies arise from the cumulative diffuse emission of early dusty star-forming galaxies at redshifts $1\lesssim z \lesssim 3$ \citep[e.g.][]{Planck_2013_XXX}. Therefore, CIB anisotropies form a diffuse extragalactic foreground to observations of $\mu$-distortion anisotropies, in particular at high frequencies.

The CIB emission is simulated by the \textsc{PSM} assuming three populations of spiral, starburst, and proto-spheroid galaxies, which are distributed across redshift shells according to dark matter distribution \citep{Delabrouille2013,Planck_2015_XII}. Each population of infrared galaxies has its own spectral energy distribution (SED), which is redshifted accordingly depending on the frequency channel of observation. Maps of each population and each redshift shell are coadded to form CIB maps across frequencies, which have been shown by \cite{Planck_2015_XII} to reproduce the auto- and cross power spectra of CIB anisotropies as measured by \textit{Planck} \citep{Planck_2013_XXX}.

%%%%%%%%%%%%%%%%%%%%%%%%%%%%%%%%%%%%%%%%%%%%%%%%%%%%%%
%%%%%%%%%%%%%%%%%%%%%%%%%%%%%%%%%%%%%%%%%%%%%%%%%%%%%%

\subsubsection{SZ effects} 

In our sky simulation we include the thermal and kinetic Sunyaev-Zeldovich (SZ) effects as extragalactic foregrounds to primordial spectral distortion anisotropies. 

Kinetic SZ (kSZ) effect is the Doppler boost of CMB photons that is caused by the proper radial velocities of galaxy clusters. In the non-relativistic limit, the spectral signature of the kSZ effect is identical to that of CMB anisotropies, i.e. the derivative of the blackbody with respect to temperature, and thus kSZ emission is achromatic across frequencies in thermodynamic temperature units \citep{Sunyaev1980}.

Inverse Compton scattering of CMB photons off the hot gas of free electrons residing in galaxy clusters also causes $y$-type spectral distortions to the CMB blackbody spectrum. This is known as thermal SZ (tSZ) effect from galaxy clusters, whose peculiar spectral signature in thermodynamic temperature units is given in the non-relativistic limit by \citep{Zeldovich:1969ff}
\begin{align}
\label{eq:y_sed}
a_{\rm tSZ}(\nu) = T_{\rm CMB}\left[ x\coth \left({x\over 2}\right) - 4 \right]\,.
\end{align}
It is worth noticing that there is a full spectral degeneracy between this type of extragalactic foreground emission in the low-redshift universe and the primordial $y$-distortion signal in the early universe. It thus makes it challenging to deal with the tSZ foreground for the search for primordial  $y$-distortions without some sort of spatial filtering, e.g. by masking most galaxy clusters in the maps.

Galaxy clusters are simulated by the  \textsc{PSM} by using both real and random cluster catalogues. A first catalogue of halos randomly distributed over the sky is generated using a Poisson distribution of the mass function from \cite{Tinker:2008ff}. The Compton-$y$ parameter of the tSZ effect is modelled using the universal pressure profile from \cite{Arnaud:2009tt}, while the modelling of the kSZ effect is done by assigning peculiar velocities to each galaxy cluster depending on their redshift using the continuity equation for linear growth of structures. 
Real clusters are also injected in the map using the \textit{Planck}, ACT, SPT, and ROSAT catalogues. In addition, large-scale diffuse tSZ emission is simulated and added to the SZ maps as a Gaussian realisation based on the theoretical tSZ power spectrum for the same aforementioned mass function, pressure profile and cosmological parameters. 

The tSZ maps across the frequency bands are obtained by scaling the simulated Compton-$y$ map using the SED Eq.~\eqref{eq:y_sed} as
\begin{align}
I_\nu^{\rm tSZ}(\hat{n}) = a_{\rm tSZ}(\nu)\, y(\hat{n})\,.
\end{align}
We neglected relativistic corrections to the SZ effect \citep[e.g.,][]{Itoh1998, Chluba:2012dv} in our simulations, although these may become important for future spectral distortion science \citep{Hill2015, Abitbol:2017vwa,Chluba:2019nxa} and SZ analyses \citep{Remazeilles2019,Remazeilles:2019mld}.

%%%%%%%%%%%%%%%%%%%%%%%%%%%%%%%%%%%%%%%%%%%%%%%%%%%%%%
\begin{table}
	\centering
	\caption{Instrumental specifications of \textit{LiteBIRD} \protect\citep{Litebird2021}. The sensitivity on intensity channels (\textit{third column}) is assumed to be a factor of $\sqrt{2}$ higher than the sensitivity on polarization channels (\textit{fourth column}).}
	\label{tab:litebird}
	\begin{tabular}{cccc} % four columns, alignment for each
		\toprule
		Frequency   & Beam FWHM & Sensitivity ($I$) & Sensitivity ($Q$,$U$) \\
		$[\rm GHz]$   & $[\rm arcmin]$  & $[\rm \mu K.arcmin]$ & $[\rm \mu K.arcmin]$\\
		\midrule 
		 40 & 70.5  & 26.46  & 37.42 \\
		 50 &  58.5 & 23.66  & 33.46 \\
		 60 &  51.1  & 15.07  & 21.31 \\
		 68 & 41.6  & 11.93 & 16.87 \\
		 78 & 36.9  & 8.53  & 12.07 \\
		 89 &  33.0  & 7.99  & 11.30 \\
		 100 &  30.2  & 4.64  & 6.56 \\
		 119 & 26.3  & 3.24  & 4.58 \\
		 140 &  23.7   & 3.39  & 4.79 \\
		 166 & 28.9   & 3.94  & 5.57 \\
		 195 & 28.0   & 4.14  & 5.85 \\
		 235 &  24.7  & 7.63  & 10.79 \\
		 280 & 22.5   & 9.76  & 13.80  \\
		 337 & 20.9  & 15.52  & 21.95 \\
		 402 & 17.9   & 33.55 & 47.45\\
		\bottomrule
	\end{tabular}
\end{table}
%%%%%%%%%%%%%%%%%%%%%%%%%%%%%%%%%%%%%%%%%%%%%%%%%%%%%%

\subsection{Experiment specifications}

 \textit{LiteBIRD} \citep{Litebird2019} is a fourth-generation CMB satellite mission selected by the Japan Aerospace Exploration Agency (JAXA) for a launch in 2029. The \textit{LiteBIRD} satellite will scan the full sky during three years through 15 frequency bands ranging from $40$ to $402$\,GHz. With a total of $4339$ transition-edge sensor (TES) detectors on its focal plane, \textit{LiteBIRD} will offer a combined sensitivity from all frequency bands of about $2.16\mu{\rm K.arcmin}$ in polarization \citep{Litebird2021}.
 
 The main instrumental specifications of \textit{LiteBIRD} that we use for our sky simulations are summarised in Table~\ref{tab:litebird}. We assumed that the sensitivities per channel for the temperature are a factor of $\sqrt{2}$ better than the sensitivities per channel for polarization given by \cite{Litebird2021}, thus providing a combined sensitivity from all frequency bands of about $1.53\mu{\rm K.arcmin}$ in temperature.

The simulated maps of the cosmological signal and astrophysical foreground components are coadded in each \textit{LiteBIRD} frequency band, assuming $\delta$-function bandpasses, and convolved with a Gaussian beam of \textit{full width at half maximum} (FWHM) values listed in Table~\ref{tab:litebird}. Using the noise RMS values listed Table~\ref{tab:litebird}, we also simulate Gaussian white noise maps for each frequency channel which we add to the convolved sky maps, thus obtaining 15 \textit{LiteBIRD}  observation maps for each of the $I$, $Q$, and $U$ fields.

\section{Expected ILC noise curves for \texorpdfstring{$\mu$}{mu}-distortion anisotropies}
\label{sec:clarification}

Here we address the exact analytical derivation of the expected noise curve, $C_\ell^{\,\mu\mu, N}$, for the recovered anisotropic $\mu$-distortion signal of a given experiment, and compare it with the actual noise curve as obtained from sky map simulations. We will show how the estimates of the $\mu$-distortion noise, and subsequent forecasts on $\sigma(f_{\rm NL})$, can differ from earlier theoretical estimates in the  literature depending on (i) which effective weighting of the frequency channels is implemented for signal reconstruction and (ii) which frequency bands are used given the effective weighting of the channel sensitivities by the SED of the $\mu$ distortion.

In the absence of foregrounds, the resulting noise RMS, $\sigma_N$, in the reconstructed $\mu$-map obtained from an internal linear combination (ILC) of $n$ frequency channels is given by the analytic expression:
\begin{equation}
\label{eq:analytic0}
{1\over \sigma_N^2} = \sum_{i=1}^{n}  {a_\mu (\nu_i)^2\over \sigma_i^2}\,,
\end{equation}
where $\{\sigma_i\}_{1\leq i \leq n}$ are the \textit{LiteBIRD} noise RMS values across the frequency channels, as listed in Table~\ref{tab:litebird}, and $a_\mu(\nu)$ is the \textit{unitless} version of the peculiar SED of the $\mu$-distortion signal Eq.~\eqref{eq:mu_sed}, i.e.
\begin{equation}
a_\mu(\nu) = {\pi^2\over 18\zeta(3)} - {1\over x}\,,
\end{equation}
evaluated at the frequencies $\nu_i$ of each channel. The weighting of the noise by the $\mu$-distortion SED across the channels is an important factor which makes the effective noise different from the simple inverse-variance weighted mean of the channel sensitivities, the latter being only relevant to achromatic CMB signals.

We emphasize that alternative techniques of $\mu$-distortion signal reconstruction, e.g. an ILC with extra constraints \citep{Remazeilles2018,Remazeilles2021} to deproject some of the foregrounds or taking the difference of a pair of frequency channels \citep{Ganc:2012ae,Mukherjee:2018fxd} to filter out CMB temperature anisotropies, would result in yet another effective weighting of the channel sensitivities different from that of  Eq.~\eqref{eq:analytic0}.

The analytic expression for the projected noise power spectrum in the $\mu$-map is then given by \citep[e.g.,][]{Knox1995, Ganc:2012ae}
\begin{equation}
\label{eq:analytic}
C_\ell^{\,\mu\mu, N} =  {1\over \left(10^6\, T_{\rm CMB}\right)^2} \left({4\pi \sigma_N^2\over N_{\rm pix}}\right) e^{\ell^2\left({\theta_{\rm rad}/ \sqrt{8\ln 2}}\right)}\,,
\end{equation}
where  $\sigma_N$ is given by Eq.~\eqref{eq:analytic0}, $\theta_{\rm rad}$ is the beam FWHM in radian of the reconstructed ILC $\mu$-map, and $N_{\rm pix}$ is the number of effective pixels (or beams) in the map, i.e. ${N_{\rm pix} = 4\pi/\theta_{\rm rad}^2}$. The normalisation factor ${1/(10^6\, T_{\rm CMB})^2}$ in Eq.~\eqref{eq:analytic} is needed for \textit{unitless} noise power since the noise RMS $\sigma_N$ in Eq.~\eqref{eq:analytic0} is in  $\mu{\rm K.arcmin}$ units.

\begin{figure}
  \begin{center}
        \includegraphics[width=\columnwidth]{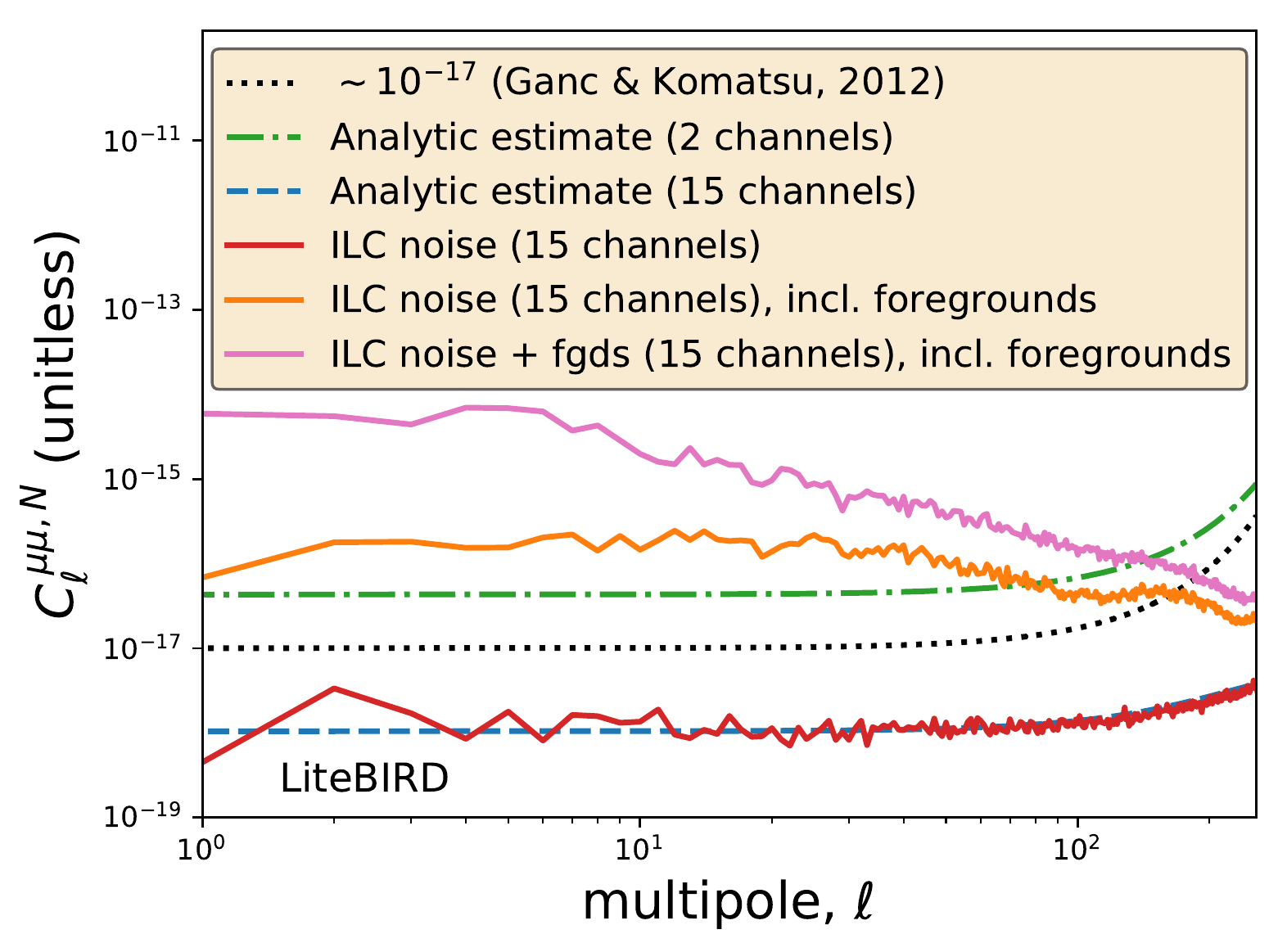}
  \end{center}
\caption{Expected noise curves $C_\ell^{\,\mu\mu, N}$ for the \textit{LiteBIRD} $\mu$-distortion map: estimate from \citet{Ganc:2012ae} based on 2 channels (\textit{dotted black}),  analytic estimate for 2 channels using  \textit{LiteBIRD} noise RMS values and $\mu$-distortion SED weighting (\textit{dash-dotted green}), same analytic estimate for the 15 \textit{LiteBIRD} channels (\textit{dashed blue}), and actual projected noise power spectrum in the ILC $\mu$-map obtained from foreground-free \textit{LiteBIRD} simulations (\textit{solid red}). Our ILC noise is consistent with our analytical estimate for 15 channels. The projected ILC noise power spectrum in the presence of foregrounds in the sky simulation is also plotted for comparison (\textit{solid orange}). The sum of projected noise  and foreground power spectra is also plotted for completeness (\textit{solid pink}).
}
\label{Fig:noise_curves}
\end{figure}

The analytic noise curve $C_\ell^{\,\mu\mu, N}$ (Eq.~\ref{eq:analytic}) is plotted in Fig.~\ref{Fig:noise_curves}  for two cases: (i)  $n=15$ \textit{LiteBIRD} channels (dashed blue line) and (ii) the $n=2$ most sensitive channels at $119$ and $140$\,GHz (dash-dotted green line). The actual noise curve as obtained from running the ILC on sky map simulations in the absence of foregrounds (i.e. only $\mu$, CMB, and instrumental noise across the 15 channels) is overplotted in red solid line and perfectly matches our analytic estimate. 

%%%%%%%%%%%%%%%%%%%%%%%%%%%%%%%%%%%%%%%%%%%%%%%%%%%%%%
%%%%%%%%%%%%%%%%%%%%%%%%%%%%%%%%%%%%%%%%%%%%%%%%%%%%%%

Using Eq.~\eqref{eq:analytic}  and Eq.~\eqref{eq:analytic0}, we emphasize that the factor of improvement on the $\mu$-distortion noise from 2 channels ($119$, $140$\,GHz) to all 15 channels is actually quite large:
\begin{equation}
{C_\ell^{\,\mu\mu, N}[15\, {\rm channels}] \over C_\ell^{\,\mu\mu, N}[2\, {\rm channels}]} = { {a_\mu (119\,{\rm GHz})^2\over \sigma_{119\,{\rm GHz}}^2} + {a_\mu (140\,{\rm GHz})^2\over \sigma_{140\,{\rm GHz}}^2} \over \sum_{i=1}^{15}  {a_\mu (\nu_i)^2\over \sigma_i^2}}  \simeq {1\over 127}\,,
\end{equation}
because of the modulation by the $\mu$-distortion SED.
In contrast, the gain on sensitivity for the achromatic CMB temperature signal, which is given by the inverse variance weighted mean of the channel sensitivities, is much less significant:
\begin{equation}
{C_\ell^{\,TT, N}[15\, {\rm channels}] \over C_\ell^{\,TT, N}[2\, {\rm channels}]} = { {1 \over \sigma_{119\,{\rm GHz}}^2} + {1 \over \sigma_{140\,{\rm GHz}}^2} \over \sum_{i=1}^{15}  {1 \over \sigma_i^2}}  \simeq {1\over 3}\,.
\end{equation}

As a reference, we added to Fig.~\ref{Fig:noise_curves} the noise estimate from \cite{Ganc:2012ae} for \textit{LiteBIRD}, ${C_\ell^{\,\mu\mu, N} \sim 10^{-17}}$ (black dotted line),  which results from taking the difference between 2 consecutive  channel maps to cancel out CMB temperature anisotropies while conserving the $\mu$-distortion signal. Although the noise estimate from \cite{Ganc:2012ae} relies on a different version of \textit{LiteBIRD} back to 2012 and on a weighting of two channels that is different from that of the ILC, it is of same order of magnitude than our analytic ILC noise estimate for two channels ($119$, $140$\,GHz), while it is about an order of magnitude larger than the ILC noise derived from all 15 channels in absence of foregrounds.

In the presence of foregrounds, the effective noise in the recovered $\mu$-distortion map is expected to increase since some of the low- and high-frequency channels are requisitioned by the ILC to subtract the foreground contamination, thus reducing the effective number of channels used to mitigate the noise. This is shown by the solid orange line in Fig.~\ref{Fig:noise_curves}, where the effective ILC noise curve from 15 channels in the presence of foregrounds is found to be of similar order of magnitude than the ILC noise for two channels in the absence of foregrounds. The addition of the power spectrum of projected foregrounds to the noise, i.e. $C_\ell^{\,\mu\mu, N}+C_\ell^{\,\mu\mu, FG}$, is also shown through the solid pink line, which corresponds to the actual contribution to the total uncertainty on anisotropic $\mu$-distortion measurements.

%%%%%%%%%%%%%%%%%%%%%%%%%%%%%%%%%%%%%%%%%%%%%%%%%%%%%%
\begin{figure}
  \begin{center}
     \includegraphics[width=\columnwidth]{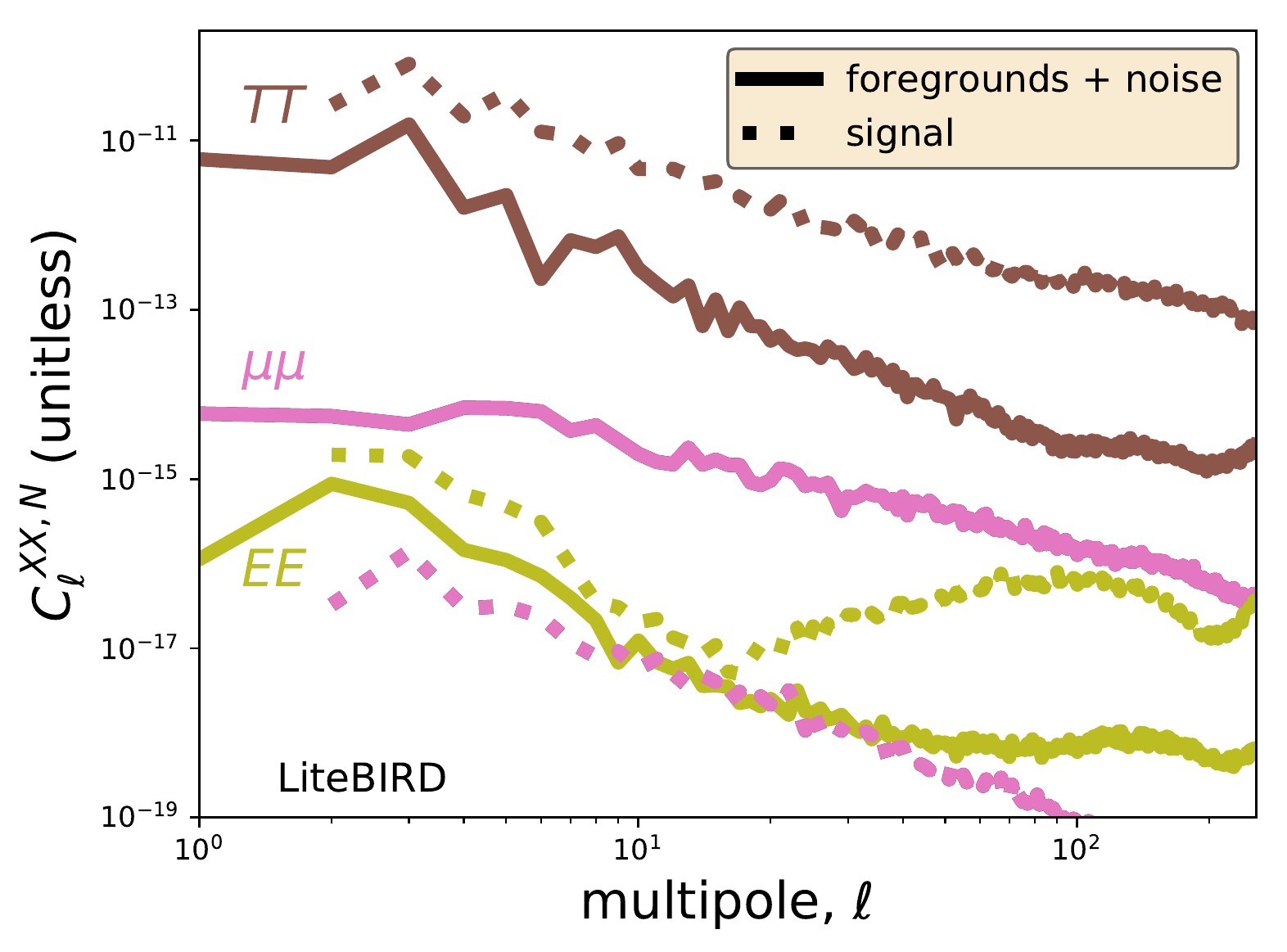}
  \end{center} 
\caption{Power spectrum of the cosmological signals (\emph{dotted lines}) versus noise curves (\emph{solid lines}) for each of the observables $\mu$ (\emph{pink}), $T$ (\emph{brown}) and $E$ (\emph{olive green}).}
\label{Fig:snr}
\end{figure}
%%%%%%%%%%%%%%%%%%%%%%%%%%%%%%%%%%%%%%%%%%%%%%%%%%%%%%

In Fig.~\ref{Fig:snr}, the overall noise curves $C_\ell^{TT, N}$ (solid brown line) and $C_\ell^{EE, N}$ (solid olive green line) accounting for residual foreground and noise contamination in the recovered ILC maps of CMB temperature and CMB $E$-mode polarization, respectively, are shown along with the overall noise curve of the $\mu$-distortion discussed before (solid pink line). On top of the noise curves, the auto-power spectra of the cosmological signals are plotted as dotted lines. This highlights that the overall uncertainty on CMB temperature and $E$-mode measurements by \emph{LiteBIRD} is mostly driven by the cosmic variance of the CMB signal, while in contrast the uncertainty on $\mu$-distortion anisotropies is largely dominated by the residual foreground contamination. Therefore, the main limiting factor in $\mu T$ and $\mu E$ cross-power spectrum measurements is not the residual foreground contamination in CMB maps but the residual foreground contamination in the $\mu$-distortion map.

%%%%%%%%%%%%%%%%%%%%%%%%%%%%%%%%%%%%%%%%%%%%%%%%%%%%%%
\begin{figure}
  \begin{center}
     \includegraphics[width=\columnwidth]{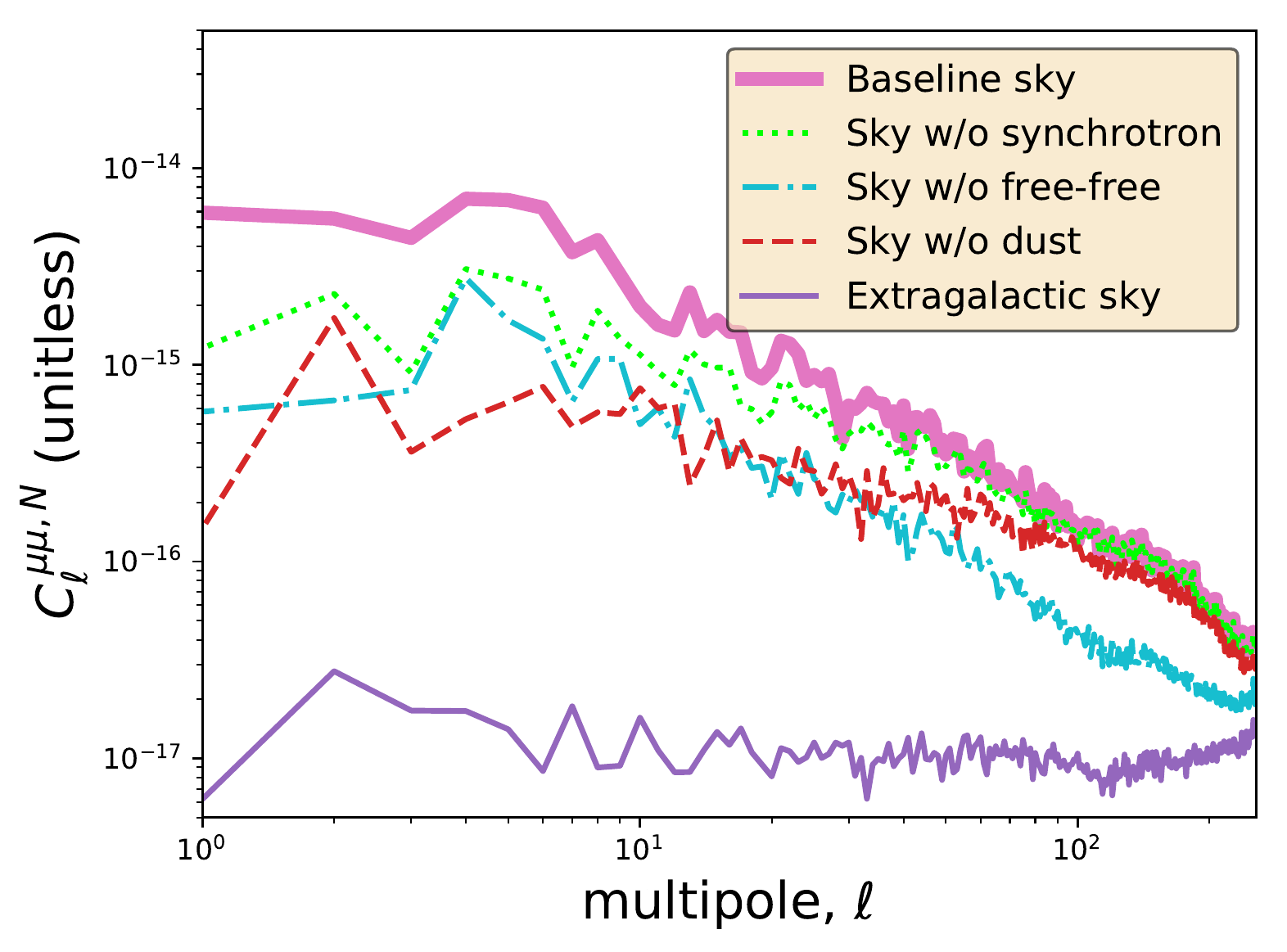}
  \end{center}
\caption{Contribution of various foregrounds to the total error budget on $\mu$-distortion anisotropies. Galactic foreground contamination, in particular thermal dust at $\ell < 20$ and free-free emission at $\ell > 30$, prevails over extragalactic foregrounds (see main text for details).}
\label{Fig:contributions}
\end{figure}
%%%%%%%%%%%%%%%%%%%%%%%%%%%%%%%%%%%%%%%%%%%%%%%%%%%%%%

In Fig.~\ref{Fig:contributions}, we investigate which foregrounds prevail in the total error budget of the recovered $\mu$-distortion anisotropies. The thick solid pink line in Fig.~\ref{Fig:contributions} is just the replicate of the one in Fig.~\ref{Fig:noise_curves} and Fig.~\ref{Fig:snr}, showing the power spectrum of the total foreground and noise contamination in the recovered $\mu$-distortion map for the baseline sky simulation which includes all the foregrounds. In contrast, the solid purple line shows the sole contribution from the extragalactic foregrounds (SZ, CIB, CMB) to the error budget, i.e. ignoring Galactic foregrounds in the sky. The significant drop off of the noise curve in this case at low multipoles, where bulk of the $\mu T$/$\mu E$  correlation and constraining power on $f^\mu_{\rm NL}$ lie, highlights that Galactic foregrounds largely prevail over extragalactic foregrounds in the total error budget. The green dotted line, cyan dash-dotted line and red dashed line show the resulting noise curve when removing from the baseline sky simulation either the synchrotron, free-free or thermal dust component, respectively. The resulting gain in sensitivity across multipoles shows that thermal dust (dashed red) was accounting for the largest contribution to the $\mu$-distortion noise at low multipoles $\ell < 20$ while Galactic free-free emission (dash-dotted cyan) is the most damaging foreground at multipoles $\ell > 30$.

Finally, while we were investigating on the actual noise curves for anisotropic $\mu$-distortions, we noticed that in our earlier paper \citep{Remazeilles2018}, we omitted the binning factor $\sqrt{\Delta\ell}$ in our estimates of the error on the binned power spectra. While this omission does not change the relevance of the component method that was developed in the paper nor the conclusion that foregrounds degrade the recovery of the $\mu$-distortion anisotropies by about one order of magnitude, our earlier forecasts on $\sigma(f^\mu_{\rm NL})$ were more pessimistic than they should be since the $\sigma(f^\mu_{\rm NL})$ values computed in \cite{Remazeilles2018} should actually be divided by a factor of $\sqrt{30}\sim 5$.
By including CMB polarization in the current analysis to leverage the detection of anisotropic $\mu$-distortions, we will see that the expected constraints on $\sigma(f_{\rm NL})$ for \textit{LiteBIRD} are even more promising than any of earlier forecasts in the literature.

\section{Forecasts}
\label{sec:forecast}

\subsection{Component separation}
\label{subsec:cilc}

For any line-of-sight $\hat{n}$ and any frequency $\nu$, the observed data in either temperature or $E$-mode polarization can be decomposed in thermodynamic temperature units as follows:
\begin{subequations}
\begin{align}
T_\nu(\hat{n}) &= a_\mu(\nu)\,\mu(\hat{n})\,+T_{\rm CMB}(\hat{n})\, +\, n_T(\nu,\hat{n})\,, \label{eq:t}\\
E_\nu(\hat{n}) &= E_{\rm CMB}(\hat{n}) + n_E(\nu,\hat{n})\,, \label{eq:e}
\end{align}
\end{subequations}
where $\mu(\hat{n})$ is the anisotropic $\mu$-distortion signal that we aim at recovering and $a_\mu(\nu)$ is the peculiar SED of the $\mu$ distortion (Eq.~\ref{eq:mu_sed}), while $T_{\rm CMB}(\hat{n})$ and $E_{\rm CMB}(\hat{n})$ are, respectively, the CMB temperature anisotropies and CMB $E$-mode polarization anisotropies, which both are achromatic, i.e. independent of frequency, in thermodynamic temperature units. Astrophysical foregrounds and instrumental noise are collected together into unparameterized terms $n_T(\nu,\hat{n})$ and $n_E(\nu,\hat{n})$ for the temperature and polarization channels, respectively.

Equations~\eqref{eq:t}-\eqref{eq:e} highlight two important aspects:  
\begin{enumerate}[label=(\roman*)]
\item CMB temperature anisotropies are a peculiar foreground to $\mu$-distortion anisotropies because they are also correlated with the $\mu$-distortion signal. Hence, residual CMB temperature anisotropies in the reconstructed $\mu$-map, $\widehat{\mu}(\hat{n})$, from temperature channels may bias the measurement of the $\mu T$ cross-power spectrum $C_\ell^{\,\mu T}$ because of residual $TT$ correlations \citep{Remazeilles2018}:
\begin{align}
\label{eq:bias}
\widehat{\mu}(\hat{n}) = \mu(\hat{n}) + \varepsilon T_{\rm CMB}(\hat{n}) + \dots  \Rightarrow  C_\ell^{\widehat{\mu}\, \widehat{T}} = C_\ell^{\,\mu T} + \varepsilon C_\ell^{ TT} + \dots
\end{align}
where $\varepsilon < 1$ is an arbitrary percentage of residual CMB contamination in the reconstructed $\mu$-map.
\item In contrast to temperature channels, polarization channels are free from $\mu$-distortion signals, so that any CMB $E$-mode map obtained from the data can be safely cross-correlated with the recovered $\mu$-map to measure $C_\ell^{\,\mu E}$ without suffering from spurious $\mu\mu$, $TT$, and $EE$ correlations.
\end{enumerate}

Following the methodology of \cite{Remazeilles2018}, we perform foreground cleaning and component separation by means of \textit{Constrained} needlet ILC (CILC) methods \citep{Remazeilles2011,Remazeilles2021} instead of standard needlet ILC \citep[NILC;][]{Delabrouille2009} methods in order to get rid of spurious correlations between the CMB foreground and the $\mu$-distortion signal (see Eq.~\ref{eq:bias}). 

\subsubsection{CMB-free $\mu$-map}

The estimate of the $\mu$-map is obtained from the constrained linear combination of the temperature channel maps as
\begin{align}
\label{eq:linear}
\widehat{\mu}(\hat{n}) = \sum_\nu w(\nu) \, T_\nu(\hat{n}) \equiv \bdw^{t}\, \bdT(\hat{n}) \,,
\end{align}
in which the CILC weights $\bdw\equiv \{w(\nu)\}_\nu$ fulfill three conditions
\begin{equation}
\label{eq:cond_cilc}
\begin{cases}
{\partial \over \partial \bdw} \left(\bdw^{t}\, \tC_T\, \bdw\right) = 0\,, \\[1.5mm]
\sum_\nu w(\nu)\, a_\mu(\nu) = 1\,,\\[1.5mm]
\sum_\nu w(\nu)\, a_{\rm CMB}(\nu) = 0\,,
\end{cases}
\end{equation}
where the CMB SED $a_{\rm CMB}(\nu) = 1$ in thermodynamic units because of achromaticity, and $\tC_T = \langle \bdT(\hat{n})\, \bdT(\hat{n})^{t}\rangle$ is the covariance matrix of the temperature data. 
The first condition of Eq.~\eqref{eq:cond_cilc} guarantees the mitigation of astrophysical foreground and noise contamination through minimisation of the variance ${\langle \widehat{\mu}(\hat{n})^2 \rangle = \bdw^{t}\, \tC_T\, \bdw}$ of the estimate. The second condition of Eq.~\eqref{eq:cond_cilc} guarantees the full conservation of the signal of interest, here $\mu(\hat{n})$, despite variance minimization. The third constraint of Eq.~\eqref{eq:cond_cilc} guarantees the full cancellation of residual CMB temperature anisotropies in the recovered $\mu$-map, i.e. $\varepsilon = 0$ in Eq.~\eqref{eq:bias}.

The CILC solution to Eqs.~\eqref{eq:linear}-\eqref{eq:cond_cilc} for the reconstructed $\mu$-map is given by \citep{Remazeilles2018}
\begin{align}
\label{eq:mu_map}
 \widehat{\mu}(\hat{n}) =
  { \left( \bda_{\rm CMB}^{t} \tC_T^{-1} \bda_{\rm CMB} \right)
  \bda_{\mu}^{t} \tC_T^{-1} - \left( \bda_{\mu}^{t}
  \tC_T^{-1} \bda_{\rm CMB} \right) \bda_{\rm CMB}^{t}
  \tC_T^{-1}
  \over
  \left(\bda_{\mu}^{t} \tC_T^{-1} \bda_{\mu}
  \right)\left(\bda_{\rm CMB}^{t} \tC_T^{-1} \bda_{\rm CMB} \right) -
  \left( \bda_{\mu}^{t} \tC_T^{-1} \bda_{\rm CMB} \right)^2 }\,
  \bdT(\hat{n}).
\end{align}
We refer to \cite{Remazeilles2011,Remazeilles2021} for more details about the derivation of the CILC Eq.~\eqref{eq:mu_map}.

We emphasize that without the extra nulling constraint of the CILC in Eq.~\eqref{eq:cond_cilc}, the solution would reduce to the standard NILC solution, for which the recovered $\mu$-map would suffer from residual CMB temperature anisotropies since the NILC weights would no longer be orthogonal to the CMB SED.

\subsubsection{$\mu$-free CMB temperature map}

By interchanging the second and third conditions in Eq.~\eqref{eq:cond_cilc}, i.e. ensuring ${\sum_\nu w(\nu)\, a_\mu(\nu) = 0}$ and ${\sum_\nu w(\nu)\, a_{\rm CMB}(\nu) = 1}$, we similarly guarantee full cancellation of residual $\mu$-distortions in the recovered CMB temperature map. With this prescription, the CILC solution for the reconstructed CMB temperature map is given by:
\begin{align}
\label{eq:t_map}
 \widehat{T}_{\rm CMB}(\hat{n}) =
  { \left( \bda_\mu^{t} \tC_T^{-1} \bda_\mu \right)
  \bda_{\rm CMB}^{t} \tC_T^{-1} - \left( \bda_{\rm CMB}^{t}
  \tC_T^{-1} \bda_\mu \right) \bda_\mu^{t}
  \tC_T^{-1}
  \over
  \left(\bda_{\rm CMB}^{t} \tC_T^{-1} \bda_{\rm CMB}
  \right)\left(\bda_\mu^{t} \tC_T^{-1} \bda_\mu \right) -
  \left( \bda_{\rm CMB}^{t} \tC_T^{-1} \bda_\mu \right)^2 }\,
  \bdT(\hat{n}).
\end{align}
%------------------
Likewise the CMB-free $\mu$-map (Eq.~\ref{eq:mu_map}) which allows us to get rid of spurious $TT$ correlations in the recovered cross-power spectrum $C_\ell^{\widehat{\mu}\, \widehat{T}}$ (Eq.~\ref{eq:bias}), the $\mu$-free CMB temperature map (Eq.~\ref{eq:t_map}) will allow us to get rid of any residual $\mu\mu$ correlations.  

\subsubsection{CMB $E$-mode map}

Since the polarization channels are naturally immune from $\mu$-distortions and CMB temperature anisotropies, we do not need to impose nulling constraints in this case. Hence, the CMB $E$-mode map is obtained from the standard NILC method:
\begin{align}
\label{eq:e_map}
 \widehat{E}_{\rm CMB}(\hat{n}) =
  { \bda_{\rm CMB}^{t} \tC_E^{-1}  \over
  \left(\bda_{\rm CMB}^{t} \tC_E^{-1} \bda_{\rm CMB}
  \right) }\,\bdE(\hat{n})\,,
\end{align}
where $\tC_E = \langle \bdE(\hat{n})\, \bdE(\hat{n})^{t}\rangle$ is the covariance matrix of the $E$-mode polarization data. 

Cross-correlating the NILC CMB $E$-mode map Eq.~\eqref{eq:e_map} with the CILC $\mu$-map Eq.~\eqref{eq:mu_map}, which is free from CMB temperature anisotropies, will allow us to get rid of residual $TE$ correlations in the recovered $\mu E$ cross-power spectrum $C_\ell^{\,\widehat{\mu}\, \widehat{E}}$.

Additional nulling constraints can in principle be added to the CILC in order to deproject e.g. thermal SZ effects and primordial $y$-distortions from either the $\mu$-distortion map or the CMB temperature map\footnote{For cross power spectrum analysis, a given foreground only needs to be deprojected from one of the cross-correlated maps, either $\widehat{\mu}$ or $\widehat{T}_{\rm CMB}$, to remove biases from $C_\ell^{\,\widehat{\mu}\, \widehat{T}}$. This also tempers the noise penalty from deprojection.}, but also to remove the bulk of Galactic foreground contamination by nulling its spectral moments \citep[see][]{Remazeilles2021}. With $m$ additional nulling constraints, ${\sum_\nu w(\nu)\, b_i(\nu)=0}$, against the SEDs $\{b_i\}_{1\leq i \leq m}$ of specific foregrounds, the expression for e.g. the recovered $\mu$-map (Eq.~\ref{eq:mu_map}) would simply be generalised as
\begin{align}
\label{eq:mu_map_plus}
 \widehat{\mu}(\hat{n}) = \bde^{\rm t}\left(\tA^t\tC_T^{-1}\tA\right)^{-1}\tA^t\tC_T^{-1}\, \bdT(\hat{n})\,,
\end{align}
where 
\begin{align}
\label{eq:sedmatrix}
\tA =
\begin{pmatrix}
\bda_\mu & \bda_{\rm CMB} & \bdb_1 &  \dots &  \bdb_m
\end{pmatrix},
\end{align}
is the $(m+2)\times n$ SED matrix for $m+2$ constraints and $n$ channels, and 
\begin{align}
\bde^t =
\begin{pmatrix}
1 & 0 & 0 & \dots & 0
\end{pmatrix}.
\end{align}
is a $(m+2)$-dimensional vector.

Finally, both CILC and NILC methods are implemented on spherical wavelets called \textit{needlets} \citep{Narcowich2006,Guilloux2009}, whose excellent properties of localization in both direct pixel space and conjugate harmonic space allow to adjust ILC filters to the local variations of the foreground and noise contamination both across the sky and across the scales. For details on the needlet implementation of the NILC and CILC filters, we refer to e.g. \cite{Delabrouille2009,Basak2012,Remazeilles2021}.

\subsection{Reconstructed \texorpdfstring{$\mu T$}{muT} and \texorpdfstring{$\mu E$}{muE} cross-power spectra}
\label{subsec:spectra}

The recovered $\mu$-distortion, CMB temperature and CMB $E$-mode all-sky maps after foreground cleaning and component separation are then masked prior to computing their cross-power spectrum. We use a Galactic mask leaving $f_{\rm sky}=65$\% of observed sky in order to mitigate residual Galactic foreground emission in the Galactic plane.

The recovered power spectra are then deconvolved from the Galactic mask, using MASTER \citep{Hivon2002}, and binned on multipoles with a binsize of $\Delta\ell = 30$. Error bars in each multipole bin are computed analytically from the recovered power spectra after component separation:
\begin{equation}
\label{eq:error}
\sigma^{\,\mu X}_{\ell_b} = \sqrt{\left(\widehat{C}_{\ell_b}^{\,\mu X}\right)^2 + \widehat{C}_{\ell_b}^{\,\mu\mu}\,\widehat{C}_{\ell_b}^{\,XX}\over\left(2\ell_b+1\right)f_{\rm sky}\,\Delta\ell}\,,
\end{equation}
where $X$ stands for either $T$ or $E$ field, $\ell_b$ is the central multipole of the bin, $\Delta\ell$ is the bin size, and $f_{\rm sky}$ is the fraction of sky outside the Galactic mask. Since the recovered maps include, on top of the signal, any residual foregrounds and noise that projected along with the signal, the uncertainty based on recovered map power spectra thus receives contributions from both signal cosmic variance and residual foreground and noise sample variance.

\begin{figure*}
  \begin{center}
    \includegraphics[width=\columnwidth]{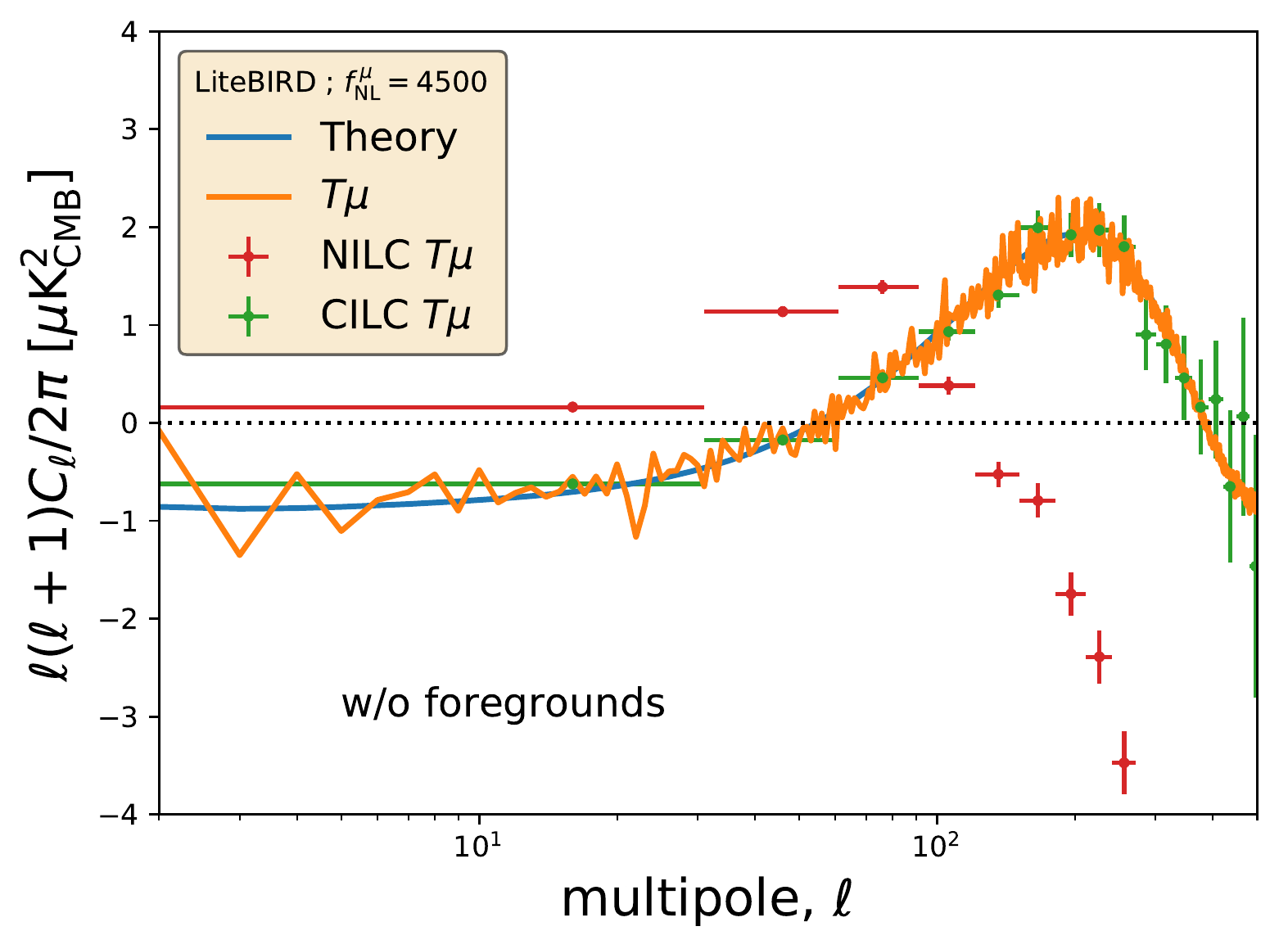}~
     \includegraphics[width=\columnwidth]{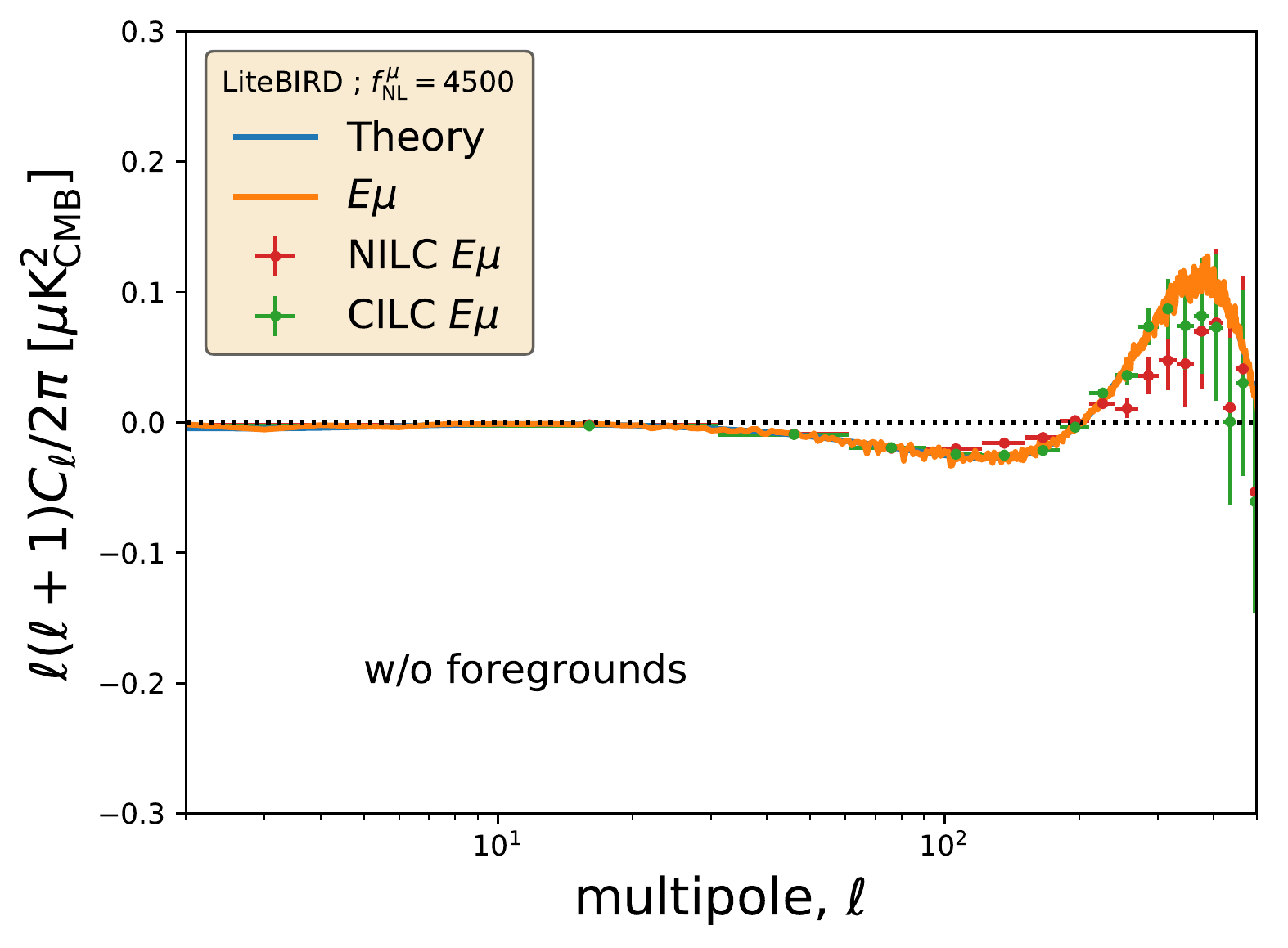}~
	\\[1.5mm]
	 \includegraphics[width=\columnwidth]{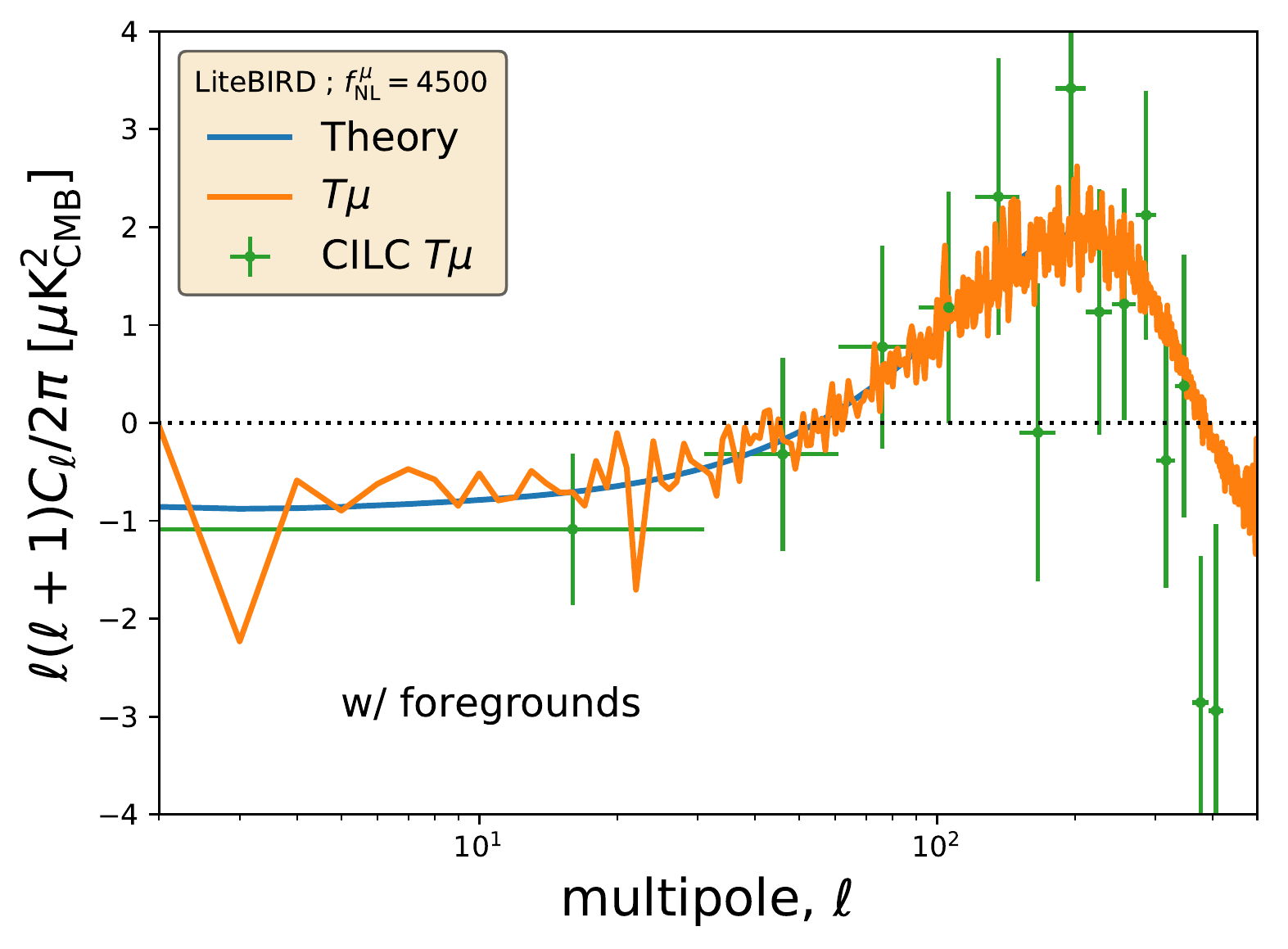}~
        	\includegraphics[width=\columnwidth]{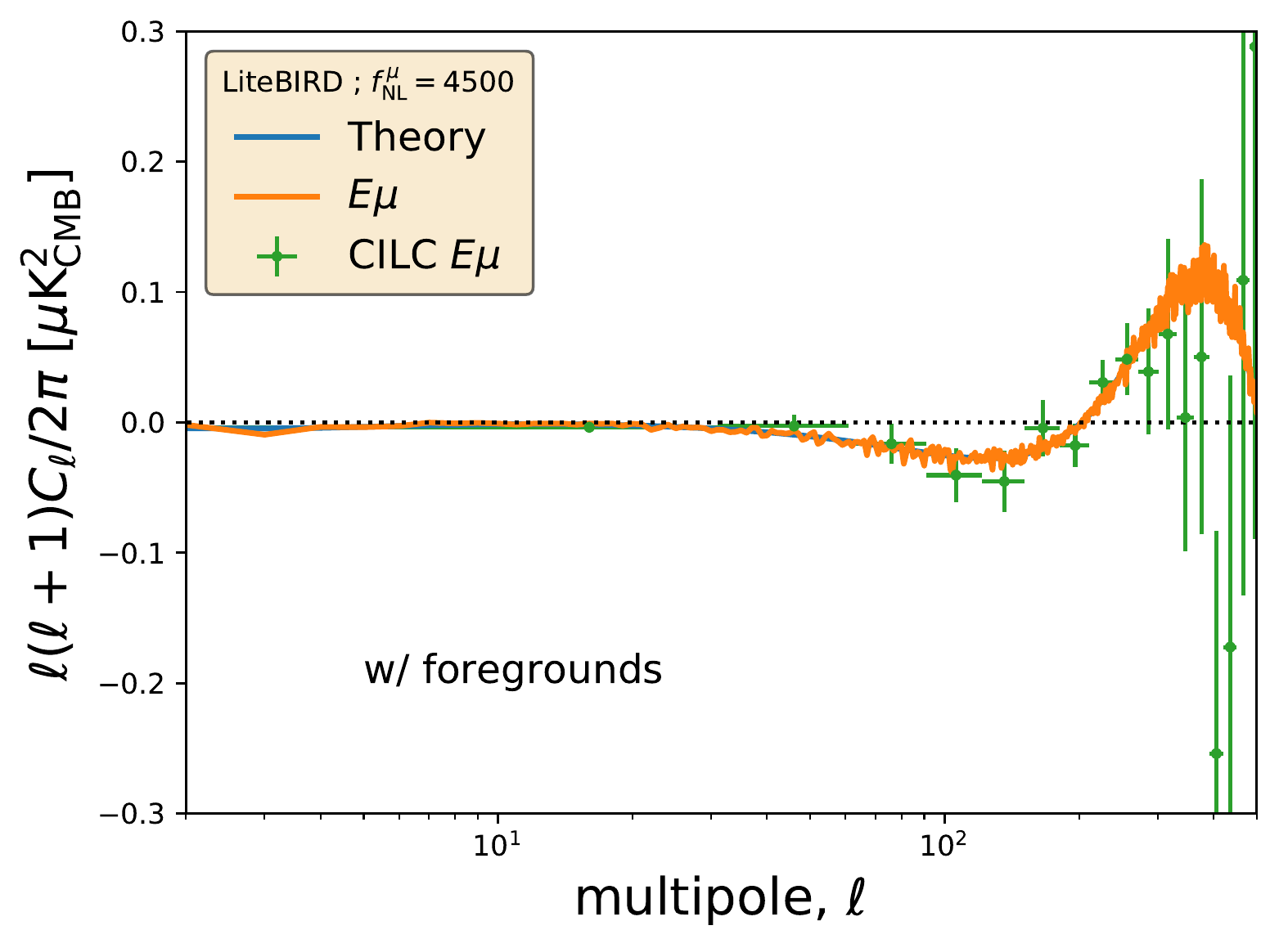}~
 \end{center}
\caption{Cross-power spectrum between the recovered anisotropic $\mu$-distortion map ($f_{\rm NL}^{\mu}=4500$) and the recovered CMB temperature and $E$-mode maps after foreground cleaning and component separation: $\widehat{C}_\ell^{\,\mu T}$ (\textit{left column}) and $\widehat{C}_\ell^{\,\mu E}$ (\textit{right column}), either without foregrounds (\textit{upper row})  or with foregrounds (\textit{lower row}). The upper row displays also NILC results (\textit{red dots}) versus CILC results  (\textit{green dots}), highlighting that NILC results suffer from spurious $TT$ and $\mu\mu$ correlations due to CMB $T$ residuals in the $\mu$-map and $\mu$ residuals in the CMB $T$ map, which biases $\widehat{C}_\ell^{\,\mu T}$. In contrast, CILC allows to fully recover $\widehat{C}_\ell^{\,\mu T}$ and $\widehat{C}_\ell^{\,\mu E}$ without bias. $\widehat{C}_\ell^{\,\mu E}$ is also less affected by spurious correlations since the CMB $E$-mode polarization is free from $\mu$- distortions.}
\label{Fig:nilc_vs_cilc}
\end{figure*}

\begin{figure*}
  \begin{center}
     \includegraphics[width=\columnwidth]{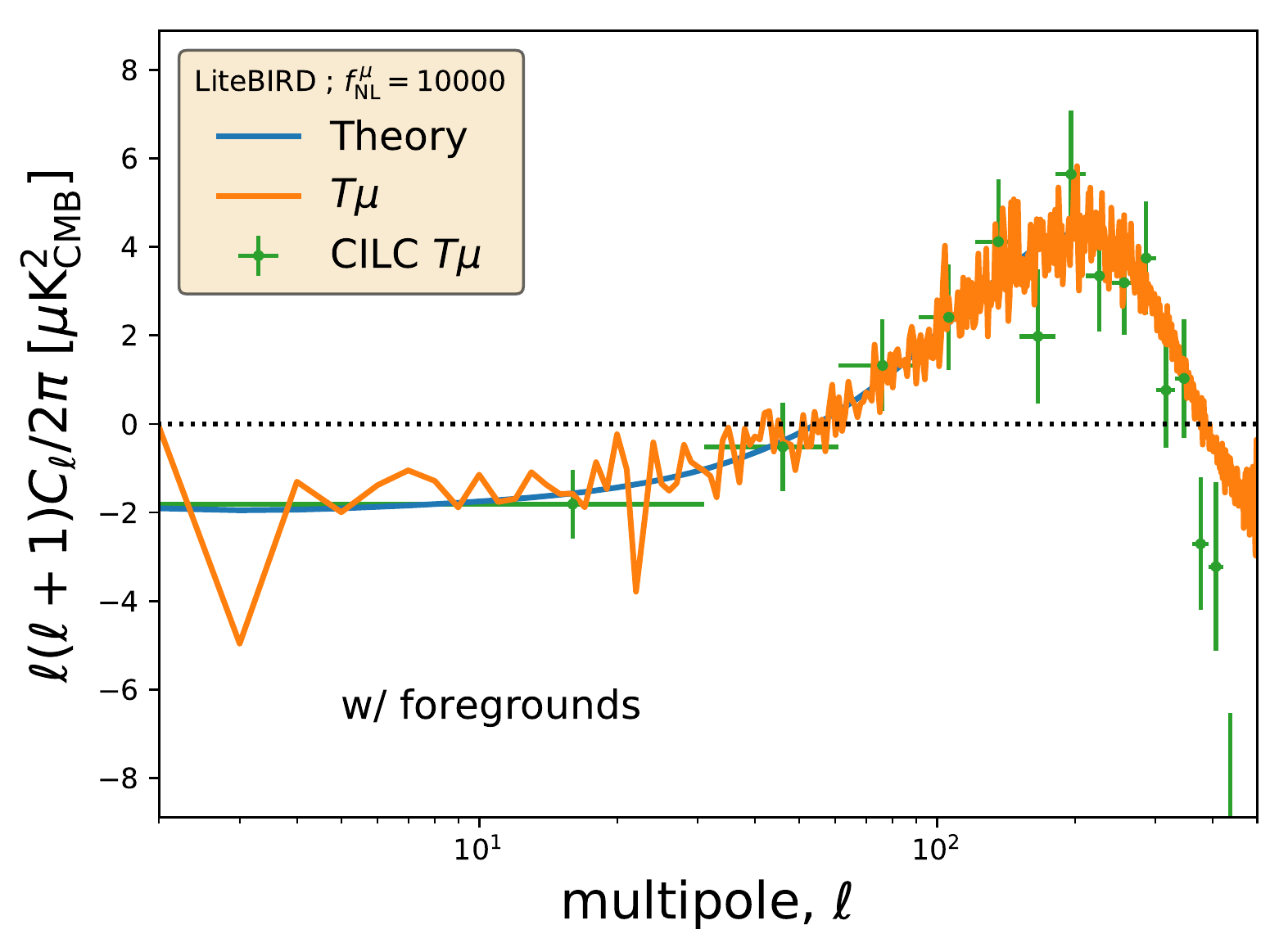}~
     \includegraphics[width=\columnwidth]{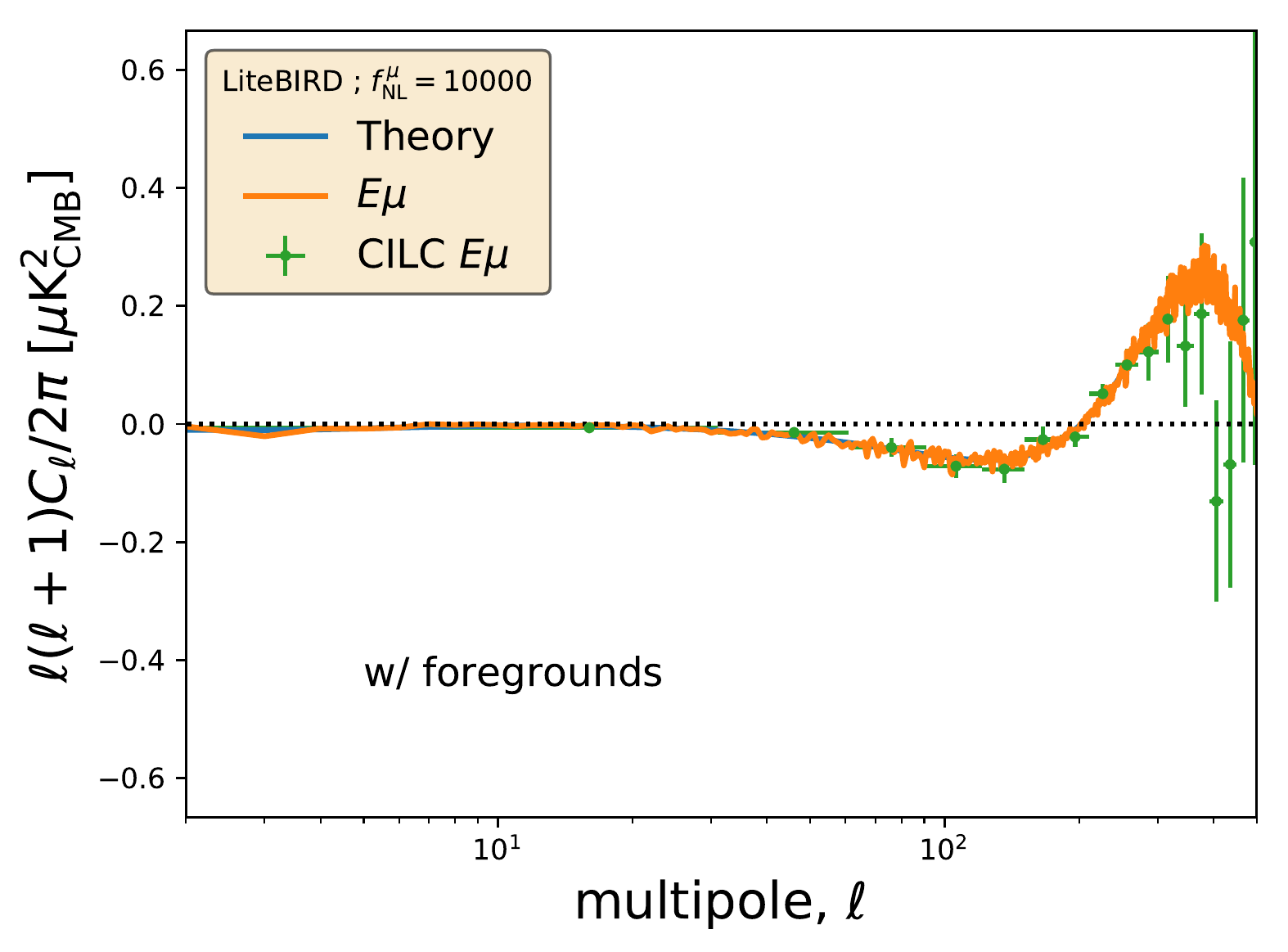}~
    \\[1.5mm]
        \includegraphics[width=\columnwidth]{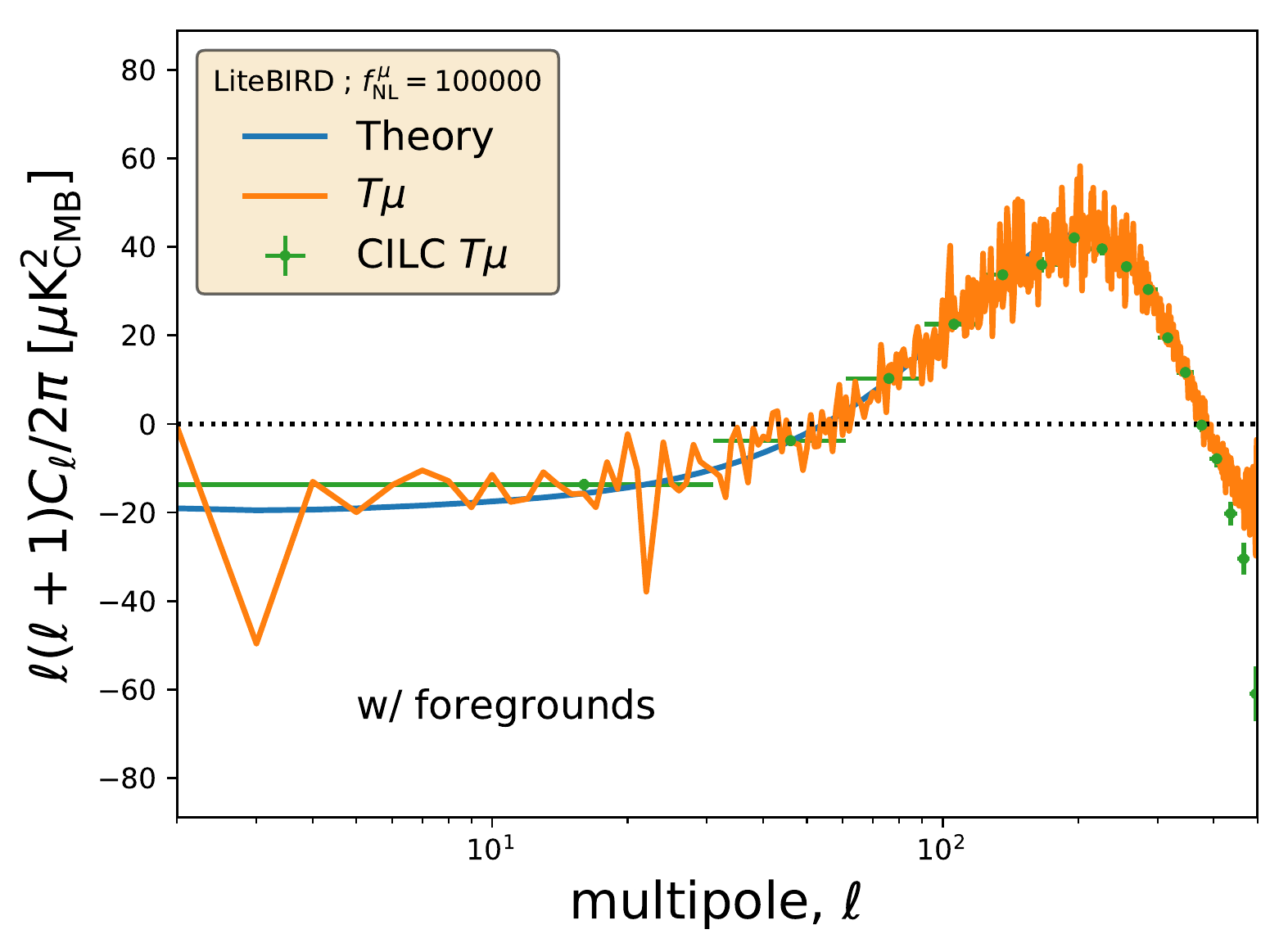}~
        	\includegraphics[width=\columnwidth]{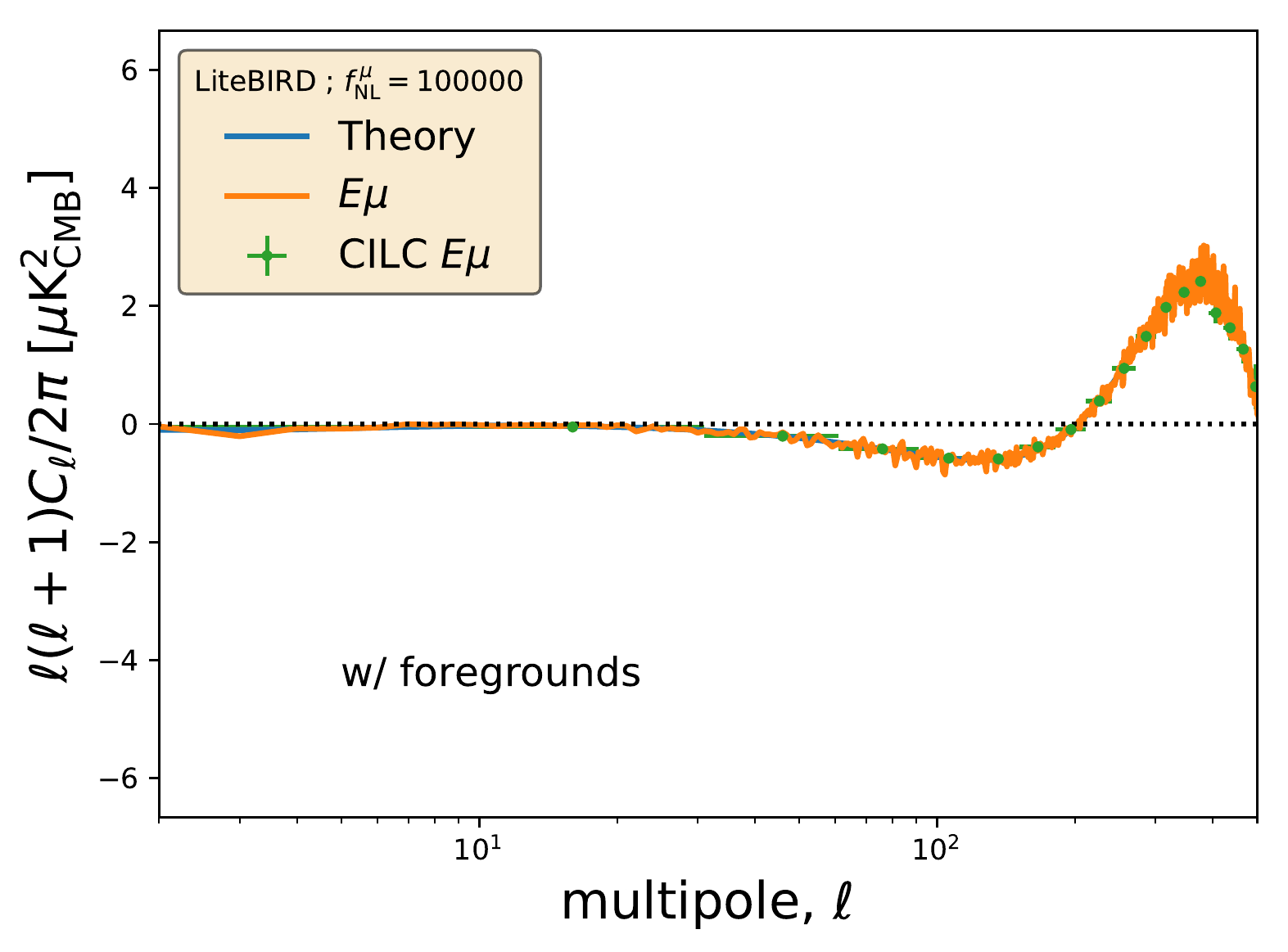}~
 \end{center}
\caption{Cross-power spectrum between the recovered anisotropic $\mu$-distortion map and the recovered CMB temperature and $E$-mode maps after foreground cleaning and component separation: $\widehat{C}_\ell^{\,\mu T}$ (\textit{left column}) and $\widehat{C}_\ell^{\,\mu E}$ (\textit{right column}), for fiducial values $f_{\rm NL}^{\mu}=10^4$ (\textit{top row}) and $f_{\rm NL}^{\mu}=10^5$ (\textit{bottom row}).}
\label{Fig:cross}
\end{figure*}

Figures~\ref{Fig:nilc_vs_cilc} and \ref{Fig:cross} show the cross-power spectrum between the recovered map of $\mu$-distortion anisotropies, $\widehat{\mu}(\hat{n})$, and the recovered maps of CMB temperature and $E$-mode anisotropies, $\widehat{T}_{\rm CMB}(\hat{n})$ and $\widehat{E}_{\rm CMB}(\hat{n})$, after foreground cleaning and component separation, for different fiducial $f_{\rm NL}^{\mu}$ values. In all figures, the left panels show $C_\ell^{\,\widehat{\mu}\, \widehat{T}}$, while the right panels show $C_\ell^{\,\widehat{\mu}\, \widehat{E}}$. The blue solid lines show the fiducial $\mu T$ and $\mu E$ cross-power spectra as expected from the theory, while the orange solid lines show the cross-power spectra from the input signal map realisations of the simulation. The binned cross-power spectra of the recovered signal maps after component separation are plotted as green dots.

%%%%%%%%%%%%%%%%%%%%%%%%%%%%%%%%%%%%%%%%%%%%%%%%%%%%%%
%%%%%%%%%%%%%%%%%%%%%%%%%%%%%%%%%%%%%%%%%%%%%%%%%%%%%%

%%%%%%%%%%%%%%%%%%%%%%%%%%%%%%%%%%%%%%%%%%%%%%%%%%%%%%
%%%%%%%%%%%%%%%%%%%%%%%%%%%%%%%%%%%%%%%%%%%%%%%%%%%%%%

Overall, we see that \textit{LiteBIRD} allows us to recover with good accuracy both the $\mu T$ and $\mu E$ correlation signals for each of the fiducial $f_{\rm NL}^{\mu}$ values considered, as long as the CILC component separation approach is used to get rid of spurious correlation biases. In the top panel of Fig.~\ref{Fig:nilc_vs_cilc}, for $f_{\rm NL}^{\mu}=4500$ in the absence of foregrounds, we displayed also the recovered signals if the standard NILC method (red dots) had been used in place of the constrained CILC method (green dots). Clearly, the recovered $C_\ell^{\,\widehat{\mu}\, \widehat{T}}$ signal from NILC is highly biased across the multipoles because of strong spurious $TT$ correlations propagated by residual CMB temperature anisotropies in the NILC $\mu$-map, thus confirming our theoretical expectations from Sect.~\ref{subsec:cilc}. The recovered $C_\ell^{\,\widehat{\mu}\, \widehat{E}}$ signal from NILC is less affected by biases because the $E$-mode channels are immune from $\mu$-distortion and CMB temperature anisotropies, but residual CMB temperature anisotropies in the NILC $\mu$-map still projects residual $TE$ correlations in the cross-power spectrum $C_\ell^{\,\widehat{\mu}\, \widehat{E}}$. In contrast to NILC (red dots), CILC (green dots) allows us to fully recovers both correlated $\mu T$ and $\mu E$ signals without bias.

By comparing the top panels (without foregrounds) and bottom panels (with foregrounds) of Fig.~\ref{Fig:nilc_vs_cilc}, we can appreciate the impact of foregrounds on the recovery of the signals. The increase of uncertainty on the recovered $\mu T$ and $\mu E$ cross-power spectra is clearly driven by residual foreground contamination, not by instrumental noise. This goes in the same direction than our earlier results in \cite{Remazeilles2018}, in which we showed that extended frequency coverage typically provides more leverage to anisotropic $\mu$-distortions than increased detector sensitivities. Finally, comparing Fig.~\ref{Fig:nilc_vs_cilc} and Fig.~\ref{Fig:cross}, we can see how the recovery of the $\mu T$ and $\mu E$ cross-power spectra improves with increasing values of $f_{\rm NL}^\mu$.

\begin{figure*}
  \begin{center}
    \includegraphics[width=\columnwidth]{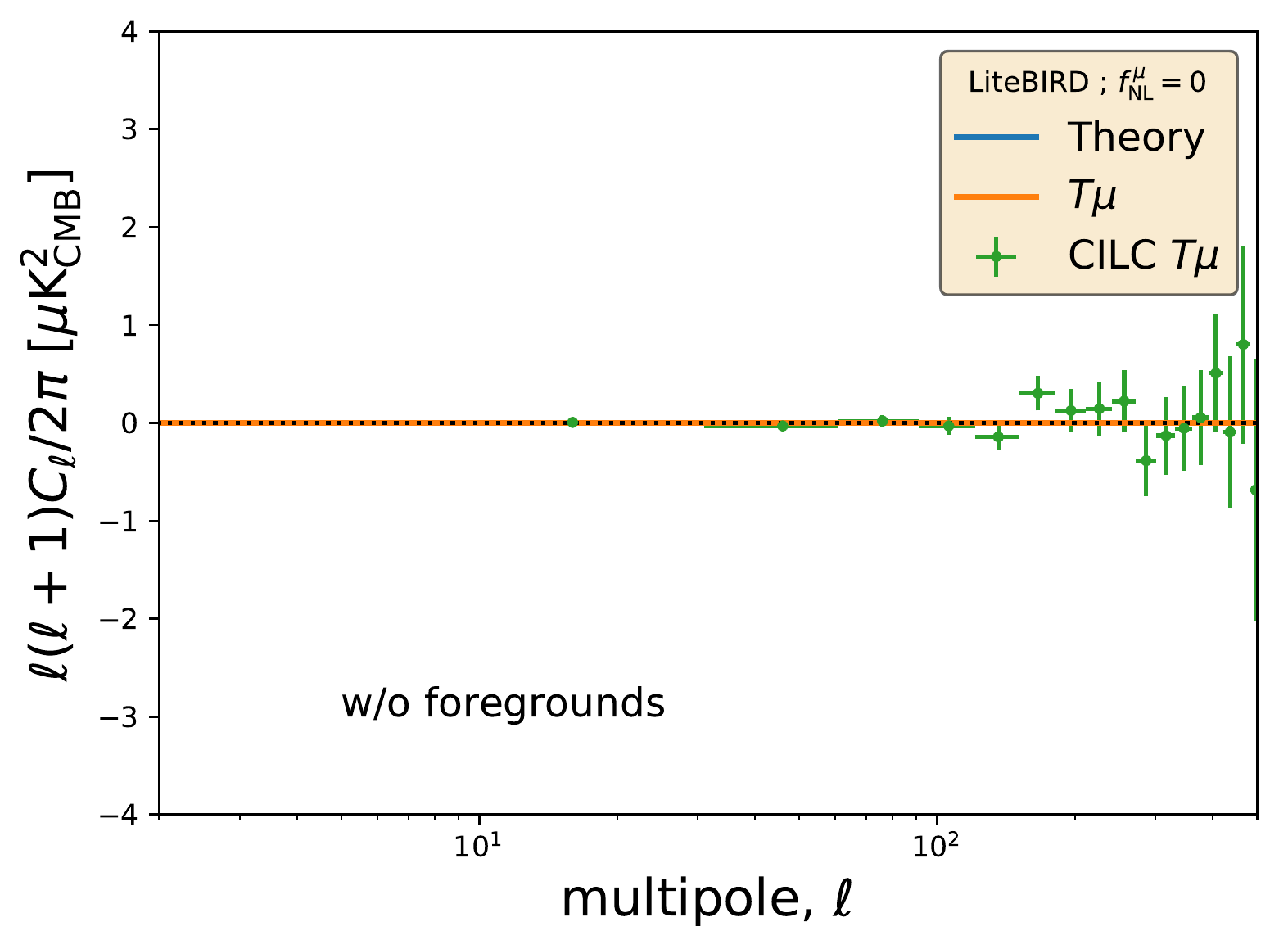}~
        	\includegraphics[width=\columnwidth]{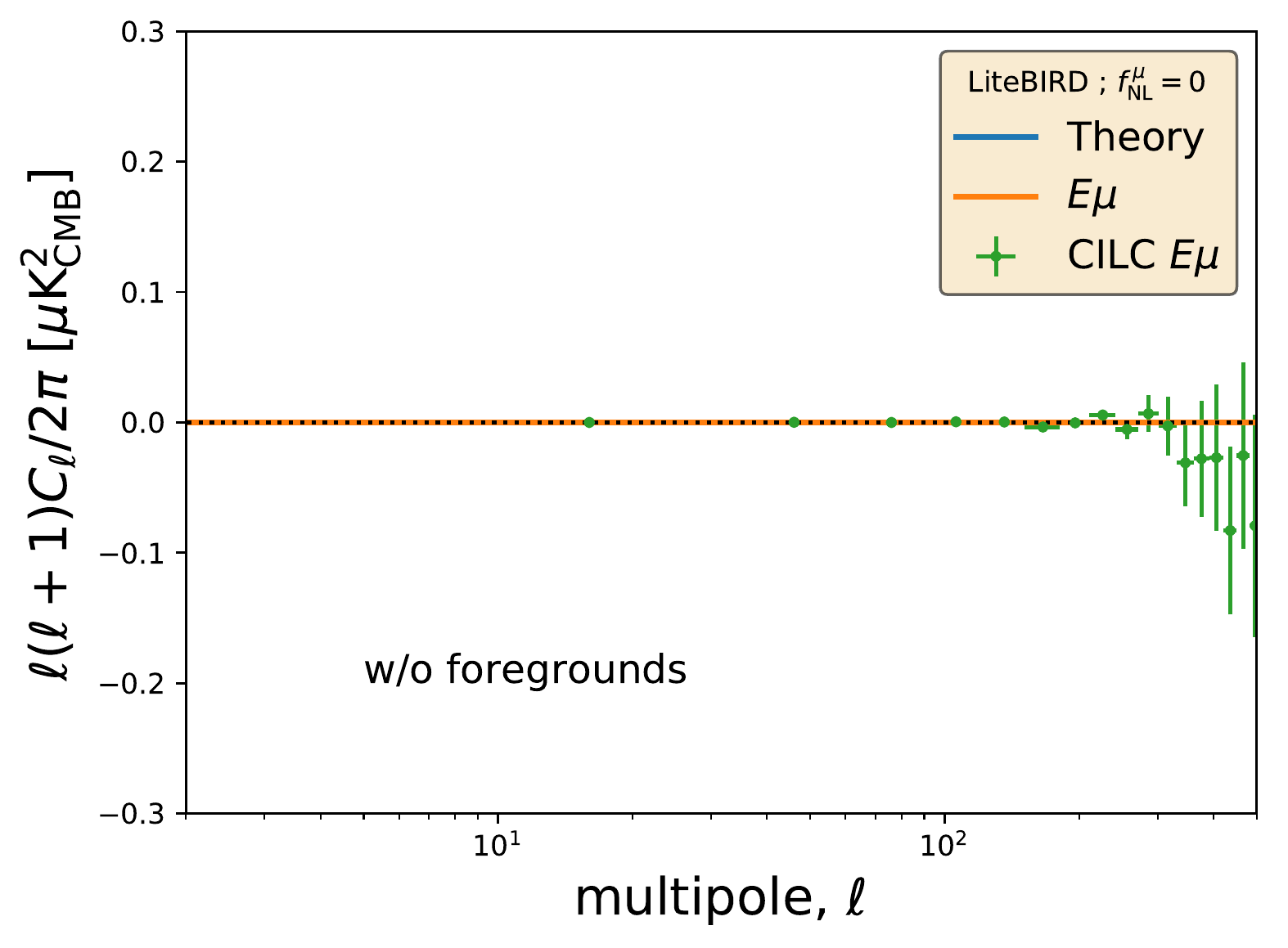}~
    \\[1.5mm]
     \includegraphics[width=\columnwidth]{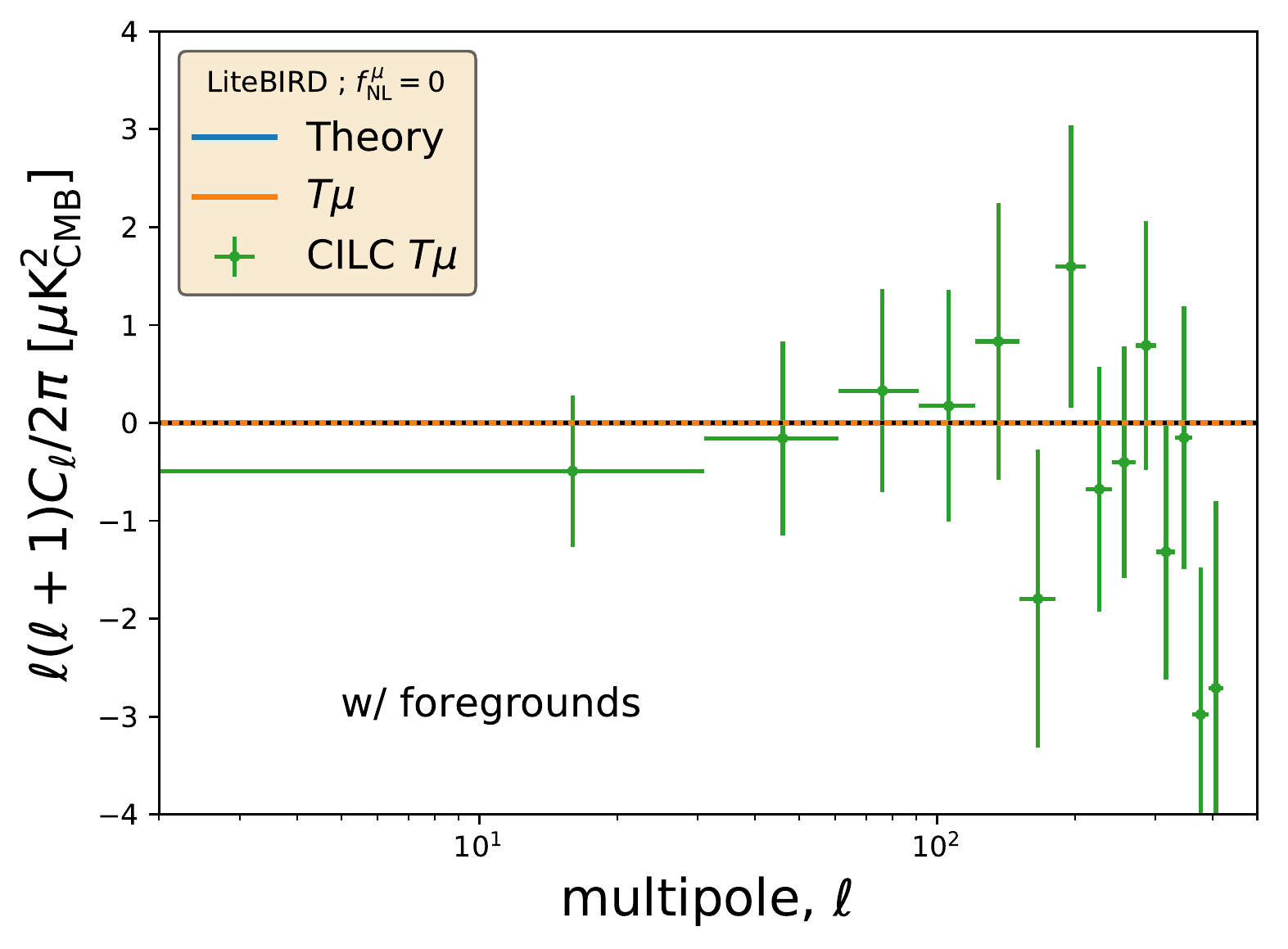}~
     \includegraphics[width=\columnwidth]{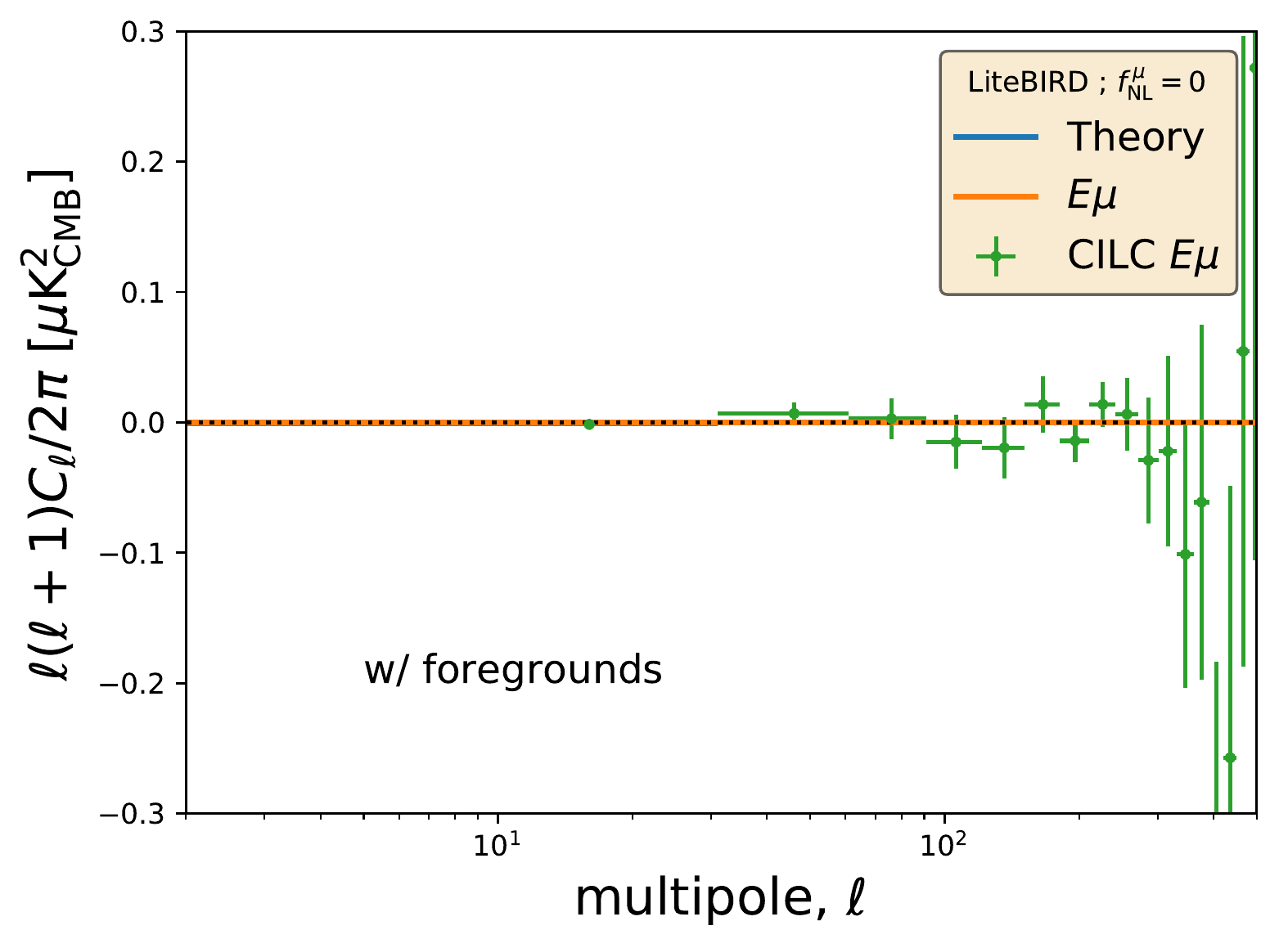}~
 \end{center}
\caption{Null-tests $(f_{\rm NL}^{\mu}=0)$, without foregrounds (\textit{top panels}) and with foregrounds (\textit{bottom panels}).}
\label{Fig:cross_fnl0}
\end{figure*}

In Fig.~\ref{Fig:cross_fnl0}, we performed a null test by running our component separation method on sky simulations in which $f_{\rm NL}^{\mu}=0$, i.e. in the absence of any anisotropic $\mu$-distortion signal in the sky. The upper row shows results in the absence of foregrounds (i.e. only CMB and noise), while the bottom row shows results with foregrounds. In all cases, the CILC reconstructions of $C_\ell^{\,\widehat{\mu}\, \widehat{T}}$ and $C_\ell^{\,\widehat{\mu}\, \widehat{E}}$ are consistent with zero, meaning that CILC would not lead to false detections of the anisotropic $\mu$-distortions.

The most important result coming out of Figs.~\ref{Fig:nilc_vs_cilc}, Fig.~\ref{Fig:cross}, and Fig.~\ref{Fig:cross_fnl0} is the better recovery of $C_\ell^{\,\mu E}$, with lower uncertainty, compared to $C_\ell^{\,\mu T}$. Although the $\mu E$ correlation signal is weaker than $\mu T$, clearly we get better sensitivity to $\mu E$ than to $\mu T$ after foreground cleaning and component separation. As a result, $C_\ell^{\,\widehat{\mu}\, \widehat{E}}$ will provide more constraining power on $f_{\rm NL}^{\mu}$ than $C_\ell^{\,\widehat{\mu}\, \widehat{T}}$, as we will show in Sect.~\ref{subsec:fnl}. 
There are several reasons for $C_\ell^{\,\mu E}$ to be a more sensitive observable than $C_\ell^{\,\mu T}$: \textit{(i)} While the CMB $E$-mode polarization signal is weaker than the CMB temperature signal, it has a higher degree of correlation with $\mu$-distortion anisotropies, as we showed in Fig.~\ref{Fig:pearson}. \textit{(ii)} Polarization foregrounds are fewer and weaker than temperature foregrounds. \textit{(iii)} Polarization channels are immune from $\mu$-distortion and CMB temperature anisotropies, so that extra nulling constraints in the ILC are no longer needed for polarization and we do not pay the noise penalty that temperature maps get from these additional constraints. \textit{(iv)} The instrumental noise in polarization channels is uncorrelated to the noise in temperature channels, hence $C_\ell^{\,\mu E}$ do not suffer from noise auto-correlation bias, unlike $C_\ell^{\,\mu T}$.\footnote{While we could perform Jackknife on $C_\ell^{\,\mu T}$ by using different data splits for the $\mu$-distortion and CMB temperature maps to get rid of the noise auto-correlation bias, the noise sample variance would still be increased by a factor of 2 because of the data splitting. In contrast, data splitting is not needed for $\mu E$ cross-correlations since noise is uncorrelated between temperature and polarization channels.}

%%%%%%%%%%%%%%%%%%%%%%%%%%%%%%%%%%%%%%%%%%%%%%%%%%%%%%
%%%%%%%%%%%%%%%%%%%%%%%%%%%%%%%%%%%%%%%%%%%%%%%%%%%%%%

\subsection{Constraints on small-scale primordial non-Gaussianity}
\label{subsec:fnl}

From the recovered cross-power spectra $\widehat{C}_\ell^{\,\mu T}$ and $\widehat{C}_\ell^{\,\mu E}$, which have linear dependence on the $f_{\rm NL}^{\mu}$ parameter (Eq.~\ref{eq:xmu}), we derive the maximum likelihood estimate and variance (see appendix \ref{sec:A_MLE} for a brief review) for $f_{\rm NL}^{\mu}(k\simeq 740\,{\rm Mpc}^{-1})$ using three different combinations: \textit{(i)} $\widehat{C}_\ell^{\,\mu T}$ only,  \textit{(ii)} $\widehat{C}_\ell^{\,\mu E}$ only, and \textit{(iii)} joint $\widehat{C}_\ell^{\,\mu T}$ and $\widehat{C}_\ell^{\,\mu E}$.

When exploiting either $T$ or $E$ independently, the maximum-likelihood estimate of $f_{\rm NL}^{\mu}$ is computed as
\begin{equation}\label{eq:mle_est}
\widehat{f}_{\rm NL}^{\,\mu} = \frac{\sum_{{\ell_b}} \widehat{C}_{\ell_b}^{\,\mu X} C_{\ell_b}^{\,\mu X}(f_{\rm NL}^{\mu}=1)/\left(\sigma_{\ell_b}^{\,\mu X}\right)^2}{\sum_{{\ell_b}} \left[C_{\ell_b}^{\,\mu X}(f_{\rm NL}=1)\right]^2/\left(\sigma_{\ell_b}^{\,\mu X}\right)^2}\,,
\end{equation}
where $\sigma_{\ell_b}^{\,\mu X}$ is given in Eq.~\eqref{eq:error}, while the corresponding $1\sigma$-uncertainty on $f_{\rm NL}^{\mu}$ is given by
\begin{equation}\label{eq:mle_sd}
\sigma\left(\widehat{f}_{\rm NL}^{\,\mu} \right)= \left[ \sum_{{\ell_b}} \left({C_{\ell_b}^{\,\mu X}\left(f_{\rm NL}^{\mu}=1\right)\over \sigma_{\ell_b}^{\,\mu X}} \right)^2\, \right]^{-1/2}\,,
\end{equation}
where $X$ stands for either $T$ or $E$, and ${C_\ell^{\,\mu X}\left(f_{\rm NL}^{\mu}=1\right)}$ is the theoretical cross-power spectrum for $f_{\rm NL}^{\mu}=1$. 

When jointly exploiting both $T$ and $E$ observables, the estimator and uncertainty are computed as
\begin{equation}
    \widehat{f}_{\rm NL}^{\,\mu}
    =
    \frac{
        \sum_{\ell_b}
            \widehat{\vec{X}}_{\ell_b}\cdot
            \mathbb{C}_{\ell_b}^{-1}
            \vec{X_{\ell_b}}
    }{
        \sum_{\ell_b}
            \vec{X_{\ell_b}}\cdot
            \mathbb{C}_{\ell_b}^{-1}
            \vec{X_{\ell_b}}
    }\,,
\end{equation}
and
\begin{equation}
    \sigma
    \left(
        \widehat{f}_\text{NL}^{\,\mu}
    \right)
    =
    \left[
        \sum_{\ell_b}
            \vec{X_{\ell_b}}\cdot
            \mathbb{C}_{\ell_b}^{-1}
            \vec{X_{\ell_b}}
    \right]^{-1/2}\,,
\end{equation}
where
\begin{equation}
\begin{split}
    \vec{X_{\ell_b}} 
    = &\ 
    \left(
    C^{\,\mu T}_{\ell_b}\left(f_{\rm NL}^{\mu}=1\right),
    C^{\,\mu E}_{\ell_b}\left(f_{\rm NL}^{\mu}=1\right)
    \right)^T\,,
\\
    \widehat{\vec{X}}_{\ell_b} 
    = &\ 
    \left(
    \widehat{C}^{\,\mu T}_{\ell_b},
    \widehat{C}^{\,\mu E}_{\ell_b}
    \right)^T\,,
\\
     \mathbb{C}_{\ell_b}
    = &\ 
    \begin{pmatrix}
        \frac{\left(\widehat{C}_{\ell_b}^{\,\mu T}\right)^2 + \widehat{C}_{\ell_b}^{\,\mu\mu}\widehat{C}_{\ell_b}^{\,TT}}{\left(2\ell_b+1\right)f_{\rm sky}\Delta\ell}  &
        \frac{\widehat{C}_{\ell_b}^{\,\mu T}\widehat{C}_{\ell_b}^{\,\mu E} + \widehat{C}_{\ell_b}^{\,\mu\mu}\widehat{C}_{\ell_b}^{\,TE}}{\left(2\ell_b+1\right)f_{\rm sky}\Delta\ell}  \\
        \frac{\widehat{C}_{\ell_b}^{\,\mu T}\widehat{C}_{\ell_b}^{\,\mu E} + \widehat{C}_{\ell_b}^{\,\mu\mu}\widehat{C}_{\ell_b}^{\,TE}}{\left(2\ell_b+1\right)f_{\rm sky}\Delta\ell}  &
        \frac{\left(\widehat{C}_{\ell_b}^{\,\mu E}\right)^2 + \widehat{C}_{\ell_b}^{\,\mu\mu}\widehat{C}_{\ell_b}^{\,EE}}{\left(2\ell_b+1\right)f_{\rm sky}\Delta\ell}
    \end{pmatrix}\,.
\end{split}
\end{equation}
These are the two-dimensional generalisation of Eq.~\eqref{eq:mle_sd} and \eqref{eq:mle_est}.

%%%%%%%%%%%%%%%%%%%%%%%%%%%%%%%%%%%%%%%%%%%%%%%%%%%%%%
\begin{table*}
	\centering
	\caption{Forecasts on $f_{\rm NL}^{\mu}(k\simeq 740\,{\rm Mpc}^{-1})$ for \textit{LiteBIRD} after foreground cleaning.}
	\label{tab:fnl}
	\begin{tabular}{lcccccc} % four columns, alignment for each
		\toprule
		$f_{\rm NL}^{\mu}$ (fiducial)  & $10^5$ & $10^4$ & $4500$  & $0$ & $4500$ & $0$\\
		  &  &  &  &  &(w/o foregrounds) &(w/o foregrounds)\\
		\midrule 
		 $\mu \times T$  & $\left(0.97\pm 0.02\right)\times 10^5$ & $\left(1.05\pm 0.12\right)\times 10^4$ & $5264 \pm 1286$ & $951 \pm 1286$ & $4348 \pm 152$  & $8.2 \pm 103$ \\
		                & [$50\sigma$]  & [$8\sigma$]   & [$3.5\sigma$]  & - & [$30\sigma$]  & - \\
		 $\mu \times E$ & $\left(0.96\pm 0.01\right)\times 10^5$ & $\left(0.91\pm 0.11\right)\times 10^4$ & $3779 \pm 1089$ &  $-534 \pm 1084$ & $4366 \pm 108$   & $0.9 \pm 76$ \\
		               & [$100\sigma$]  &  [$9\sigma$] & [$4\sigma$]  & - & [$42\sigma$] & - \\
		 \midrule     
		 $\mu \times T, E$  (joint) & $\left(0.97\pm 0.01\right)\times 10^5$ & $\left(0.97\pm 0.08\right)\times 10^4$ & $4425 \pm 827$ & $95 \pm 824$ & $4329 \pm 90$ & $-2.8 \pm 62$\\
		               & [100$\sigma$]  & [11$\sigma$]  & [5$\sigma$] & - & [48$\sigma$]   & - \\
		\bottomrule
	\end{tabular}
\end{table*}
%%%%%%%%%%%%%%%%%%%%%%%%%%%%%%%%%%%%%%%%%%%%%%%%%%%%%%

Our Fisher forecasts on $f_{\rm NL}^{\mu}(k\simeq 740\,{\rm Mpc}^{-1})$ derived from the recovered $\widehat{C}_\ell^{\,\mu T}$ and $\widehat{C}_\ell^{\,\mu E}$ after foreground cleaning are listed in Table~\ref{tab:fnl}, from which we can draw several conclusions. First, as it was already stressed in \cite{Remazeilles2018}, astrophysical foregrounds degrade the sensitivity to $f_{\rm NL}^{\mu}$ by about one order of magnitude for any combination of observables. Second, the $\mu E$ observable provides more constraining power on $f_{\rm NL}^{\mu}$ than the $\mu T$ observable, effectively increasing the detection significance of $f_{\rm NL}^{\mu}=4500$ by at least 14\% in the presence of foregrounds. Third, the joint combination of temperature and polarization adds even more leverage to the detection of anisotropic $\mu$-distortions and small-scale primordial non-Gaussianity, with further increase of the detection significance of $f_{\rm NL}^{\mu}=4500$ by about 40\% as compared to $\mu T$ in the presence of foregrounds. Fourth, with a joint analysis of $\mu T$ and $\mu E$ correlations, \textit{LiteBIRD} would detect $f_{\rm NL}^{\mu}(k\simeq 740\,{\rm Mpc}^{-1})=4500$ at $5\sigma$ significance after foreground cleaning. 
Finally, the smallest uncertainty on $f_{\rm NL}^{\mu}$ that \textit{LiteBIRD} would achieve from the joint combination of $\mu T$ and $\mu E$ observables is about $\sigma(f_{\rm NL}^{\mu}=0) \simeq 800$ after foreground removal.

Aside from the baseline results in Table~\ref{tab:fnl}, we explored two directions of possible optimization of the analysis:

\noindent{\bf Moment expansion of the foregrounds:} We investigated adding extra nulling constraints to the CILC filter Eqs.~\eqref{eq:cond_cilc}-\eqref{eq:mu_map} in order to deproject moments of the foreground emission in addition to the CMB temperature, i.e. by implementing the cMILC methodology \citep{Remazeilles2021} outlined in the end of Sect.~\ref{subsec:cilc}. Extra constraints to null out foreground moments effectively reduce residual foreground biases in the recovered $\mu$-map at the expense of overall variance increase because of larger parameter space for the same of set of frequency channels. Since our current results on the recovered $\mu T$, $\mu E$ and $f_{\rm NL}^\mu$ obtained with the baseline CILC approach are actually unbiased within one standard deviation, there is no gain in adding extra nulling constraints on foreground moments in our case as it only increases current uncertainties on $f_{\rm NL}^\mu$.

\vspace{2mm}
\noindent{\bf Extra frequency channels:} We also explored the impact on $f_{\rm NL}^\mu$ forecasts of adding external low- or high-frequency channels to the baseline \textit{LiteBIRD} configuration in order to assess which part of the frequency spectrum adds more leverage. We first considered three extra low-frequency channels at $10$, $20$ and $30$\,GHz from a futuristic ground-based full-sky survey. The sensitivity for these extra low-frequency channels was scaled as $\sigma_{100}\times\left(\nu / 100\, {\rm GHz}\right)^{-3}$, where $\sigma_{100}$ is the \textit{LiteBIRD} sensitivity at $100$\,GHz (Table~\ref{tab:litebird}), thus following the frequency scaling of the  synchrotron emission in order to keep a constant signal-to-noise ratio across the channels. The beam FWHM for these extra low-frequency channels was computed assuming a scaling of $20'\times\left(\nu / 10\, {\rm GHz}\right)^{-1}$. As an alternative, we also considered having three extra high-frequency channels at $500$, $650$ and $800$\, GHz for which beam FWHMs and sensitivities were obtained by linear extrapolation of the current \textit{LiteBIRD} beam and sensitivity curves.

Adding extra high-frequency channels to \textit{LiteBIRD} was found to reduce the uncertainty $\sigma(f_{\rm NL}^\mu)$ by 6\% for $\mu T$ and $7\%$ for $\mu E$, while adding extra low-frequency channels reduces the uncertainties by 9\% for $\mu T$ and $10\%$ for $\mu E$. Therefore, low-frequency channels actually add a bit more leverage than high-frequency channels for $\mu$-distortion anisotropies. This important outcome is somewhat consistent with our finding in Fig.~\ref{Fig:contributions} that low-frequency Galactic free-free emission is more damaging as a foreground than dust is at multipoles $\ell > 30$. Since $\ell \simeq 200$ (resp. $\ell \simeq 400$) is where the dominant peak of the $\mu T$ (resp. $\mu E$) cross-correlation signal lies (see e.g. Fig.~\ref{Fig:cross}),  this multipole range provides more constraining power on $f_{\rm NL}^\mu$ than low multipoles $\ell < 30$ while it also coincides with the range where free-free is the most limiting factor. Using external low-frequency channels in conjunction with \emph{LiteBIRD} to further clean low-frequency foregrounds is thus more helpful for $\mu$-distortion anisotropies.

\section{Conclusions}
\label{sec:conclusions}

We investigated the capability of a future CMB satellite imager like \textit{LiteBIRD} to detect $\mu$-type spectral distortion anisotropies in the presence of foregrounds through cross-correlations with CMB temperature and $E$-mode polarization, thereby testing the ability to constrain primordial non-Gaussianity at small scales $k\simeq 740\,{\rm Mpc^{-1}}$.

First, in the ideal case of absence of foregrounds, or perfect foreground cleaning, \textit{LiteBIRD} would allow to detect ${f_{\rm NL}^{\mu}(k\simeq 740\,{\rm Mpc}^{-1})=4500}$ at about $50\sigma$ significance and achieve a minimum uncertainty of about $\sigma(f_{\rm NL}^{\mu}=0) \simeq 60$ at 68\% CL by combining $\mu T$ and $\mu E$ observables (Table~\ref{tab:fnl}).
However, using comprehensive sky simulations (Sect.~\ref{sec:sky_sim}), \textit{LiteBIRD} instrument specifications (Table~\ref{tab:litebird}) and a tailored component separation method (Sect.~\ref{subsec:cilc}), we performed the reconstruction of both $\mu T$ and $\mu E$ cross-power spectra in the presence of foregrounds (Figs.~\ref{Fig:nilc_vs_cilc}-\ref{Fig:cross}), showing that astrophysical foregrounds degrade the sensitivity to the inferred value of $f^\mu_{\rm NL}(\simeq 740\,{\rm Mpc^{-1}})$ by about a factor of 10 (Table~\ref{tab:fnl}).

We found that the main degradation factor to measuring $\mu T$ and $\mu E$ cross-power spectra arises from residual Galactic foreground contamination not in CMB fields but in the reconstructed $\mu$-distortion map (Fig.~\ref{Fig:snr}), in particular from thermal dust at $\ell < 20$ and free-free emission at $\ell > 30$ (Fig.~\ref{Fig:contributions}), while the CMB temperature and CMB $E$-mode polarization maps are signal-dominated at all multipoles for \textit{LiteBIRD}. 
The effective noise curve for $\mu$-distortion anisotropies that we computed in Sect.~\ref{sec:clarification} for \textit{LiteBIRD} in the presence of foregrounds can be used as benchmark for future studies.

We emphasized the importance of \textit{constrained} ILC approaches (Sect.~\ref{subsec:cilc}) for component separation to simultaneously null CMB temperature anisotropies in the reconstructed $\mu$-distortion map and $\mu$-distortion anisotropies in the CMB temperature map, and thereby get rid of spurious correlation biases on $C_\ell^{\,\mu T}$ and $C_\ell^{\,\mu E}$ (Fig.~\ref{Fig:nilc_vs_cilc}).

We showed that $\mu E$ cross-correlations provide slightly more constraining power than $\mu T$ cross-correlations on $f_{\rm NL}^{\mu}$ in the presence of foregrounds (Fig.~\ref{Fig:pearson}, Figs.~\ref{Fig:nilc_vs_cilc}-\ref{Fig:cross}-\ref{Fig:cross_fnl0}, and Table~\ref{tab:fnl}), while the joint combination of $\mu T$ and $\mu E$ observables adds even further leverage to the detection of $f_{\rm NL}^{\mu}$.

By combining both temperature and polarization, \textit{LiteBIRD} will be able to detect $f_{\rm NL}^{\mu}(k\simeq 740\,{\rm Mpc}^{-1})=4500$ at $5\sigma$ significance, and achieve a minimum uncertainty of about $\sigma(f_{\rm NL}^{\mu}=0) \simeq 800$ at 68\% CL after foreground removal (Table~\ref{tab:fnl}), 
a large value which is allowed by multi-field inflation models at large wavenumbers whilst still consistent with \emph{Planck} CMB constraints at  $k\simeq 0.05\,{\rm Mpc^{-1}}$ in the case of very mild scale-dependence of $f_{\rm NL}$.

We anticipate even higher detection significance for non-Bunch-Davies (NBD) initial condition models of inflation \citep{Ganc:2012ae}, since the signal-to-noise ratio in such models is expected to be larger than that of multi-field inflation models which we considered here. The investigation of NBD models is left for future work.

Given the dependence of $\mu T$ and $\mu E$ cross-correlation signals on the value of the monopole distortion $\langle \mu \rangle$, differential measurements of $\mu$-distortion anisotropies from an imager would still benefit from absolute measurement of the monopole distortion by a spectrophotometer like \textit{PIXIE} \citep{Kogut:2011xw} and the ESA Voyage 2050 vision \citep{Chluba:2019nxa} instead of relying on theoretical $\Lambda{\rm CDM}$ estimates  \citep[e.g.,][]{Chluba:2016bvg}. We also emphasize that precise inter-channel cross-calibration is needed for future imagers to avoid biasing the reconstruction of the $\mu T$ and $\mu E$ cross-correlation signals \citep[see][]{Ganc:2012ae, Remazeilles2018}. Nevertheless, studies of SDA and their correlations with PA provide a powerful new window into the physics of the early Universe.

\section*{Acknowledgements}

This work was supported by the ERC Consolidator Grant {\it CMBSPEC} (No.~725456) as part of the European Union's Horizon 2020 research and innovation program.
JC was also supported by the Royal Society as a Royal Society URF at the University of Manchester, UK.
We thank Eiichiro Komatsu for very interesting suggestions and discussions about this project.
We also thank David Zegeye and Tom Crawford for independent cross-check of our earlier predictions on $\sigma(f_{\rm NL})$ in \citet{Remazeilles2018} and for having confirmed the omission of the binning factor correction in our earlier forecasts.
Some of the results in this paper have been derived using the \texttt{HEALPix} package \citep{Gorski2005} and the \texttt{PSM} package \citep{Delabrouille2013}.

%%%%%%%%%%%%%%%%%%%%%%%%%%%%%%%%%%%%%%%%%%%%%%%%%%
\section*{Data Availability}
 
The simulations underlying this article will be shared on reasonable request to the authors.

\bibliographystyle{mnras}
\bibliography{bibliografia} 

%%%%%%%%%%%%%%%%%%%%%%%%%%%%%%%%%%%%%%%%%%%%%%%%%%

%%%%%%%%%%%%%%%%% APPENDICES %%%%%%%%%%%%%%%%%%%%%

\appendix

\section{Maximum likelihood estimator}
\label{sec:A_MLE}

Assuming a Gaussian distribution for the reconstructed power spectrum $\widehat{C}_\ell^{\,\mu T}$, the Gaussian likelihood reads
\begin{equation}\label{eq:lkl}
-2\ln \mathcal{L} = \sum_\ell {\left(\widehat{C}_\ell^{\,\mu T} - C_\ell^{\,\mu T} \left( f_{\rm NL}\right)\right)^2 \over \widehat{\sigma}_\ell^2} + {\rm constant}\,,
\end{equation}
where the theory spectrum  $C_\ell^{\,\mu T}\left( f_{\rm NL}\right)$ of a certain $f_{\rm NL}$ value is the expectation value of $\widehat{C}_\ell^{\,\mu T}$
\begin{equation}\label{eq:mean}
\langle \widehat{C}_\ell^{\,\mu T}\rangle = C_\ell^{\,\mu T} \left( f_{\rm NL}\right)
\end{equation}
and $\widehat{\sigma}_\ell^2$ is the variance of $\widehat{C}_\ell^{\,\mu T}$:
\begin{equation}\label{eq:variance}
\widehat{\sigma}_\ell^2 \equiv {\rm var}\left(\widehat{C}_\ell^{\,\mu T}\right)= {\left(\widehat{C}_\ell^{\,\mu T}\right)^2 + \widehat{C}_\ell^{\mu \mu} \widehat{C}_\ell^{\,TT}\over 2\ell +1}\,.
\end{equation}

The likelihood Eq.~\eqref{eq:lkl} can be Taylor expanded up to second-order around some pivot $f_{\rm NL}^*$ as:
\begin{align}\label{eq:taylor}
-2\ln \mathcal{L} &\simeq -2\ln \mathcal{L} \left (f_{\rm NL}^* \right) \cr
                           & + \left( f_{\rm NL} - f_{\rm NL}^*\right){\partial \left( -2\ln \mathcal{L}\right)\over \partial f_{\rm NL} }\vert_{f_{\rm NL} = f_{\rm NL}^*} \cr
                           & + {1\over 2}\left( f_{\rm NL} - f_{\rm NL}^*\right)^2{\partial ^2\left( -2\ln \mathcal{L}\right)\over \partial f_{\rm NL} ^2}\vert_{f_{\rm NL} = f_{\rm NL}^*}\,.
\end{align}

The maximum likelihood estimate (MLE) is the value $f_{\rm NL}^*=\widehat{f}_{\rm NL}$  which cancels the first derivative of the likelihood:
\begin{equation}\label{eq:mle}
{\partial \left( -2\ln \mathcal{L}\right)\over \partial f_{\rm NL} }\vert_{f_{\rm NL} = \widehat{f}_{\rm NL}} = 0\,,
\end{equation}
so that the likelihood Eq.~\eqref{eq:taylor} reduces to
\begin{equation}\label{eq:lkl2}
-2\ln \mathcal{L} \simeq \sum_\ell {\left( f_{\rm NL} - \widehat{f}_{\rm NL}\right)^2 \over \sigma^2\left(\widehat{f}_{\rm NL}\right)} + {\rm constant}\,,
\end{equation}
where the error on $\widehat{f}_{\rm NL}$ is thus given by
\begin{equation}\label{eq:variance_fnl}
\sigma\left(\widehat{f}_{\rm NL}\right) = \left({1\over 2}{\partial ^2\left( -2\ln \mathcal{L}\right)\over \partial f_{\rm NL} ^2}\vert_{f_{\rm NL} = \widehat{f}_{\rm NL}}\right)^{-1/2}\,.
\end{equation}

Equations~\eqref{eq:mle} and ~\eqref{eq:variance_fnl} allow us to derive analytic expressions of the MLE $\widehat{f}_{\rm NL}$ and the error $\sigma\left(\widehat{f}_{\rm NL}\right)$, respectively, by using the likelihood expression Eq.~\eqref{eq:lkl}. 
\newline

First, we have the first derivative of the likelihood Eq.~\eqref{eq:lkl} giving
\begin{equation}\label{eq:first}
{\partial \left( -2\ln \mathcal{L}\right) \over \partial f_{\rm NL} }  = -2 \sum_\ell {\partial C_\ell^{\,\mu T} \over \partial f_{\rm NL} } {\left(\widehat{C}_\ell^{\,\mu T} - C_\ell^{\,\mu T} \left( f_{\rm NL}\right)\right)\over \widehat{\sigma}_\ell^2}\,,
\end{equation}
so that the MLE equation~\eqref{eq:mle} implies that
\begin{align}\label{eq:mle2}
 -2 \sum_\ell {\partial C_\ell^{\,\mu T} \over \partial f_{\rm NL} } {\left(\widehat{C}_\ell^{\,\mu T} - C_\ell^{\,\mu T} \left( \widehat{f}_{\rm NL}\right)\right)\over \widehat{\sigma}_\ell^2} &= 0\,,\cr
-2 \sum_\ell {C_\ell^{\,\mu T}\left(f_{\rm NL}=1\right)\over \widehat{\sigma}_\ell^2}\left(\widehat{C}_\ell^{\,\mu T} - \widehat{f}_{\rm NL}\,C_\ell^{\,\mu T} \left( f_{\rm NL}=1\right)\right) &= 0\,,
\end{align}
where in the second line we used that $C_\ell^{\,\mu T} \left( f_{\rm NL}\right) = f_{\rm NL}\,C_\ell^{\,\mu T} \left( f_{\rm NL}=1\right)$. Equation~\eqref{eq:mle2} then gives
\begin{equation}\label{eq:mle3}
\widehat{f}_{\rm NL} \sum_\ell {\left( C_\ell^{\,\mu T} \left( f_{\rm NL}=1\right)\right)^2\over \widehat{\sigma}_\ell^2} = \sum_\ell \widehat{C}_\ell^{\,\mu T} { C_\ell^{\,\mu T} \left( f_{\rm NL}=1\right) \over \widehat{\sigma}_\ell^2}.
\end{equation}
Therefore,
\begin{equation}\label{eq:themle}
\widehat{f}_{\rm NL} = {\sum_\ell \widehat{C}_\ell^{\,\mu T} { C_\ell^{\,\mu T} \left( f_{\rm NL}=1\right) \over \widehat{\sigma}_\ell^2}\over \sum_\ell {\left( C_\ell^{\,\mu T} \left( f_{\rm NL}=1\right)\right)^2\over \widehat{\sigma}_\ell^2} }\,,
\end{equation}
which is the formula Eq.~\eqref{eq:mle_est} in our paper.

Second, we have the second derivative of the likelihood Eq.~\eqref{eq:lkl} giving
\begin{equation}\label{eq:second}
{\partial^2 \left( -2\ln \mathcal{L}\right) \over \partial f_{\rm NL}^2 }  = 2 \sum_\ell {\left( C_\ell^{\,\mu T} \left( f_{\rm NL}=1\right)\right)^2\over \widehat{\sigma}_\ell^2}\,.
\end{equation}
Therefore, the error on $\widehat{f}_{\rm NL}$ is 
\begin{equation}\label{eq:thevariance}
\sigma\left(\widehat{f}_{\rm NL}\right) = \left( \sum_\ell {\left( C_\ell^{\,\mu T} \left( f_{\rm NL}=1\right)\right)^2\over \widehat{\sigma}_\ell^2} \right)^{-1/2}\,,
\end{equation}
which is the formula Eq.~\eqref{eq:mle_sd} in our paper.
\newline

Note that the MLE Eq.~\eqref{eq:themle}/\eqref{eq:mle_est} is nothing else than the inverse variance weighted mean of $\widehat{C}_\ell^{\,\mu T} / C_\ell^{\,\mu T} \left( f_{\rm NL}=1\right)$. Indeed, by using the variance formula~\eqref{eq:thevariance}/\eqref{eq:mle_sd} the MLE  Eq.~\eqref{eq:themle}/\eqref{eq:mle_est} can be rewritten as
\begin{equation}\label{eq:ps}
\widehat{f}_{\rm NL} = {\sum_\ell  \left({\widehat{C}_\ell^{\,\mu T} \over C_\ell^{\,\mu T} \left( f_{\rm NL}=1\right)}\right)/\sigma^2\left(\widehat{f}_{\rm NL}\right) \over \sum_\ell 1 / \sigma^2\left(\widehat{f}_{\rm NL}\right)}\,.
\end{equation}
While these equations were explicitly written for the $\mu T$ cross-correlation, they extend trivially to the $\mu E$ cross-correlation simply taking $T \rightarrow E$.

A further step can be taken by jointly analysing the two cross-correlation signals.
Taking into account that the $T$ and $E$ fields are also correlated, the likelihood --- under the same assumption of Gaussianity used before --- is
\begin{equation}
    -2 \ln \mathcal{L} 
    = 
    \sum_{\ell}
    (\widehat{\vec{X}}_{\ell} - f_{\rm NL}\,\vec{X_{\ell}})
    \cdot
    \mathbb{C}_{\ell}^{-1}
    (\widehat{\vec{X}}_{\ell} - f_{\rm NL}\,\vec{X_{\ell}})
    + {\rm constant}\,,
\end{equation}
where we used the quantities (already defined in the main text)
\begin{equation*}
\begin{split}
    \vec{X_{\ell}} 
    = &\ 
    \left(
    C^{\,\mu T}_{\ell}\left(f_{\rm NL}=1\right),
    C^{\,\mu E}_{\ell}\left(f_{\rm NL}=1\right)
    \right)^T\,,
\\
    \widehat{\vec{X}}_{\ell} 
    = &\ 
    \left(
    \widehat{C}^{\,\mu T}_{\ell},
    \widehat{C}^{\,\mu E}_{\ell}
    \right)^T\,,
\\
     \mathbb{C}_{\ell}
    = &\ 
    \begin{pmatrix}
        \frac{\left(\widehat{C}_{\ell}^{\,\mu T}\right)^2 + \widehat{C}_{\ell}^{\,\mu\mu}\widehat{C}_{\ell}^{\,TT}}{2\ell+1}  &
        \frac{\widehat{C}_{\ell}^{\,\mu T}\widehat{C}_{\ell}^{\,\mu E} + \widehat{C}_{\ell}^{\,\mu\mu}\widehat{C}_{\ell}^{\,TE}}{2\ell+1}  \\
        \frac{\widehat{C}_{\ell}^{\,\mu T}\widehat{C}_{\ell}^{\,\mu E} + \widehat{C}_{\ell}^{\,\mu\mu}\widehat{C}_{\ell}^{\,TE}}{2\ell+1}  &
        \frac{\left(\widehat{C}_{\ell}^{\,\mu E}\right)^2 + \widehat{C}_{\ell}^{\,\mu\mu}\widehat{C}_{\ell}^{\,EE}}{2\ell+1}
    \end{pmatrix}\,.
\end{split}
\end{equation*}

Retracing the same steps, we thus find
\begin{equation}
    \widehat{f}_{\rm NL}
    =
    \frac{
        \sum_{\ell}
            \widehat{\vec{X}}_{\ell}\cdot
            \mathbb{C}_{\ell}^{-1}
            \vec{X_{\ell}}
    }{
        \sum_{\ell}
            \vec{X_{\ell}}\cdot
            \mathbb{C}_{\ell}^{-1}
            \vec{X_{\ell}}
    }\,,
\end{equation}
and
\begin{equation}
    \sigma
    \left(
        \widehat{f}_\text{NL}
    \right)
    =
    \left[
        \sum_{\ell}
            \vec{X_{\ell}}\cdot
            \mathbb{C}_{\ell}^{-1}
            \vec{X_{\ell}}
    \right]^{-1/2} .
\end{equation}
These formulas can be modified using $2\ell+1 \rightarrow \left(2\ell_b+1\right)f_{\rm sky}\,\Delta_\ell$ to account for partial sky coverage and multipole bins $\ell_b$ of width $\Delta\ell$.

%%%%%%%%%%%%%%%%%%%%%%%%%%%%%%%%%%%%%%%%%%%%%%%%%%

% Don't change these lines
\bsp	% typesetting comment
\label{lastpage}
\end{document}